\documentclass[12pt]{article}
\newcommand{\filename}{BH-cyclic\_arXiv\_v5}
\usepackage{authblk}
\usepackage{amsfonts}
\usepackage{amssymb}
\usepackage{amsmath}
\usepackage{amsthm}
\usepackage{framed}

\newtheorem{theorem}{Theorem}

\usepackage{enumitem}
\usepackage{geometry}
\usepackage{url}
\usepackage{mdframed}
\usepackage{mathtools}
\usepackage{algorithm}  
\usepackage{algorithmic}

\newcommand{\sco}{\sigma_\mathrm{co}} 
\newcommand{\snc}{\sigma_\mathrm{nc}} 
\newcommand{\sab}{\sigma_\mathrm{\alpha \beta}} 
\newcommand{\sch}{\sigma_\mathrm{ch}}

\newcommand{\sX}{\sigma^\mathrm{X}}

\newcommand{\Ez}{E_{(0/1)}}
\newcommand{\Ew}{E_{(\geq 1)}}
\newcommand{\Et}{E_{(\geq 2)}}
\newcommand{\Eew}{E_{(=1)}}

\newcommand{\Iz}{I_{(0/1)}}
\newcommand{\Iw}{I_{(\geq 1)}}
\newcommand{\It}{I_{(\geq 2)}}
\newcommand{\Iew}{I_{(=1)}}
\newcommand{\IVCs}{I_\mathrm{VC*}}

\newcommand{\Gac}{\Gamma_\mathrm{ac}} 
\newcommand{\Gacs}{\Gamma_\mathrm{ac,<}} 
\newcommand{\Gace}{\Gamma_\mathrm{ac,=}} 
\newcommand{\Gacl}{\Gamma_\mathrm{ac,>}}

\newcommand{\tGacex}{\widetilde{\Gamma}_\mathrm{ac}^\mathrm{ex}}  
  
\newcommand{\tGacC}{\widetilde{\Gamma}_\mathrm{ac}^\mathrm{C}}  
\newcommand{\tGacT}{\widetilde{\Gamma}_\mathrm{ac}^\mathrm{T}} 
\newcommand{\tGacF}{\widetilde{\Gamma}_\mathrm{ac}^\mathrm{F}}  
\newcommand{\tGacCT}{\widetilde{\Gamma}_\mathrm{ac}^\mathrm{CT}}  
\newcommand{\tGacTC}{\widetilde{\Gamma}_\mathrm{ac}^\mathrm{TC}}  
\newcommand{\tGacCF}{\widetilde{\Gamma}_\mathrm{ac}^\mathrm{CF}}  
\newcommand{\tGacTF}{\widetilde{\Gamma}_\mathrm{ac}^\mathrm{TF}}

\newcommand{\tLdgX}{\widetilde{\Lambda}_\mathrm{dg}^\mathrm{X}}   
\newcommand{\tLdgC}{\widetilde{\Lambda}_\mathrm{dg}^\mathrm{C}}   
\newcommand{\tLdgT}{\widetilde{\Lambda}_\mathrm{dg}^\mathrm{T}}   
\newcommand{\tLdgF}{\widetilde{\Lambda}_\mathrm{dg}^\mathrm{F}}     
\newcommand{\tLdgCnc}{\widetilde{\Lambda}_\mathrm{dg}^\mathrm{C,nc}}   
\newcommand{\tLdgTnc}{\widetilde{\Lambda}_\mathrm{dg}^\mathrm{T,nc}}   
\newcommand{\tLdgFnc}{\widetilde{\Lambda}_\mathrm{dg}^\mathrm{F,nc}}   
\newcommand{\tLdgXnc}{\widetilde{\Lambda}_\mathrm{dg}^\mathrm{X,nc}}

\newcommand{\tGecex}{\widetilde{\Gamma}_\mathrm{ec}^\mathrm{ex}}  
  
\newcommand{\tGecC}{\widetilde{\Gamma}_\mathrm{ec}^\mathrm{C}}  
\newcommand{\tGecT}{\widetilde{\Gamma}_\mathrm{ec}^\mathrm{T}} 
\newcommand{\tGecF}{\widetilde{\Gamma}_\mathrm{ec}^\mathrm{F}}  
\newcommand{\tGecCT}{\widetilde{\Gamma}_\mathrm{ec}^\mathrm{CT}}  
\newcommand{\tGecTC}{\widetilde{\Gamma}_\mathrm{ec}^\mathrm{TC}}  
\newcommand{\tGecCF}{\widetilde{\Gamma}_\mathrm{ec}^\mathrm{CF}}  
\newcommand{\tGecTF}{\widetilde{\Gamma}_\mathrm{ec}^\mathrm{TF}}

\newcommand{\typ}{\mathrm{t}}

\newcommand{\f}{\pmb{f}}
\newcommand{\w}{\pmb{w}}
\newcommand{\x}{\pmb{x}}

\newcommand{\z}{\pmb{z}}

\newcommand{\1}{\pmb{1}}

\newcommand{\tbc}{{\tt bc}}
\newcommand{\ta}{{\tt a}}
\newcommand{\tb}{{\tt b}}
\newcommand{\Ldg}{\Lambda_{\mathrm{dg}}}

\newcommand{\mUB}{m_{\mathrm{UB}}}

\newcommand{\dmax}{d_{\max}}
\newcommand{\val}{\mathrm{val}}

\newcommand{\inl}{\mathrm{inl}}
\newcommand{\en}{\mathrm{end}}
\newcommand{\pair}{\mathrm{pair}}

\newcommand{\code}{\mathrm{code}}

\newcommand{\Pt}{\mathcal{P}}
\newcommand{\F}{\mathcal{F}}
\newcommand{\Sb}{\mathcal{S}}
\newcommand{\Cr}{\mathrm{Cr}}  
\newcommand{\G}{\mathcal{G}}

\newcommand{\W}{\mathrm{W}}  
\newcommand{\T}{\mathcal{T}}

\newcommand{\h}{\mathrm{ht}}
\newcommand{\co}{\mathrm{co}}
\newcommand{\nc}{\mathrm{nc}}
\newcommand{\cs}{\mathrm{cs}}
\newcommand{\ch}{\mathrm{ch}}

\newcommand{\dg}{\mathrm{dg}}

\newcommand{\na}{\mathrm{na}}

\newcommand{\naXp}{\mathrm{na}_{\mathrm{X}(p)}}
\newcommand{\naC}{\mathrm{na}_\mathrm{C}}
\newcommand{\naT}{\mathrm{na}_\mathrm{T}}

\newcommand{\ecX}{\mathrm{ec}_\mathrm{X}}

\newcommand{\ecXp}{\mathrm{ec}_{\mathrm{X}(p)}}

\newcommand{\ecC}{\mathrm{ec}_\mathrm{C}}
\newcommand{\ecT}{\mathrm{ec}_\mathrm{T}}
\newcommand{\ecF}{\mathrm{ec}_\mathrm{F}}
\newcommand{\ecCT}{\mathrm{ec}_\mathrm{CT}}
\newcommand{\ecTC}{\mathrm{ec}_\mathrm{TC}}
\newcommand{\ecTF}{\mathrm{ec}_\mathrm{TF}}
\newcommand{\ecCF}{\mathrm{ec}_\mathrm{CF}}
 
\newcommand{\acX}{\mathrm{ac}_\mathrm{X}}

\newcommand{\acXp}{\mathrm{ac}_{\mathrm{X}(p)}}
\newcommand{\acC}{\mathrm{ac}_\mathrm{C}}
\newcommand{\acT}{\mathrm{ac}_\mathrm{T}}
\newcommand{\acF}{\mathrm{ac}_\mathrm{F}}
\newcommand{\acCT}{\mathrm{ac}_\mathrm{CT}}
\newcommand{\acTC}{\mathrm{ac}_\mathrm{TC}}
\newcommand{\acTF}{\mathrm{ac}_\mathrm{TF}}
\newcommand{\acCF}{\mathrm{ac}_\mathrm{CF}}

\newcommand{\bdX}{\mathrm{bd}_\mathrm{X}}

\newcommand{\bdXp}{\mathrm{bd}_{\mathrm{X}(p)}}
\newcommand{\bdC}{\mathrm{bd}_\mathrm{C}}
\newcommand{\bdT}{\mathrm{bd}_\mathrm{T}}
\newcommand{\bdF}{\mathrm{bd}_\mathrm{F}}
\newcommand{\bdCT}{\mathrm{bd}_\mathrm{CT}}
\newcommand{\bdTC}{\mathrm{bd}_\mathrm{TC}}
\newcommand{\bdTF}{\mathrm{bd}_\mathrm{TF}}
\newcommand{\bdCF}{\mathrm{bd}_\mathrm{CF}}

\newcommand{\ns}{\mathrm{ns}}

\newcommand{\dgXp}{\mathrm{dg}_{\mathrm{X}(p)}}

\newcommand{\dgC}{\mathrm{dg}_\mathrm{C}}
\newcommand{\dgT}{\mathrm{dg}_\mathrm{T}}

\newcommand{\ec}{\mathrm{ec}}
\newcommand{\ac}{\mathrm{ac}}

\newcommand{\bl}{\mathrm{bl}}

\newcommand{\bc}{\mathrm{bc}}
\newcommand{\bd}{\mathrm{bd}}
\newcommand{\bn}{\mathrm{bn}}
\newcommand{\br}{\mathrm{br}}
\newcommand{\bh}{\mathrm{bh}}

\newcommand{\UB}{\mathrm{UB}}
\newcommand{\LB}{\mathrm{LB}}

\newcommand{\inn}{\mathrm{in}}
\newcommand{\ex}{\mathrm{ex}}

\newcommand{\GC}{G_\mathrm{C}}

\newcommand{\mC}{m_\mathrm{C}}

\newcommand{\jF}{j^\mathrm{F}} 
\newcommand{\jX}{j^\mathrm{X}} 
\newcommand{\hC}{h^\mathrm{C}}
\newcommand{\hT}{h^\mathrm{T}} 
 
\newcommand{\hX}{h^\mathrm{X}}

\newcommand{\PT}{P_\mathrm{T}} 
\newcommand{\PF}{P_\mathrm{F}} 
 
\newcommand{\VF}{V_\mathrm{F}}
\newcommand{\VT}{V_\mathrm{T}}
\newcommand{\VC}{V_\mathrm{C}} 
 
\newcommand{\VX}{V_\mathrm{X}}

\newcommand{\ET}{E_\mathrm{T}}
\newcommand{\EC}{E_\mathrm{C}}

\newcommand{\EF}{E_\mathrm{F}}
\newcommand{\ECT}{E_\mathrm{CT}}
\newcommand{\ETC}{E_\mathrm{TC}}
\newcommand{\ETF}{E_\mathrm{TF}}

\newcommand{\ECF}{E_\mathrm{CF}}

\newcommand{\EX}{E_\mathrm{X}}
\newcommand{\nT}{n_\mathrm{T}}
\newcommand{\nC}{n_\mathrm{C}} 
 
\newcommand{\nX}{n_\mathrm{X}}

\newcommand{\nF}{n_\mathrm{F}}

\newcommand{\vT}{{v^\mathrm{T}}}
\newcommand{\vC}{{v^\mathrm{C}}} 
  
\newcommand{\vX}{{v^\mathrm{X}}}

\newcommand{\vF}{{v^\mathrm{F}}}

\newcommand{\eF}{{e^\mathrm{F}}}
\newcommand{\eT}{{e^\mathrm{T}}}
\newcommand{\eC}{{e^\mathrm{C}}} 
 
\newcommand{\eX}{{e^\mathrm{X}}}

\newcommand{\eCF}{{e^\mathrm{CF}}}

\newcommand{\eCT}{{e^\mathrm{CT}}}
\newcommand{\eTC}{{e^\mathrm{TC}}}
\newcommand{\eTF}{{e^\mathrm{TF}}}

\newcommand{\tT}{{t_\mathrm{T}}}
\newcommand{\tC}{{t_\mathrm{C}}} 
\newcommand{\tF}{{t_\mathrm{F}}} 
 
\newcommand{\tX}{{t_\mathrm{X}}}

\newcommand{\Cld}{\mathrm{Cld}}
\newcommand{\prt}{\mathrm{prt}}

\newcommand{\CldT}{\mathrm{Cld}_{\mathrm{T}}}
\newcommand{\CldC}{\mathrm{Cld}_{\mathrm{C}}} 
  
\newcommand{\CldX}{\mathrm{Cld}_{\mathrm{X}}}

\newcommand{\CldF}{\mathrm{Cld}_{\mathrm{F}}}

\newcommand{\Tprc}{\mathcal{T}_\mathrm{prc}}
\newcommand{\Pprc}{P_\mathrm{prc}}

\newcommand{\PprcX}{P_\mathrm{prc,X}}

\newcommand{\degCex}{{\deg_\mathrm{C}^\mathrm{ex}}}

\newcommand{\degF}{{\deg^\mathrm{F}}}
\newcommand{\degT}{{\deg^\mathrm{T}}}
\newcommand{\degC}{{\deg^\mathrm{C}}} 
 
\newcommand{\degX}{{\deg^\mathrm{X}}}

\newcommand{\degCT}{\deg_\mathrm{CT}}
\newcommand{\degTC}{\deg_\mathrm{TC}}

\newcommand{\degCTT}{\deg^\mathrm{CT}_\mathrm{T}}
\newcommand{\degTCT}{\deg^\mathrm{TC}_\mathrm{T}}
\newcommand{\degCFF}{\deg^\mathrm{CF}_\mathrm{F}}
\newcommand{\degTFF}{\deg^\mathrm{TF}_\mathrm{F}}

\newcommand{\tldgC}{{\widetilde{\deg}_\mathrm{C}} }

\newcommand{\cF}{{c_\mathrm{F}}}

\newcommand{\kC}{{k_\mathrm{C}}}

\newcommand{\chiF}{{\chi^\mathrm{F}}} 
\newcommand{\dclrF}{\delta_\mathrm{\chi}^\mathrm{F}}
\newcommand{\clrF}{\mathrm{clr}^{\mathrm{F}}}

\newcommand{\chiT}{{\chi^\mathrm{T}}}
\newcommand{\dclrT}{\delta_\mathrm{\chi}^\mathrm{T}}
\newcommand{\clrT}{\mathrm{clr}^{\mathrm{T}}}

\newcommand{\tail}{\mathrm{tail}} 
\newcommand{\hd}{\mathrm{head}} 

\newcommand{\tailF}{\mathrm{tail}^{\mathrm{F}}} 
\newcommand{\hdC}{\mathrm{head}^{\mathrm{C}}} 
\newcommand{\tailC}{\mathrm{tail}^{\mathrm{C}}}

\newcommand{\prtX}{\mathrm{prt}_{\mathrm{X}}}

\newcommand{\ddgF}{\delta_\mathrm{dg}^\mathrm{F}}
\newcommand{\ddgT}{\delta_\mathrm{dg}^\mathrm{T}}
\newcommand{\ddgC}{\delta_\mathrm{dg}^\mathrm{C}} 
 
\newcommand{\ddgX}{\delta_\mathrm{dg}^\mathrm{X}}

\newcommand{\bF}{\beta^\mathrm{F}}
\newcommand{\bT}{\beta^\mathrm{T}}
\newcommand{\bC}{\beta^\mathrm{C}} 
\newcommand{\bX}{\beta^\mathrm{X}}
  
\newcommand{\bCT}{\beta^\mathrm{CT}}
\newcommand{\bTC}{\beta^\mathrm{TC}} 
\newcommand{\bTF}{\beta^\mathrm{TF}} 
\newcommand{\bCF}{\beta^\mathrm{CF}}

\newcommand{\delb}{\delta_{\beta}}
\newcommand{\delbF}{\delta_{\beta}^\mathrm{F}}
\newcommand{\delbT}{\delta_{\beta}^\mathrm{T}}
\newcommand{\delbC}{\delta_{\beta}^\mathrm{C}}

\newcommand{\delbX}{\delta_{\beta}^\mathrm{X}}

\newcommand{\aF}{{\alpha}^\mathrm{F}}
\newcommand{\aT}{{\alpha}^\mathrm{T}}
\newcommand{\aC}{{\alpha}^\mathrm{C}}  
 
\newcommand{\aX}{{\alpha}^\mathrm{X}}
\newcommand{\aCT}{{\alpha}^\mathrm{CT}}
\newcommand{\aTC}{{\alpha}^\mathrm{TC}}
\newcommand{\aCF}{{\alpha}^\mathrm{CF}}  
\newcommand{\aTF}{{\alpha}^\mathrm{TF}}

\newcommand{\delaC}{\delta_\mathrm{\alpha}^{\mathrm{C}}}
\newcommand{\delaT}{\delta_\mathrm{\alpha}^{\mathrm{T}}}
\newcommand{\delaF}{\delta_\mathrm{\alpha}^{\mathrm{F}}}
\newcommand{\delaX}{\delta_\mathrm{\alpha}^{\mathrm{X}}}

\newcommand{\dlnsF}{\delta_{\mathrm{ns}}^\mathrm{F}}
\newcommand{\dlnsT}{\delta_{\mathrm{ns}}^\mathrm{T}}
\newcommand{\dlnsC}{\delta_{\mathrm{ns}}^\mathrm{C}} 
\newcommand{\dlnsX}{\delta_{\mathrm{ns}}^\mathrm{X}}

\newcommand{\dlacF}{\delta_{\mathrm{ac}}^\mathrm{F}}
\newcommand{\dlacT}{\delta_{\mathrm{ac}}^\mathrm{T}}
\newcommand{\dlacC}{\delta_{\mathrm{ac}}^\mathrm{C}}

\newcommand{\dlacCT}{\delta_{\mathrm{ac}}^\mathrm{CT}}
\newcommand{\dlacTC}{\delta_{\mathrm{ac}}^\mathrm{TC}}

\newcommand{\dlacCF}{\delta_{\mathrm{ac}}^\mathrm{CF}} 
\newcommand{\dlacTF}{\delta_{\mathrm{ac}}^\mathrm{TF}} 
\newcommand{\dlacX}{\delta_{\mathrm{ac}}^\mathrm{X}}

\newcommand{\DlacFp}{\Delta_{\mathrm{ac}}^\mathrm{F+}}
\newcommand{\DlacTp}{\Delta_{\mathrm{ac}}^\mathrm{T+}}
\newcommand{\DlacCp}{\Delta_{\mathrm{ac}}^\mathrm{C+}}

\newcommand{\DlacCTp}{\Delta_{\mathrm{ac}}^\mathrm{CT+}}
\newcommand{\DlacTCp}{\Delta_{\mathrm{ac}}^\mathrm{TC+}}

\newcommand{\DlacCFp}{\Delta_{\mathrm{ac}}^\mathrm{CF+}} 
\newcommand{\DlacTFp}{\Delta_{\mathrm{ac}}^\mathrm{TF+}} 
\newcommand{\DlacXp}{\Delta_{\mathrm{ac}}^\mathrm{X+}}

\newcommand{\DlacFm}{\Delta_{\mathrm{ac}}^\mathrm{F-}}
\newcommand{\DlacTm}{\Delta_{\mathrm{ac}}^\mathrm{T-}}
\newcommand{\DlacCm}{\Delta_{\mathrm{ac}}^\mathrm{C-}}

\newcommand{\DlacCTm}{\Delta_{\mathrm{ac}}^\mathrm{CT-}}
\newcommand{\DlacTCm}{\Delta_{\mathrm{ac}}^\mathrm{TC-}}

\newcommand{\DlacCFm}{\Delta_{\mathrm{ac}}^\mathrm{CF-}} 
\newcommand{\DlacTFm}{\Delta_{\mathrm{ac}}^\mathrm{TF-}} 
\newcommand{\DlacXm}{\Delta_{\mathrm{ac}}^\mathrm{X-}}

\newcommand{\dlecF}{\delta_{\mathrm{ec}}^\mathrm{F}}
\newcommand{\dlecT}{\delta_{\mathrm{ec}}^\mathrm{T}}
\newcommand{\dlecC}{\delta_{\mathrm{ec}}^\mathrm{C}}

\newcommand{\dlecCT}{\delta_{\mathrm{ec}}^\mathrm{CT}}
\newcommand{\dlecTC}{\delta_{\mathrm{ec}}^\mathrm{TC}}

\newcommand{\dlecCF}{\delta_{\mathrm{ec}}^\mathrm{CF}} 
\newcommand{\dlecTF}{\delta_{\mathrm{ec}}^\mathrm{TF}} 
\newcommand{\dlecX}{\delta_{\mathrm{ec}}^\mathrm{X}}

\newcommand{\DlecFp}{\Delta_{\mathrm{ec}}^\mathrm{F+}}
\newcommand{\DlecTp}{\Delta_{\mathrm{ec}}^\mathrm{T+}}
\newcommand{\DlecCp}{\Delta_{\mathrm{ec}}^\mathrm{C+}}

\newcommand{\DlecCTp}{\Delta_{\mathrm{ec}}^\mathrm{CT+}}
\newcommand{\DlecTCp}{\Delta_{\mathrm{ec}}^\mathrm{TC+}}

\newcommand{\DlecCFp}{\Delta_{\mathrm{ec}}^\mathrm{CF+}} 
\newcommand{\DlecTFp}{\Delta_{\mathrm{ec}}^\mathrm{TF+}} 
\newcommand{\DlecXp}{\Delta_{\mathrm{ec}}^\mathrm{X+}}

\newcommand{\DlecFm}{\Delta_{\mathrm{ec}}^\mathrm{F-}}
\newcommand{\DlecTm}{\Delta_{\mathrm{ec}}^\mathrm{T-}}
\newcommand{\DlecCm}{\Delta_{\mathrm{ec}}^\mathrm{C-}}

\newcommand{\DlecCTm}{\Delta_{\mathrm{ec}}^\mathrm{CT-}}
\newcommand{\DlecTCm}{\Delta_{\mathrm{ec}}^\mathrm{TC-}}

\newcommand{\DlecCFm}{\Delta_{\mathrm{ec}}^\mathrm{CF-}} 
\newcommand{\DlecTFm}{\Delta_{\mathrm{ec}}^\mathrm{TF-}} 
\newcommand{\DlecXm}{\Delta_{\mathrm{ec}}^\mathrm{X-}}

\newcommand{\DsnX}{\mathrm{Dsn}_\mathrm{X}}

\begin{document} 
 
\begin{center}
   {\Large\bf 
   A Novel Method for Inference of Chemical Compounds 
   with Prescribed Topological Substructures Based on Integer Programming}
\end{center} 

\begin{center}
 Tatsuya Akutsu$^1$, 
Hiroshi Nagamochi$^2$
\end{center} 
%
%
{\small 
1.   Bioinformatics Center,  Institute for Chemical Research, 
 Kyoto University, Uji 611-0011, Japan \\
2.  Department of Applied Mathematics and Physics, Kyoto University, Kyoto 606-8501, Japan\\
}

\begin{quote}  
{\bf Abstract}\\ 
Analysis of chemical graphs is becoming a major research topic
in computational molecular biology due to its potential applications
to drug design.
One of the major approaches in such a study is
inverse quantitative structure activity/property relationships
(inverse QSAR/QSPR) analysis, which is to infer chemical structures
from given chemical activities/properties.
Recently, a novel framework has been proposed for inverse QSAR/QSPR
using both artificial neural networks (ANN) and
mixed integer linear programming (MILP).
This method consists of a prediction phase and an inverse prediction phase.
In the first phase,  
a feature vector $f(G)$ of a chemical graph $G$ is introduced and 
a prediction function $\psi_{\mathcal{N}}$ on a chemical property $\pi$
is constructed with an ANN $\mathcal{N}$.  
In the second phase, given a target value $y^*$ of the chemical property $\pi$,
   a feature vector $x^*$ is inferred  
   by solving an MILP formulated from  the trained ANN    $\mathcal{N}$
  so that  $\psi_{\mathcal{N}}(x^*)$ is equal to  $y^*$ 
  and   then a set of chemical structures $G^*$
 such that $f(G^*)= x^*$  is enumerated by a graph enumeration algorithm. 
The framework has been applied to 
chemical compounds with a rather abstract topological structure such as
 acyclic or monocyclic graphs and
graphs with a specified polymer topology 
with cycle index up to 2.

In this paper, we propose a new flexible modeling method to the framework
so that we can specify  
  a  topological substructure of  graphs  and
   a partial assignment of chemical elements and bond-multiplicity
  to  a target graph.

\noindent 
{\bf Keywords: } QSAR/QSPR,  Molecular Design, 
             Artificial Neural Network, Mixed  Integer Linear Programming, 
             Enumeration of Graphs  

\noindent 
{\bf  Mathematics Subject Classification: } 
Primary  
05C92,  
92E10, 
Secondary
05C30, 
68T07, 
90C11,  
92-04 
\end{quote}

  
\section{Introduction}\label{sec:introduction}

Graphs are a fundamental data structure in information science.
Recently, design of novel graph structures has become a hot topic
in artificial neural network (ANN) studies.
In particular, extensive studies have been done on designing
chemical graphs having desired chemical properties
because of its potential application to drug design.
For example, variational autoencoders~\cite{Gomez18}, 
recurrent neural networks~\cite{Segler18,Yang17}, 
grammar variational autoencoders~\cite{Kusner17},
generative adversarial networks~\cite{DeCao18},
and
invertible flow models~\cite{Madhawa19,Shi20}
have been applied.

On the other hand, 
computer-aided design of chemical graphs has a long 
history in the field of chemo-informatics,
under the name of inverse quantitative structure activity/property
relationships (inverse QSAR/QSPR).
In this framework,
chemical compounds are usually represented 
as vectors of real or integer numbers,
which are often called \emph{descriptors} and
correspond to \emph{feature vectors} in machine learning.
Using these chemical descriptors,
various heuristic and statistical methods have been developed for
finding optimal or near optimal chemical graphs~\cite{Ikebata17,Miyao16,Rupakheti15}.
In many of such methods,
inference or enumeration of graph structures from a given set of
descriptors is a crucial subtask, and thus
various methods have been developed \cite{Fujiwara08,Kerber98,Li18,Reymond15}.
However, enumeration in itself is a challenging task, 
since the number of
molecules (i.e., chemical graphs) with up to 30 atoms (vertices)
{\tt C}, {\tt N}, {\tt O}, and {\tt S},
may exceed~$10^{60}$~\cite{BMG96}.
Furthermore, even inference is a challenging task since it is NP-hard
except for some simple cases~\cite{Akutsu12,Nagamochi09}.
Indeed, most existing methods including ANN-based ones
do not guarantee optimal or exact solutions.

In order to guarantee the optimality mathematically,
a novel approach has been proposed~\cite{AN19} for ANNs, 
 using mixed integer linear programming (MILP).
However, this method outputs feature vectors only, not chemical structures.
To overcome this issue, a new framework has been proposed
\cite{ACZSNA20,CWZSNA20,ZZCSNA20} 
by combining two previous approaches; efficient enumeration
of tree-like graphs~\cite{Fujiwara08}, and 
MILP-based formulation of the inverse problem on ANNs~\cite{AN19}.
This combined framework for  inverse QSAR/QSPR mainly consists of two phases.
The first phase solves (I) {\sc Prediction Problem}, where 
a feature vector $f(G)$ of a chemical graph $G$ is introduced
and 
a prediction function $\psi_{\mathcal{N}}$ on a chemical property $\pi$
is constructed with an ANN $\mathcal{N}$ 
using a data set of  chemical compounds $G$ and their values $a(G)$ of $\pi$.
The second phase solves (II) {\sc Inverse Problem},
  where (II-a) given a target value $y^*$ of the chemical property $\pi$,
   a feature vector $x^*$ is inferred from the trained ANN  $\mathcal{N}$
  so that  $\psi_{\mathcal{N}}(x^*)$ is close to  $y^*$ 
  and (II-b) then a set of chemical structures $G^*$
 such that $f(G^*)= x^*$  is enumerated by a graph search algorithm.  
In (II-a) of the above-mentioned previous methods~\cite{ACZSNA20,CWZSNA20,ZZCSNA20},
an MILP is formulated for acyclic chemical compounds.
 Afterwards, Ito et~al. \cite{IAWSNA20} and Zhu et~al. \cite{ZCSNA20}
 designed a method of  inferring  chemical graphs 
 with  rank (or cycle index) 1 and 2, respectively 
by formulating a new MILP and using
an efficient algorithm for enumerating  chemical graphs with rank 1 
\cite{Suzuki14} and rank 2 \cite{2A1B20,2A2B20}. 
The computational results conducted on instances with
$n$ non-hydrogen atoms show that
a feature vector $x^*$ can be inferred for up to around $n=40$
whereas  graphs $G^*$ can be enumerated  for up to around $n=15$. 
Recently  Azam~et~al.~\cite{AZSSSZNA20} 
 introduced a new characterization of acyclic graph structure,
called ``branch-height'' to define a class of acyclic graphs
with a restricted structure that still covers the most of the acyclic
chemical compounds in the database. 
They also employed the dynamic programming method
to design a new algorithm for generating chemical acyclic graphs
which now works for instances with size $n(G^*)=50$. 

The framework has been applied so far to a case of 
chemical compounds with a rather abstract topological structure such as
 acyclic or monocyclic graphs and
 graphs with a specified polymer topology 
with rank up to 2.
When there is a more specific requirement on some part of
the graph structure and the assignment of chemical elements
in a chemical graph to be inferred,
none of the above-mentioned methods can be used directly.
The main reason is that generating chemical graphs
from a given feature vector is a considerably hard problem: 
 an efficient algorithm needed to be newly designed
for each of different classes of graphs. 
In this paper, we discover a new mechanism of
generating chemical graphs that can avoid the necessity that
we design a  new algorithm whenever a graph class changes
(see Section~\ref{sec:graph_search} for the details). 
Based on this, we propose a new method based on the framework
so that a target chemical graph  to be inferred can be specified 
in a more flexible way.
With our specification, we can include a prescribed 
  substructure of  graphs such as a benzene ring into a target chemical graph
  while imposing constraints on a global topological structure of a target
  graph  at the same time.

The paper is organized as follows.  
Section~\ref{sec:preliminary} introduces some notions on graphs,
 a modeling of chemical compounds and a choice of descriptors. 
Section~\ref{sec:inverse_process} reviews the framework for inferring
chemical compounds based on ANNs and MILPs. 
Section~\ref{sec:specification} introduces a method of 
specifying topological substructures of target chemical graphs to be inferred.
Section~\ref{sec:graph_MILP} presents a formulation
of an MILP that can infer a chemical graph
under a given specification to target chemical graphs. 
Section~\ref{sec:graph_search} describes a new
dynamic programming type of algorithm 
that generates chemical graphs that are 
isomorphic to a given chemical graph $G^\dagger$
in the sense that all generated chemical graphs
$G^*$ have the same feature vector $f(G^*)=f(G^\dagger)$. 
%
%
Section~\ref{sec:conclude} makes some concluding remarks. 
Appendix~\ref{sec:full_milp} describes the details of all variables and constraints
in our MILP formulation.

 
\section{Preliminary}\label{sec:preliminary}

This section  introduces some notions and terminology on graphs,
 a modeling of chemical compounds and our choice of descriptors. 
 
Let $\mathbb{R}$, $\mathbb{Z}$  and $\mathbb{Z}_+$ 
denote the sets of reals, integers and non-negative integers, respectively.
For two integers $a$ and $b$, let $[a,b]$ denote the set of 
integers $i$ with $a\leq i\leq b$.

\subsection{Graphs} 
\medskip\noindent{\bf Multi-digraphs}    
A multi-digraph $G$ is defined to be a pair of a set $V$ of vertices and
a set $E$ of directed edges such that each edge $e\in E$ corresponds to
an ordered pair $(u,v)$ of vertices, where $u$ and $v$ are
called the {\em tail} and a {\em head} of $e$ and denoted
by $\hd(e)$ and $\tail(e)$, respectively.
A multi-digraph $G$ may contain an edge $e\in E$ with $\hd(e)=\tail(e)$,
which is called a {\em self-loop}; 
or two edges $e,e'\in E$ with the same pair of tail and head,
which are called {\em multiple edges}.

Let $G=(V,E)$ be a multi-digraph.
For each vertex $v\in V$, we define the sets as follows:
\[ E_G^{-}(v)\triangleq \{  e\in E\mid  \hd(e)=v\}, 
    E^+_G(v)\triangleq \{  e\in E\mid  \tail(e)=v\},  \]
\[ N_G^{-}(v)\triangleq \{  \tail(e)\in V\mid e\in E_G^{-}(v) \}, 
    N^+_G(v)\triangleq \{ \hd(e)\in V\mid  e\in E_G^{+}(v)\}. \] 
The {\em in-degree} $\deg_G^-(v)$ and   {\em out-degree} $\deg_G^+(v)$ 
of a vertex $v\in V$ are defined to be
$\deg_G^-(v)\triangleq |E_G^{-}(v)|$ and  
$\deg_G^+(v)\triangleq |E^+_G(v)|$, respectively. 
Given a multi-digraph $G$, let $V(G)$ and $E(G)$ denote the sets
of vertices and edges, respectively.

\medskip\noindent{\bf Muligraphs}     
The graph obtained from a multi-digraph by ignoring
the order between the head and the tail of each edge
is called a  multigraph, 
where the head and the tail of an edge are called
the {\em end-vertices} of the edge.
Let $V_G(e)$ denote the set of end-vertices of an edge $e$.  
A multigraph $G$ may contain an edge $e\in E$ with only one end-vertex,
which is called a {\em self-loop}; 
or two edges $e,e'\in E$ with the same pair of end-vertices,
which are called {\em multiple edges}.
A multigraph with  no self-loops 
and no multiple edges is called {\em simple}. 

Let $G=(V,E)$ be a  multigraph.
For each vertex $v\in V$, we define the sets as follows:
\[ E_G (v)\triangleq \{  e\in E\mid  v\in V_G(e) \},  
  N_G (v)\triangleq \{ u\in V_G(e)\setminus \{v\} \mid    e\in E_G(v)\}  . \] 
The {\em degree} $\deg_G(v)$ of a vertex $v\in V$ is  defined to be
$\deg_G(v)\triangleq |E_G(v)|$, where 
$\deg_G(v)=p + 2q$ for the number $p$ of non-loop edges incident to $v$
and the number $q$ of self-loops incident to $v$. 

The length of a path is defined to be the number of edges in the path. 
Denote by $\ell(P)$ the length of a path $P$.
A simple connected graph  is called a {\em tree} if it contains no cycle
and is called {\em cyclic} otherwise. 
For two multigraphs $G_1$ and $G_2$, we denote by $G_1\simeq G_2$
when they are isomorphic. 
Given a multigraph $G$, let $V(G)$ and $E(G)$ denote the sets
of vertices and edges, respectively.

\medskip\noindent{\bf Rank of Multigraphs}     
The {\em rank} $\mathrm{r}(G)$ of a multigraph $M$  is defined to be 
the minimum number of edges to be removed to make the multigraph
a tree (a simple and connected graph).   
We call a multigraph $G$ with $\mathrm{r}(G)=k$ a {\em rank-$k$  graph}.

\medskip\noindent{\bf Rooted Trees}     
A {\em rooted tree} is defined to be a tree where a   vertex
 (or a  pair of adjacent vertices) is designated as the {\em root}.
Let $T$ be a rooted tree, where for two adjacent vertices
$u$ and $v$, vertex $u$ is called the parent of $v$ if
$u$ is closer to the root than $v$ is. 
The {\em height} $\mathrm{height}(v)$ of a vertex $v$  in $T$
is defined to be the maximum length of a path
from $v$ to a leaf $u$ in the  descendants of $v$,
where  $\mathrm{height}(v)=0$ for each leaf $v$ in $T$. 
The {\em height} $\h(T)$ of a rooted tree $T$ is defined
to be the $\mathrm{height}(r)$ of the root $r$.

\medskip\noindent{\bf Bi-rooted Trees}     
As an extension of rooted trees, we define
a {\em bi-rooted tree}
to be a tree $T$ with two designated vertices $r_1(T)$ and $r_2(T)$,
called    {\em terminals}.
Let $T$ be a  bi-rooted tree.
Define the {\em backbone path} $P_T$ to be the path of $T$
between terminals $r_1(T)$ and $r_2(T)$,
and denote by $\F(T)$ (or by $\F(P_T)$) the set of
 subtrees of $T$ in the graph $T-E(P_T)$ obtained from
 $T$ by removing the edges  in $P_T$,
 where we regard each tree $T'\in \F(T)$
 as a tree rooted at the unique vertex in $V(T')\cap V(P_T)$.
 The {\em height} $\h(T)$ of $T$ is defined to be 
the maximum  of the heights of rooted trees in  $\F(T)$.

We may regard a rooted tree $T$ as a bi-rooted tree $T$ with
$r_1(T)=r_2(T)$.
 
\begin{figure}[h!] \begin{center}
\includegraphics[width=.75\columnwidth]{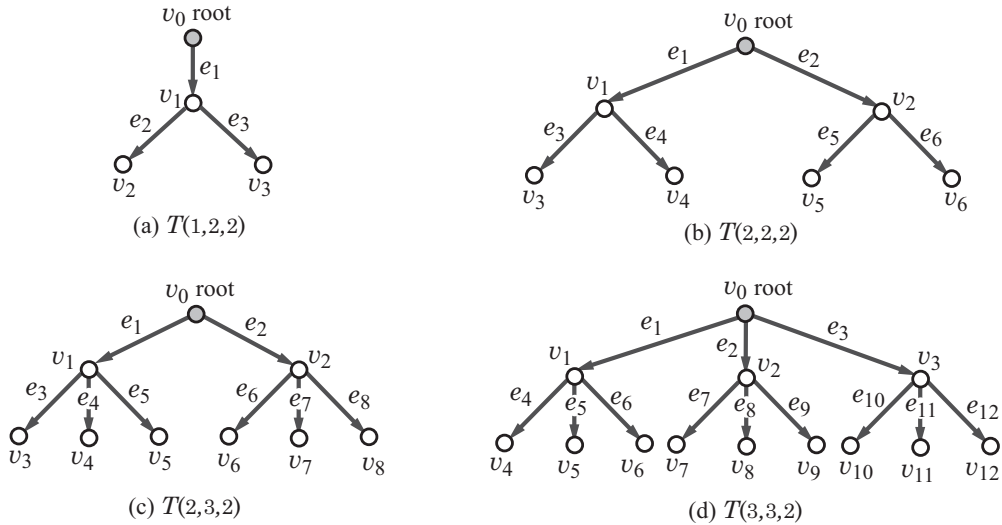}
\end{center} \caption{An illustration of rooted trees $T(a,b,c)$:
(a) $T(1,2,2)$; 
(b)$T(2,2,2)$; 
(c) $T(2,3,2)$; 
(d) $T(3,3,2)$. }
\label{fig:regular-tree} \end{figure}

\medskip\noindent{\bf Degree-bounded Trees}     
For positive integers $a,b$ and $c$ with $b\geq 2$, 
let $T(a,b,c)$ denote the rooted tree
such that the number of children of the root is $a$,
the number of children of each non-root internal vertex is $b$
and the distance from the root to each leaf is $c$.
Figure~\ref{fig:regular-tree}(a)-(d) illustrate rooted trees
$T(a,b,c)$ with $(a,b,c)\in \{(1,2,2),(2,2,2),(2,3,2),(3,3,2)\}$. 

We see that  the number of vertices in $T(a,b,c)$ is
$a(b^c-1)/(b-1)+1$, and 
the number of non-leaf vertices in $T(a,b,c)$
is $a(b^{c-1}-1)/(b-1)+1$.
%
In the rooted tree  $T(a,b,c)$, we denote the vertices 
by $v_0,v_1,\ldots,v_{n-1}$ with a breadth-first-search order,
and denote the edge between a vertex $v_i$ with $i\in [1,n-1]$ 
and its parent by $e_i$, 
where $n=a(b^c-1)/(b-1)+1$ and each vertex 
$v_i$ with $i\in [1, a(b^{c-1}-1)/(b-1)+1]$ is a non-leaf vertex.
For each vertex $v_i$ in   $T(a,b,c)$, 
let  $\Cld(i)$ denote the set of indices
$j$ such that $v_j$ is a child of $v_i$, 
 and $\prt(i)$ denote the index $j$
 such that $v_j$ is the parent of $v_i$ when $i\in [1,n-1]$. 
To reduce the number of graph-isomorphic rooted trees 
presented as a solution in an MILP, we use 
a precedence constraint introduced by  Zhang~et~al.~\cite{ZZCSNA20}.
We say that two rooted trees are {\em isomorphic} if
they admits a graph isomorphism such that the two roots 
correspond to each other.  
Let $\mathcal{T}(a,b,c)$ denote the set of subtrees of $T(a,b,c)$
that have the same root of $T(a,b,c)$.
Let $\Pprc(a,b,c)$ be a set of ordered index pairs $(i,j)$
of vertices $v_i$ and $v_j$ in   $T(a,b,c)$ and
$\Tprc(a,b,c)$ denote the set of 
subtree $T\in \mathcal{T}(a,b,c)$ such that,
for each pair $(i,j)\in  \Pprc(a,b,c)$,
$T$ contains vertex $v_i$ if it contains vertex $v_j$. 
We call $\Pprc(a,b,c)$ {\em proper}
if the next conditions hold:  
\begin{enumerate}
\item[(a)] Each  subtree $T_1\in \mathcal{T}(a,b,c)$
is isomorphic to a  subtree $T_2\in \Tprc(a,b,c)$
such that \\ 
~~~  for each pair $(i,j)\in \Pprc(a,b,c)$, 
 if $v_j\in V(T_2)$ then $v_i\in V(T_2)$; and

\item[(b)] For each pair of vertices $v_i$ and $v_j$ in $T(a,b,c)$
such that $v_i$ is the parent of $v_j$,
there is a sequence $(i_1,i_2),(i_2,i_3),\ldots,(i_{k-1},i_{\rho})$
of index pairs in $\Pprc(a,b,c)$ 
such that $i_1=i$ and $i_{\rho}=j$. 
\end{enumerate} 
Condition (b) can be used to reduce the size of a proper set  $\Pprc(a,b,c)$ 
by omitting  some pair $(i,j)$ of indices of a vertex $v_j$
and the parent $v_i$ of $v_j$.  
Note that a proper set $\Pprc(a,b,c)$ is not necessarily unique.    

For the rooted trees in Figure~\ref{fig:regular-tree},
we obtain proper sets of ordered index pairs as follows. \\
~~~ $\Pprc(1,2,2)=\{(0,1),(1,2),(2,3)\}$, \\
~~~ $\Pprc(2,2,2)=\{(0,1),(1,2),(1,3),(2,5),(3,4),(3,5),(4,6),(5,6)\}$, \\
~~~ $\Pprc(2,3,2)=\{(0,1),(1,2),(1,3),(2,6),(3,4),(3,6),(4,5),(4,7),(5,8),(6,7),(7,8)\}$
and \\
~~~ $\Pprc(3,3,2)=\{(0,1),(1,2),(1,4),(2,3),(2,7),(3,10),(4,5),$ 
$(4,7),(5,6),(5,8),(6,9),(7,8),$ \\
~~~~ $(7,10),(8,9),(8,11),$ $(9,12),(10,11),(11,12)\}$.\\ 
With these proper sets, 
we see that every rooted tree $T\in  \Tprc(a,b,c)$ satisfies
a special property that the leftmost path (or the path that visits 
children with the smallest index) from the root is 
alway  of the length of the height $\h(T)$.

\medskip\noindent{\bf Branch-height in   Trees}    
 Azam~et~al.~\cite{AZSSSZNA20} 
  introduced ``branch-height'' of a tree 
as a new measure to the ``agglomeration degree'' of trees.
We specify a non-negative integer ${\rho}$, called a {\em branch-parameter} 
to define branch-height. 
 First we regard   $T$ as a rooted tree by choosing
 the center of $T$ as the root.

We introduce the following terminology on a rooted tree $T$.
\begin{itemize}
\item[-] A {\em leaf ${\rho}$-branch}:  a non-root vertex $v$ in $T$  such that
  $\mathrm{height}(v)= {\rho}$.
  
\item[-] A  {\em non-leaf ${\rho}$-branch}:  
 a vertex $v$ in $T$ such that $v$ has at least two children $u$  
with  $\mathrm{height}(u)\geq {\rho}$.
We call a leaf or non-leaf ${\rho}$-branch a {\em  ${\rho}$-branch}. 

\item[-] A {\em ${\rho}$-branch-path}: a path $P$ in $T$
that joins two vertices $u$ and $u'$ such that 
each of $u$ and $u'$ is the root or a ${\rho}$-branch and
$P$ does not contain   the root or a ${\rho}$-branch
as an internal vertex. 
 
\item[-]
The  {\em ${\rho}$-branch-subtree} of $T$:
 the subtree of $T$ that consists of 
the edges in all ${\rho}$-branch-paths of $T$. 
 We call a vertex (resp., an edge) in $T$ 
 a {\em ${\rho}$-internal vertex} (resp., a {\em ${\rho}$-internal edge})
  if it is contained in the ${\rho}$-branch-subtree  of $T$
 and a {\em ${\rho}$-external vertex}   (resp., a {\em ${\rho}$-external edge}) otherwise.
 Let $V^\inn$ and $V^\ex$ (resp., $E^\inn$ and $E^\ex$)  
 denote the sets of  ${\rho}$-internal and ${\rho}$-external vertices (resp., edges) in $T$.
 
\item[-]
The  {\em ${\rho}$-branch-tree} of $T$: the rooted tree 
obtained from the ${\rho}$-branch-subtree  of $T$
by replacing each ${\rho}$-branch-path with a single edge.  
 
\item[-] 
 A {\em ${\rho}$-fringe-tree}: One of the connected components
that consists of the edges not in any  ${\rho}$-branch-subtree.
Each ${\rho}$-fringe-tree $T'$ contains exactly one vertex $v$ in a  ${\rho}$-branch-subtree,
where $T'$ is regarded as a   tree rooted at $v$.
Note that the height of any ${\rho}$-fringe-tree is at most ${\rho}$. 
 
\item[-]
The {\em ${\rho}$-branch-number} $\bn_{\rho}(T)$: 
the number of ${\rho}$-branches  in $T$. 

\item[-]
The {\em ${\rho}$-branch-height} $\bh_{\rho}(T)$ of $T$: 
the maximum number of non-root ${\rho}$-branches along a path
from the root  to a leaf of $T$; i.e., $\bh_{\rho}(T)$ is 
the height of the ${\rho}$-branch-tree $T^*$ (the maximum length of a path
from the root to a leaf in $T^*$).   
\end{itemize}

\begin{figure}[h!] \begin{center}
\includegraphics[width=.78\columnwidth]{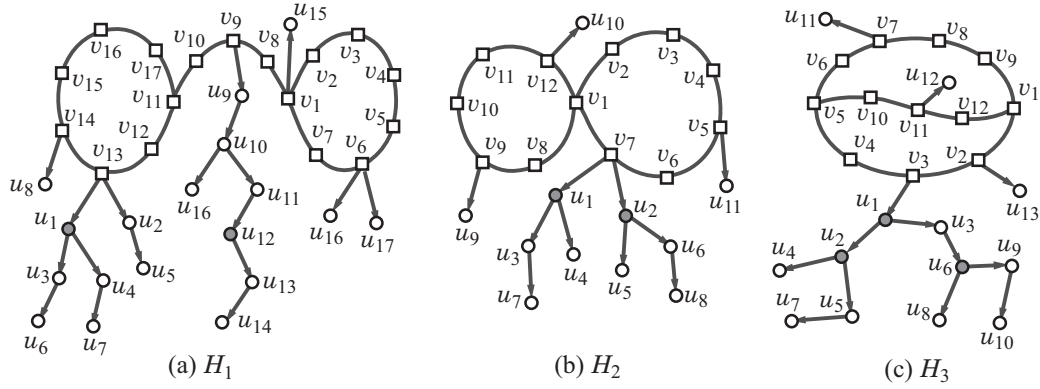}
\end{center} \caption{An illustration of rank-2 graphs $H_i$, $i=1,2,3$,
 where the core-vertices are depicted wtih squares, 
  the $2$-branch vertices are depicted with gray circles
and non-core edges are depicted as directed edges with arrows:
(a) $H_1$ is 2-lean, 
 $\cs(H_1)=17$, $\ch(H_1)=6$, $\bh_2(H_1)=1$
 and $\bc(H_1)_2=\bl_2(H_1)=2$; 
(b)  $H_2$ is not 2-lean,  $\cs(H_2)=12$, $\ch(H_2)=3$, 
  $\bh_2(H_2)=1$, 
$\bc_2(H_2)=1$ and $\bl_2(H_2)=2$; 
(c)  $H_3$ is not 2-lean,  $\cs(H_3)=12$, $\ch(H_3)=5$, 
  $\bh_2(H_3)=2$, $\bc_2(H_3)=1$ and $\bl_2(H_3)=2$. }
\label{fig:rank2_graph_examples} \end{figure}

\medskip\noindent{\bf Core  in Cyclic Graphs}     
Let $H$ be a connected simple graph with rank $\mathrm{r}(H)\geq 1$.

The {\em core} $\Cr(H)$ of $H$ is defined to be 
an induced subgraph $\Cr(H)=(V'=V'_1\cup V'_2,E')$
such that $V'_1$ is the set of vertices in a cycle of $H$ 
and $V'_2$ is the set of verices each of which is 
in a path between two vertices $u,v\in V'_1$.
A vertex (resp., an edge) in $H$ is called a {\em core-vertex}
(resp., {\em core-edge}) if it is contained in the core $\Cr(H)$
and is called  a {\em non-core-vertex}
(resp., {\em non-core-edge}) otherwise. 
 We denote by   $V^\co$ (resp., $V^\nc$) and  $E^\co$ (resp., $E^\nc$)
 the set of core-vertices (resp., non-core-vertices)
  and the set of core-edges (resp., non-core-edges)   in $H$.  
The {\em core size} $\cs(H)$ is defined 
to be the number $|V^\co|$ of  core-vertices in the core of $H$. 

\begin{figure}[!htb]
\begin{center}
\includegraphics[width=.78\columnwidth]{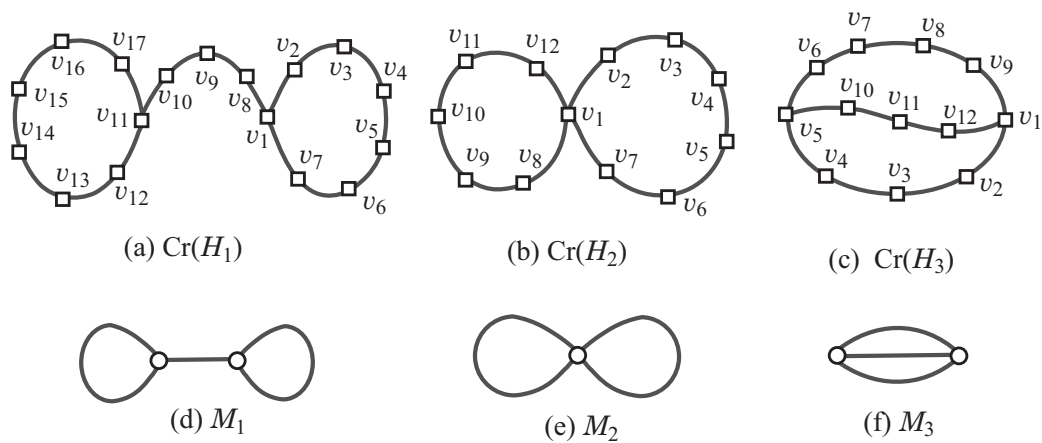}
\end{center} \caption{An illustration of rank-2 graphs and rank-2 multigraphs:
(a), (b), (c)  the cores  $\Cr(H_i)$, $i=1,2,3$ 
 of the  rank-2 graphs $H_i$, $i=1,2,3$ in Figure~\ref{fig:rank2_graph_examples};
(d), (e), (f) the three  polymer topologies 
$M_i\in\mathcal{PT}(2,4)$, $i=1,2,3$, 
where $M_i\simeq \mathrm{Pt}(H_i)$.  }
\label{fig:rank_2_polymer} 
\end{figure}

 Figure~\ref{fig:rank2_graph_examples} illustrates
 three examples of rank-2 graphs $H_i$, $i=1,2,3$ and 
  Figure~\ref{fig:rank_2_polymer} illustrates their cores $\Cr(H_i)$, 
 where $\cs(H_1)=17$, $\ch(H_1)=6$, 
  $\cs(H_2)=12$, $\ch(H_2)=3$, $\cs(H_3)=12$ and $\ch(H_3)=5$.
 
A connected component in the subgraph induced 
by the non-core-vertices 
of $H$ is called a {\em non-core component} of $H$.
Each non-core component $T$ contains exactly one non-core-vertex
$v_T\in V^\nc$ that is adjacent to  a core-vertex $u_T\in V^\co$,
where the tree $T'$ that consists of $T$ and edge $v_T u_T\in E^\nc$
is called a {\em pendant-tree} of $H$ regarded as a tree rooted
at the core-vertex $u_T$.
The {\em core height} $\ch(H)$ is defined to be
the maximum height $\h(T)$ of a pendant-tree $T$ of $H$. 
 
A {\em core-path} $P$ of a graph $H$ is defined 
to be a subgraph of the core $\Cr(H)=(V^\co, E^\co)$ 
such that the degree of each internal vertex $v$ of $P$
 is 2 in the core; i.e., $\deg_P(v)=\deg_{\Cr(H)}(v)=2$. 
A {\em path-partition} $\Pt$ of  the core $\Cr(H)$
is defined to be a collection of core-paths $P_i$
with $\ell(P_i)\geq 1$, $i=[1,p]$ 
such that each core-edge belongs to exactly one core-path in $\Pt$;
i.e., 
\[\bigcup_{i\in[1,p]}E(P_i)=E^\co, ~~
     E(P_i)\cap E(P_j)=\emptyset, 1\leq i<j\leq p. \]
For example, the core $\Cr(H_1)$ in 
Figure~\ref{fig:rank2_graph_examples}(a) admits a  path-partition 
$\Pt=\{P_1,P_2,\ldots,P_5\}$ such that 
$P_1=(v_1,v_2,v_3,v_4,v_5)$, 
$P_2=(v_1,v_7,v_6,v_5)$, 
$P_3=(v_1,v_8,v_9,v_{10},v_{11})$, 
$P_4=(v_{11},v_{12},v_{13},v_{14},v_{15})$ and  
$P_5=(v_{15},v_{16},v_{17},v_{11})$.


\medskip\noindent{\bf Branch-height in Cyclic Graphs}     
Let $H$ be a connected simple graph with rank $\mathrm{r}(H)\geq 1$.
 
  For a branch parameter ${\rho}\geq 0$,
 we define ${\rho}$-fringe-tree, leaf ${\rho}$-branch, ${\rho}$-branch, ${\rho}$-branch-path, 
  ${\rho}$-branch-subtree, 
 ${\rho}$-internal/${\rho}$-external vertex/edges, ${\rho}$-branch-tree and ${\rho}$-branch-height 
 in each pendant-tree of $H$ analogously,
 where we do not regard any core-vertex as a ${\rho}$-branch.
 A non-core-vertex (resp., non-core-edge) in $H$
 is called a ${\rho}$-internal vertex (resp., edge) or 
 a ${\rho}$-external vertex (resp., edge)
 if   it is in some ${\rho}$-fringe-tree of $H$. 
 Let $V^{\inn}$ and $V^{\ex}$ (resp., $E^{\inn}$ and $E^{\ex}$) 
 denote the sets of
  ${\rho}$-internal and ${\rho}$-external vertices (resp., edges) in $H$,
  where $V^\nc=V^{\inn} \cup V^{\ex}$ and
  $E^\nc =E^{\inn} \cup E^{\ex}$.

  Define the {\em ${\rho}$-branch-leaf-number} $\bl_{\rho}(H)$ of $H$
 to the number of leaf ${\rho}$-branches  in $H$ and  
 the {\em ${\rho}$-branch-height} $\bh_{\rho}(H)$
   to be  the maximum  ${\rho}$-branch-height  $\bh_{\rho}(T)$ 
 over all pendant-trees $T$  of $H$. 
 %
 We call a pendant-tree  of $H$ a {\em ${\rho}$-pendant-tree}
 if it contains  at least one ${\rho}$-branch. 
 %
 We call a  core-vertex adjacent to a ${\rho}$-pendant-tree
 a {\em ${\rho}$-branch-core-vertex}, 
  denote by $V_{\rho}^{\bc}$ the set of ${\rho}$-branch-core-vertices
  and 
   define the {\em ${\rho}$-branch-core-size} $\bc_{\rho}(H)$  to be $|V_{\rho}^{\bc}|$. 
Note that 
$\cs(H)\geq \bc_{\rho}(H)$, $\bl_{\rho}(H)\geq \bc_{\rho}(H)$ and 
either 
${\rho}>\ch(H)$ or 
${\rho} \leq \ch(H)=\bh_{\rho}(H)+ {\rho}$. 
 
 We call a graph $H$ {\em ${\rho}$-lean} if $\bl_{\rho}(H)=\bc_{\rho}(H)$;
 i.e., all ${\rho}$-branches in $H$ are leaf ${\rho}$-branches and
 no two ${\rho}$-pendant-trees share the same ${\rho}$-branch-core-vertex.
 Note that the ${\rho}$-branch height of any ${\rho}$-lean graph is at most 1.
 Figure~\ref{fig:rank2_graph_examples} 
  illustrates  three examples  of   rank-2 graphs.
In  the first example, 
  $u_1$ and $u_{12}$ are the leaf 2-branches, 
$v_{13}$ and $v_9$ are the $2$-branch-core-vertices, 
 $\bc_2(H_1)=\bl_2(H_1)=2$ holds and  $H_1$ is $2$-lean.
 In the second example,
 $u_1$ and $u_2$ are the leaf 2-branches, 
$v_7$ is the $2$-branch-core-vertex, 
$\bc_2(H_2)=1<\bl_2(H_2)=2$ holds and
$H_2$ is not 2-lean.
In the third example, $u_2$ and $u_6$ are the leaf 2-branches,
 $u_1$ is the non-leaf 2-branch,
$v_3$ is the $2$-branch-core-vertex, 
  $\bc_2(H_3)=1<\bl_2(H_3)=2$ holds  and 
  $H_3$ is not $2$-lean.
 
We here show some statical feature of
the chemical graphs in PubChem in terms of 
rank of graphs and ${\rho}$-branch height
(see  \cite{AZSSSZNA20}  for more details). 
\begin{enumerate} 
\item[-]
Nearly 87\% (resp., 99\%) of  rank-4 chemical compounds  with up to
   100 non-hydrogen atoms in  PubChem
   have the maximum degree 3 (resp., 4) of non-core-vertices.
      
\item[-]
Nearly 84\% of the chemical compounds in the chemical database PubChem
have rank at most 4.

\item[-]
Over 87\% (resp., 96\%) of  rank-1 or rank-2
(resp., rank-3 or rank-4)  chemical compounds with up to
   50 non-hydrogen atoms in  PubChem
   have the $2$-branch height $\bh_2(G)$ at most 1.
   
\item[-] 
Over 92\% of 2-fringe-trees of  chemical compounds with up to
   100 non-hydrogen atoms in  PubChem obey the following 
   size constraint: 
\begin{equation}\label{eq:fringe-size}
   \mbox{$n \le 2d + 2$ for each 2-fringe-tree  $T$ 
  with $n$ vertices and $d$ children of the root. }\end{equation}

\item[-] For ${\rho}=2$, nearly 97\% of cyclic chemical compounds with up to
100 non-hydrogen atoms in  PubChem  are 2-lean. 
\end{enumerate}  

\medskip\noindent{\bf Polymer Topology}     
A multigraph is called a {\em polymer topology} if
it is connected and the degree of every vertex is at least 3.
Tezuka and Oike~\cite{TO02} pointed out that a
classification of  polymer topologies will lay a foundation
for elucidation of structural relationships 
between different macro-chemical molecules and their synthetic pathways.
For integers $r\geq0$ and $d\geq 3$,
let $\mathcal{PT}(r,d)$ denote the set of all rank-$r$ polymer topologies
with  maximum degree at most $d$.
For example, there are three polymer topologies in $\mathcal{PT}(2,4)$, 
 as illustrated in  Figure~\ref{fig:rank_2_polymer}(d)-(f).

 The {\em polymer topology} $\mathrm{Pt}(H)$ of a multigraph $H$ 
 with $\mathrm{r}(H)\geq 2$ is defined to be
a multigraph $H'$ of degree at least 3 that is obtained 
from the core $\mathrm{Cr}(H)$ 
by contracting all vertices of degree 2. 
Note that $\mathrm{r}(\mathrm{Pt}(H))= \mathrm{r}(H)$.

\subsection{Modeling of Chemical Compounds}\label{sec:chemical_model}

We represent the graph structure of a chemical compound as a graph $H$
with labels on vertices and multiplicity on edges in a hydrogen-suppressed model.
We treat a cyclic graph $H$ as a  {\em mixed graph} (a graph possibly
with undirected and directed edges) 
  by regarding each non-core-edge $uv\in E^\nc$
as a directed edge $(u,v)$ such that 
$u$ is the parent of $v$ in some pendant-tree of $H$.    
Each of the examples of   rank-2 graphs 
in Figure~\ref{fig:rank2_graph_examples}  is 
represented as a mixed graph where non-core-edges are regarded
as directed edges. 
 
Let $\Lambda$ be a set of  labels each of which represents a chemical element
 such as
 {\tt C} (carbon), {\tt O} (oxygen), {\tt N} (nitrogen)
 and so on,
 where we assume that $\Lambda$ does not contain {\tt H} (hydrogen).
Let $\mathrm{mass}(\ta)$ and $\val(\ta)$ 
denote the mass and    valence of a  chemical element $\ta\in \Lambda$,
respectively.  
In our model, we   use integers
  $\mathrm{mass}^*(\ta)=\lfloor 10\cdot \mathrm{mass}(\ta)\rfloor$, 
  $\ta\in \Lambda$ and assume that
  each chemical element $\ta\in \Lambda$ has a unique 
  valence  $\val(\ta)\in [1,4]$. 
  
We introduce a total order $<$ over the elements in $\Lambda$
according to their mass values; i.e., we write ${\tt a<b}$
for chemical elements ${\tt a,b}\in \Lambda$ with
 $\mathrm{mass}(\ta)<\mathrm{mass}(\tb)$.
 To represent how two atoms $\ta$ and $\tb$ are joined in a chemical graph,
 we define some notions.
 A tuple $(\ta,\tb,m)$ with chemical elements $\ta,\tb$
 and a bond-multiplicity $m$, called an {\em adjacency-configuration}
 was used to represent a pair of atoms $\ta$ and $\tb$
 joined by a bond-multiplicity $m$~\cite{CWZSNA20}.
 In this paper, we introduce ``edge-configuration,'' 
 a refined notion of adjacency-configuration.

 We represent an atom $\ta\in \Lambda$ with $i$ neighbors
  in  a chemical compound by a pair $(\ta, i)$
  of the chemical element $\ta$ and the degree $i$, 
  which we call a {\em chemical symbol}. 
For a notational convenience,
we write a chemical symbol $(\ta, i)$ (resp., $(\ta, i+j)$) 
 as $\ta i$ (resp., $\ta\{i+j\}$).
 Define  the set of the chemical symbols to be 
 \[ \Ldg(\Lambda,\val) \triangleq
 \{\ta i \mid \ta\in \Lambda, i\in [1,\val(\ta)]\}. \]
 We extend the total order $<$ over $\Lambda$ to one over the elements in $\Ldg(\Lambda,\val)$
 so that $\ta i< \tb j$ if and only if 
  ``$\ta <\tb$'' or ``$\ta=\tb$ and $i<j $.''  
 
 To represent how two atoms $\ta$ and $\tb$ are joined in  a chemical compound, 
 we use a tuple $(\ta i, \tb j, m)$, 
  $\ta i, \tb j \in \Ldg(\Lambda,\val)$, $m\in [1,3]$ such that
 $i$ (resp., $j$) are the number of neighbors of the atom $\ta$ (resp., $\tb$)
 and  $m$ is the bond-multiplicity between these atoms.
 We call the tuple $(\ta i, \tb j, m)$ the 
   {\em edge-configuration} of the pair of adjacent atoms.
Let $\Ldg'$ be a subset of $\Ldg(\Lambda,\val)$. 
We denote by  $\Gamma(\Ldg')$ ` the set of all tuples 
 $\gamma=(\ta i, \tb j, m)\in \Ldg' \times \Ldg' \times [1,3]$
  such that   
  \[  m\leq \min\{ \val(\ta)-i ,  \val(\tb)-j\} .\] 
For a tuple  $\gamma=(\ta i, \tb j, m)\in \Gamma(\Ldg')$,
let $\overline{\gamma}$ denote the tuple $(\tb j,  \ta i, m)$.
Define sets  
\[ \Gamma_{<}(\Ldg')\triangleq 
\{  \gamma=(\ta i, \tb j, m)\in \Gamma(\Ldg') \mid  \ta i < \tb j \}, \]
\[ \Gamma_{=}(\Ldg')\triangleq 
\{  \gamma=(\ta i, \tb j, m)\in \Gamma(\Ldg') \mid  \ta i = \tb j \}, \]
\[ \Gamma_{>}(\Ldg')\triangleq 
\{  \gamma=(\ta i, \tb j, m)\in \Gamma(\Ldg') \mid  \ta i > \tb j \}. \]

As components of a chemical graph to be inferred,
we choose  sets 
 \[  \Ldg^\co \subseteq \Ldg(\Lambda,\val), ~~~
   \Ldg^\nc \subseteq \Ldg(\Lambda,\val), \] 
\[ \Gamma^\co\subseteq \Gamma_{<}(\Ldg^\co)\cup \Gamma_{=}(\Ldg^\co), 
~~ \Gamma^\nc\subseteq \Gamma(\Ldg^\nc)  \]
 such that   $i\geq 2$ for any symbol $\ta i\not\in \Ldg^\co $,
 where the degree of any vertex in the core of a cyclic graph is at least 2. 

Let $e=uv$ be an edge in a chemical graph $G$
such that $\ta,\tb\in \Lambda$ are assigned to
the vertices $u$ and $v$ with   $\deg_G(u)=i$ and  $\deg_G(v)=j$, respectively and
the bond-multiplicity between them is $m$.
When $uv$ is a  core-edge  which is regarded  as an undirected edge,  
 the edge-configuration $\tau(e)$ of edge $e$ is defined to be
 $(\ta i, \tb j, m)$ if $(\ta i, \tb j, m) \in \Gamma^\co$
 (or  $(\tb j,\ta i,m)$ otherwise).
When $uv$ is a non-core-edge  which is regarded  as a directed edge
$(u,v)$ where $u$ is the parent of $v$ in some pendant-tree, 
 the  edge-configuration $\tau(e)$ of edge $e$ is defined to be 
 $(\ta i, \tb j, m)  \in \Gamma^\nc$.
 
When a branch-parameter ${\rho}$ is specified,
we choose  sets 
\[ \Gamma^\inn\subseteq \Gamma^\nc, ~~~
  \Gamma^\ex\subseteq \Gamma^\nc  \]
 such that   $i,j\geq 2$ for any tuple $(\ta i, \tb j ,m)\not\in \Gamma^\inn$,
 where the degree of any ${\rho}$-internal vertex is at least 2. 
  
 We use  a hydrogen-suppressed model because hydrogen atoms can be
added at the final stage.
A {\em chemical cyclic graph} over $\Lambda$ and
 $\Gamma=\Gamma^\co\cup\Gamma^\nc$
  is defined to be 
a  tuple $G=(H,\alpha,\beta)$
of a cyclic graph $H=(V,E)$, a function   $\alpha:V\to \Lambda$ 
and a function $\beta: E\to [1,3]$ 
such that 
\begin{enumerate}
\item[(i)] $H$ is connected;  
\item[(ii)]  $\sum_{uv\in E}\beta(uv)\leq  \val(\alpha(u))$ 
   for each vertex $u\in V$; and
\item[(iii)] $\tau(e) \in \Gamma^\co$
 for each core-edge $e\in E$; and 
 $\tau(e) \in \Gamma^\nc$
 for each directed  non-core-edge $e\in E$. 
\end{enumerate} 
For a notational convenience, we denote the sum of bond-multiplicities
of edges incident to a vertex $u$ as follows:
\[    \beta(u) \triangleq \sum_{uv\in E}\beta(uv) 
\mbox{ for each vertex $u\in V$.}\]

When a branch-parameter ${\rho}$ is given, 
the condition (iii) is given as follows. \\
~~ $\tau(e) \in \Gamma^\co$  for each core-edge $e\in E$; \\
~~  $\tau(e) \in \Gamma^\inn$  
 for each directed ${\rho}$-internal non-core-edge $e\in E$; and \\
~~  $\tau(e) \in \Gamma^\ex$
 for each directed ${\rho}$-external non-core-edge $e\in E$. 

We represent the graph structure of a chemical compound as a graph
with labels on vertices and multiplicity on edges in a hydrogen-suppressed model.    
    

\subsection{Descriptors}\label{sec:descriptors}
In our method, we use only graph-theoretical descriptors for defining a feature vector,
 which facilitates our designing an algorithm for constructing graphs.    
We choose a branch-parameter ${\rho}\geq 1$, 
   sets $ \Ldg^\co $ and $\Ldg^\nc$  of chemical symbols
  and   sets  $\Gamma^\co,\Gamma^\inn$ and $\Gamma^\ex$ 
   of edge-configurations.
 Let   $G=(H=(V,E),\alpha,\beta)$
  be a  chemical cyclic graph  with the chemical symbols 
  and the edge-configurations.

  We define a {\em feature vector} $f(G)$
that consists of the following 16
 kinds of descriptors. 
 
\begin{itemize} 
\item[-] 
$n(G)$: the number $|V|$ of vertices.
 
\item[-] 
$\cs(G)$:  the core size of $G$.
  
\item[-] 
$\ch(G)$:  the  core height of $G$.  

\item[-] $\bl_{{\rho}}(G)$:  the ${\rho}$-branch-leaf-number of $G$. 

\item[-] 
$\overline{\mathrm{ms}}(G)$: the average mass$^*$ of atoms in $G$; 
 i.e., $\overline{\mathrm{ms}}(G)\triangleq 
\sum_{v\in V}\mathrm{mass}^*(\alpha(v))/n(G)$. 

\item[-] 
$\dg_i^\co(G)$,  $i\in [1,4]$: 
the number of core-vertices of degree $i$ in $G$; \\
~~~ i.e., $\dg_i^\co(G)\triangleq |\{v\in V^\co\mid \deg_{H}(v)=i\}|$.

\item[-] 
$\dg_i^\nc(G)$,  $i\in [1,4]$: 
the number of non-core-vertices of degree $i$ in $G$; \\
~~~ i.e., $\dg_i^\nc(G)\triangleq |\{v\in V^\nc\mid \deg_{H}(v)=i\}|$.
    
   
\item[-] $\bd_m^\co(G)$, $m\in[2,3]$: 
the number of  core-edges with bond multiplicity $m$; \\
~~~ i.e., $\bd_m^\co(G)\triangleq \{e\in E^\co\mid \beta(e)=m\}$. 


\item[-] $\bd_m^\inn(G)$,  $m\in[2,3]$: 
the number of  ${\rho}$-internal  edges 
with bond multiplicity $m$; \\
~~~ i.e., $\bd_m^\inn(G)\triangleq \{e\in E^\inn \mid \beta(e)=m\}$.  

\item[-] $\bd_m^\ex(G)$,  $m\in[2,3]$:  
the number of   ${\rho}$-external edges with bond multiplicity $m$; \\
~~~ i.e., $\bd_m^\ex(G)\triangleq \{e\in E^\ex \mid \beta(e)=m\}$.  

\item[-] $\ns_\mu^\co(G)$,  $\mu\in \Ldg^\co$: 
the number of core-vertices $v$ with $\alpha(v)=\ta$ and $\deg_G(v)=i$
for  $\mu=\ta i$.

\item[-] $\ns_\mu^\nc(G)$,  $\mu\in \Ldg^\nc$: 
the number of non-core-vertices $v$ with $\alpha(v)=\ta$ and $\deg_G(v)=i$
for  $\mu=\ta i$.

   
\item[-] $\ec_{\gamma}^\co(G)$, $\gamma \in \Gamma^\nc$: 
 the number of undirected core-edges $e\in E^\co$ such that  $\tau(e)= \gamma$.  
 
\item[-] $\ec_{\gamma}^\inn(G)$,  $\gamma \in \Gamma^\inn $:
 the number of   directed ${\rho}$-internal edges 
   $e\in E^\inn$ such that  $\tau(e)= \gamma$.  
 
\item[-] $\ec_{\gamma}^\ex(G)$, $\gamma \in \Gamma^\ex$:
 the number of  directed ${\rho}$-external edges $e\in E^\ex$ 
 with  $\tau(e)= \gamma$.  
 
\item[-] $\ns_{\tt H}(G)$:  the number of hydrogen atoms; i.e., 
$\displaystyle{ 
     \ns_{\tt H}(G) \triangleq 
      \sum_{v\in V}\val( \alpha(v))  - 2\sum_{e\in E} \beta(e) .    } $
\end{itemize}

 The number $K$ of descriptors in our feature vector $x=f(G)$  
 is $K = |\Ldg^\co|+  |\Ldg^\nc|+ |\Gamma^\co|
  + |\Gamma^\inn|+ |\Gamma^\co|+20$.  
 Note that the set of the above $K$ descriptors is not independent in the sense that
 some descriptor depends on the combination of other descriptors in the set.
 For example, descriptor $\bd_m^\ex(G)$ can be determined by 
$\sum_{\gamma=(\mu, \mu', i)\in \Gamma: i=m }\ec_{\gamma}^\ex(G)$.


\section{A Method for Inferring Chemical Graphs}\label{sec:inverse_process}

\subsection{Framework for the Inverse QSAR/QSPR}

\begin{figure}[!ht]  \begin{center}
\includegraphics[width=.98\columnwidth]{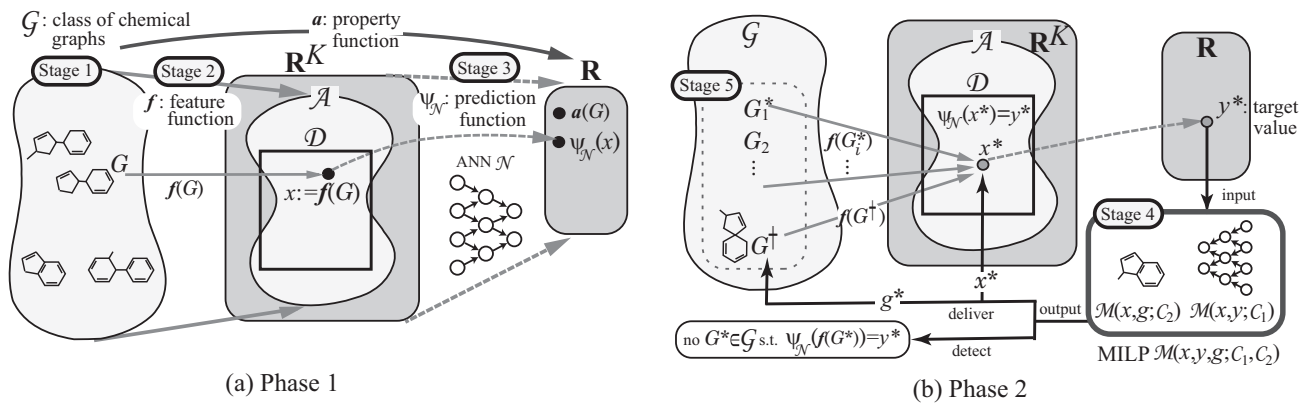}
\end{center} \caption{  (a) An illustration of Phase~1: 
Stage~1 for preparing  a data set $D_{\pi}$ for  a graph class $\G$
and a specified chemical property $\pi$;   
Stage~2 for introducing a feature function $f$ with descriptors;  
Stage~3 for constructing  a prediction function $\psi_\mathcal{N}$ 
with an ANN $\mathcal{N}$;  
(b) An illustration of Phase~2: 
 Stage~4 for formulating 
an  MILP  $\mathcal{M}(x,y,g;\mathcal{C}_1,\mathcal{C}_2)$
and finding   a feasible solution $(x^*,g^*)$ of the MILP
for a target value $y^*$ so that  $\psi_\mathcal{N}(x^*)=y^*$
(possibly detecting that no target graph $G^*$ exists);   
Stage~5 for enumerating graphs $G^*\in \G$ 
such that $f(G^*)=x^*$.  }
\label{fig:framework}  \end{figure}

We review the framework that solves the inverse QSAR/QSPR
by using MILPs ~\cite{CWZSNA20,IAWSNA20,ZCSNA20}, 
which is illustrated in Figure~\ref{fig:framework}.
%
For a specified chemical property $\pi$  such as boiling point,
we denote by $a(G)$ the observed value of the property $\pi$ for a chemical compound $G$.
As the first phase, we solve (I) {\sc Prediction Problem} 
with the following three steps.
 \\
 
\noindent {\bf Phase~1.} \\
\smallskip\noindent 
{\bf Stage~1:}~ 
Let $\mathrm{DB}$ be a set of chemical graphs.
For a specified chemical property $\pi$, choose a class $\G$ of graphs 
such as acyclic graphs or graphs with a given rank $r$.
Prepare a data set $D_{\pi}=\{G_i\mid i=1,2,\ldots,m\}\subseteq 
\G\cap \mathrm{DB}$ such that 
  the  value $a(G_i)$ of each chemical graph
$G_i$, $i=1,2,\ldots,m$ is available.
Set reals  $\underline{a}, \overline{a}\in \mathbb{R}$
so that $\underline{a}\leq  a(G_i)\leq \overline{a}$, $i=1,2,\ldots,m$.  

\smallskip\noindent 
{\bf Stage~2:}~ 
Introduce a feature function $f: \G\to \mathbb{R}^K$ for a positive integer $K$.
We call $f(G)$ the {\em feature vector} of $G\in \G$, and
 call each entry of  a vector $f(G)$  a {\em descriptor} of $G$.   

\smallskip\noindent 
{\bf Stage~3:}~ 
Construct a prediction function $\psi_\mathcal{N}$ 
with an ANN $\mathcal{N}$ that,  
given a   vector  in $\mathbb{R}^K$, 
returns a real in the range $[\underline{a},\overline{a}]$  
so that $\psi_\mathcal{N}(f(G))$ takes a value nearly equal to $a(G)$ 
for many chemical graphs  in  $D$. 

 See Figure~\ref{fig:framework}(a)  for an illustration 
 of  Stages~1, 2 and 3 in Phase~1.  
 
For the set of descriptors $\kappa_j$, $j\in [1,K]$
 in a feature vector $x\in \mathbb{R}^K$,
we can choose lower and upper bounds  $\kappa_j^\LB$ and $\kappa_j^\UB$
on each descriptor $\kappa_j$, and denote by
 $\mathcal{D}$ the set of vectors $x\in \mathbb{R}^K$
 such that   $\kappa_j^\LB\leq x_j \leq \kappa_j^\UB$, $j\in [1,K]$.
 For example, we can use the range-based method 
to define an applicability domain (AD)~\cite{Netzeva05}
 to inverse  QSAR/QSPR by using such a restricted set $\mathcal{D}$. 
Compute  the minimum value $\underline{x_j}$  and the maximum
value $\overline{x_j}$ of the $j$-th descriptor $x_j$ in $f(G_i)$
 over all graphs $G_i$, $i=1,2,\ldots,m$ in a data set $D_\pi$. 
Choose   lower and upper bounds  $\kappa_j^\LB$ and $\kappa_j^\UB$
so that 
 $\underline{x_j}\leq \kappa_j^\LB$ and
 $\kappa_j^\UB\leq \overline{x_j}$, $j\in [1,K]$.

In the second phase,  we try to find a vector  $x^*\in \mathbb{R}^K$  
from a target value $y^*$ of  the chemical propery $\pi$ 
 such that $\psi_\mathcal{N}(x^*)=y^*$. 
Based on the  method due to Akutsu and Nagamochi~\cite{AN19},
  Chiewvanichakorn~et~al.~\cite{CWZSNA20} 
  showed that this problem can be formulated as an MILP. 
By including a set of linear constraints such that $x\in  \mathcal{D}$ 
into their MILP,  we obtain the next result. 

\begin{theorem} \label{Th1}{\rm (\cite{IAWSNA20,ZCSNA20})}
Let   $\mathcal{N}$ be an ANN with  a piecewise-linear activation function
for an input vector $x\in \mathbb{R}^K$, 
 $n_A$ denote the number of   nodes in the architecture 
  and   $n_B$ denote the total number of break-points
over all  activation functions. 
Then there is an MILP $\mathcal{M}(x,y;\mathcal{C}_1)$  
that consists of variable vectors
$x\in \mathcal{D}~(\subseteq \mathbb{R}^K)$, 
$y\in \mathbb{R}$, 
and an auxiliary variable vector $z\in \mathbb{R}^p$ 
for some integer  $p=O(n_A+n_B)$
and a set $\mathcal{C}_1$ of $O(n_A+n_B)$ constraints on these variables 
such that:  $\psi_{\mathcal{N}}(x^*)=y^*$ if and only if
 there is a vector    $(x^*,y^*)$   feasible to  $\mathcal{M}(x,y;\mathcal{C}_1)$.
\end{theorem}
 

In the second phase, we  solve  (II) {\sc Inverse Problem},
 wherein    given a target chemical value $y^*$,
 we are asked to generate chemical graphs $G^*\in\G$
 such that $f(G^*)=x^*$.  
For this, we first  find a vector $x^*\in \mathbb{R}^K$
 such that $\psi_{\mathcal{N}}(x^*)=y^*$ and
 then generate chemical graphs $G^*\in\G$  such that $f(G^*)=x^*$.  
 However, the resulting vector $x^*$ may not admit
 such a chemical graph  $G^*\in\G$.  
 Azam~et~al.~\cite{ACZSNA20} called 
  a vector $x\in \mathbb{R}^K$  {\em admissible}
 if   there is a graph $G\in \G$  such that $f(G)=x$.  
 Let $\mathcal{A}$ denote the set of admissible vectors $x\in \mathbb{R}^K$. 
 To ensure that a vector $x^*$ inferred from a given target value $y^*$
  becomes admissible, we introduce
    a new vector variable $g\in \mathbb{R}^{q}$ for an integer $q$
 so that a feasible solution $(x^*,g^*)$ of the MILP
 for a target value $y^*$ delivers a vector $x^*$ with 
$\psi_{\mathcal{N}}(x^*)=y^*$ and
 a vector $g^*$ that represents  a  chemical  graph 
 $G^\dagger\in \G$  with $f(G^\dagger)=x^*$.  
In the second phase, we treat the next two problems.\\

\smallskip
\noindent (II-a)   Inference of Vectors \\
{\bf Input:} A real $y^*$ with $\underline{a}\leq y^*\leq \overline{a}$. \\
{\bf Output:} Vectors $x^*\in  \mathcal{A}\cap   \mathcal{D}$ 
and $g^*\in  \mathbb{R}^{q}$ such that $\psi_\mathcal{N}(x^*)=y^*$
and  $g^*$ forms a chemical graph $G^\dagger\in \G$ with 
$f(G^\dagger)=x^*$.

\bigskip  \noindent
 (II-b)  Inference of Graphs \\
{\bf Input:} A vector $x^*\in \mathcal{A}\cap \mathcal{D}$.    \\
{\bf Output:} All graphs $G^*\in \G$ such that
$f(G^*)=x^*$.    
\smallskip

The second phase consists of the next two steps.

\medskip \noindent {\bf Phase~2.}  \\
\smallskip\noindent 
{\bf Stage~4:}~  Formulate Problem (II-a)    
as the above MILP  $\mathcal{M}(x,y,g;\mathcal{C}_1,\mathcal{C}_2)$ 
based on $\G$ and $\mathcal{N}$. 
Find a feasible solution $(x^*,g^*)$ of the MILP 
such that  
 \[\mbox{
  $x^*\in \mathcal{A}\cap \mathcal{D}$  and  $\psi_\mathcal{N}(x^*)=y^*$ }\]
(where the second requirement may be replaced with inequalities  
 $(1-\varepsilon)y^* \leq \psi_\mathcal{N}(x^*) \leq(1+\varepsilon)y^*$
 for a tolerance $\varepsilon>0$).

\smallskip\noindent 
{\bf Stage~5:}~ To solve Problem (II-b),
enumerate all (or a specified number) of graphs $G^*\in \G$ 
such that $f(G^*)=x^*$ for the inferred vector $x^*$. \\

See Figure~\ref{fig:framework}(b) for an illustration of  Stages~4 and 5 in Phase~2.
  
\bigskip
\subsection{A New Mechanism for Stage~5}
Execution of Stage~5; i.e.   generating chemical  graphs $G^*$
that satisfy $f(G^*)=x^*$ for a given feature vector $x^*\in\mathbb{Z}_+^K$
  is a challenging issue  for a relatively large instance with
  size $n(G^*)\geq 20$.
There have been proposed  algorithms for Stage~5 
 for  classes of graphs with rank 0 to 2
~\cite{Fujiwara08,Suzuki14,2A1B20,2A2B20}.
All of these are designed based on the branch-and-bound method
where an enormous number of chemical graphs are
constructed by repeatedly appending and removing a vertex  one by one
until a target chemical graph is constructed.
These algorithms can generate a target chemical graph with size $n(G^*)\leq 20$.
To break this barrier,  Azam~et~al.~\cite{AZSSSZNA20} recently 
 employed the dynamic programming method
for designing a new algorithm in Stage~5  
and showed that  chemical acyclic graphs $G^*$ 
with a bounded branch-height
can be generated for size $n(G^*)=50$. 
However, for a class of graphs with a different rank, 
we may need to design again a new algorithm
 by the dynamic programming method.
 Moreover, algorithms for higher ranks can be more complicated
 and do not run as fast as the algorithm for acyclic graphs due to 
Azam~et~al.~\cite{AZSSSZNA20}.

\begin{figure}[!ht]  \begin{center}
\includegraphics[width=.98\columnwidth]{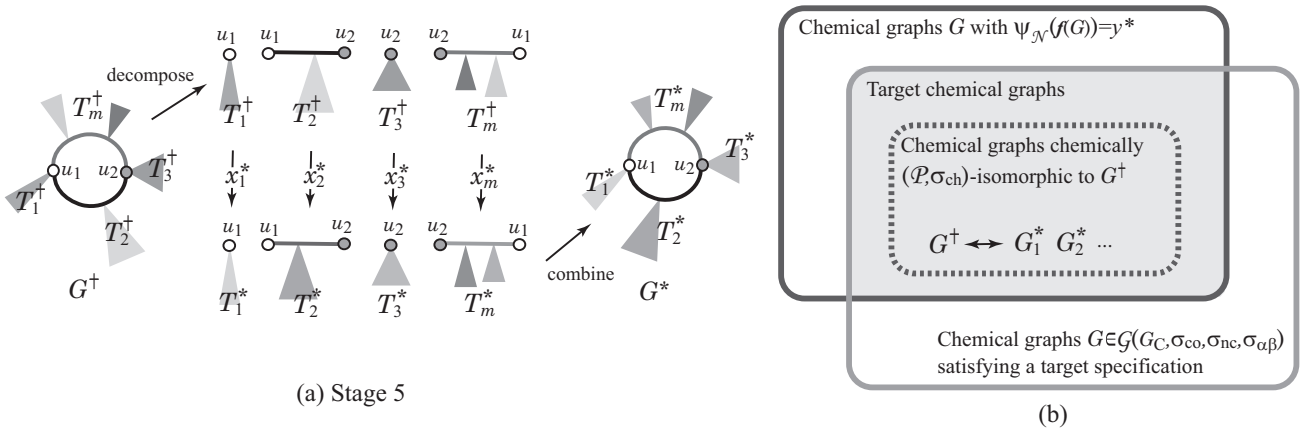}
\end{center} \caption{ An illustration of a process of Stage~5
and a set of target isomorphic graphs: 
 (a) 
A new mechanism to Stage~5,
where a given target chemical graph $G^\dagger$ is decomposed
into chemical trees $T^\dagger_i$, $i=1,2,\ldots,m$
into chemical trees $T^\dagger_i$, $i=1,2,\ldots,m$ based on
 a set $V_B=\{u_1,u_2\}$ of core-vertices
and for each feature vector $x^*_i=f(T^\dagger_i)$,
 a chemical tree $T^*_i$ such that
$f(T^*_i)=x^*_i$ is constructed before 
a new target graph $G^*_1$ is obtained as a combination of
the resulting chemical trees $T^*_1,\ldots,T^*_m$; 
(b) 
 Given a target value $y^*$,
 an MILP $\mathcal{M}(x,y,g;\mathcal{C}_1,\mathcal{C}_2)$
 delivers a target chemical graph $G^\dagger$, if
 the intersection of the set of chemical graphs $G$ with  
 $\psi_\mathcal{N}(f(G))=y^*$ and the set of
 chemical graphs $G$
 that satisfy a target specification $(\GC,\sco,\snc,\sab)$ is not empty,
 where all chemically $(\Pt,\sch)$-isomorphic graphs to $G^\dagger$
 belong to the intersection. 
  }
\label{fig:chemical-isomorphism}  \end{figure}

In this paper,  
as a new mechanism of Stage~5, 
we adopt an idea of utilizing the chemical graph
$G^\dagger\in \G$  obtained as part of 
a feasible solution of an MILP in Stage~4.
In other words, we modify the chemical graph 
$G^\dagger$ to generate other chemical graphs $G^*$ that 
are ``chemically isomorphic'' to $G^\dagger$ 
in the sense that  $f(G^*)=f(G^\dagger)$ holds.
Informally speaking, we reduce the problem of finding such a graph $G^*$
into a problem of generating chemical acyclic graphs, to which
we have obtained an efficient dynamic programming algorithm~\cite{AZSSSZNA20}.
We first decompose $G^\dagger$ into a collection of
chemical trees $T^\dagger_1, T^\dagger_2, \ldots, T^\dagger_m$
such that for a subset $V_B$ of the core-vertices of  $G^\dagger$,
any  tree  $T^\dagger_i$ contains at most two   vertices in $V_B$, 
as illustrated in Figure~\ref{fig:chemical-isomorphism}(a).
Let $x^*_i$ denote  the feature vector $f(T^\dagger_i)$. 
For each index $i$, we generate chemical acyclic graphs $T^*_i$
such that $f(T^*_i)=x^*_i$.
Finally we combine the generated chemical trees 
 $T^*_1, T^*_2, \ldots, T^*_m$ to construct
 a chemical cyclic graph $G^*$  such that 
 $f(G^*)=\sum_{i\in[1,m]}x^*_i =f(G^\dagger)$.
 See Section~\ref{sec:graph_search} for the details. 
Although a family of chemical graphs $G^*$ chemically isomorphic to
$G^\dagger$ depends on a choice of decomposition into trees $T^\dagger_i$
and covers only part of the entire set of target graphs $G^*$
with $f(G^*)=x^*$, the new method can be applied to any class of graphs
or even to a graph with a specific substructure. 
 
\subsection{A Flexible Specification to Target Chemical Graphs} 
In the previous application of the framework,
a target chemical graph $G$ to be inferred is specified with a small number of
parameters such as the number $n(G)$ of vertices,
the core size $\cs(G)$ and the core height $\ch(G)$.

In this paper, we  also introduce
a more flexible way of specifying a target graph so that
our new algorithm for generating chemically isomorphic graphs $G^*$
can be used.
Suppose that we are given a requirement $R$ on
a target graph specified other than the feature vector $f$.
Now a target graph is defined to be a chemical graph $G$
 that satisfies 
 $\psi_\mathcal{N}(f(G))=y^*$ for a given target value $y^*$
 and the requirement at the same time.
In general, a chemical graph $G^*$ such that $f(G^*)=f(G^\dagger)$ 
 may not satisfy such an additional requirement $R$. 
Recall that $G^*$ in Stage~5  is obtained as a combination
of chemical trees $T^*_i$ each of which is chemically isomorphic to
the corresponding tree $T^\dagger_i$
 of the given graph $G^\dagger$.  
Hence if  the requirement $R$ on a target chemical graph is 
independent among such chemical trees $T^*_i$ to be inferred
from a vector $x^*_i$, then any combination $G^*$ of
inferred chemical trees $T^*_i$ still satisfies the requirement $R$,
whenever the original graph $G^\dagger$ satisfies $R$.  
See Figure~\ref{fig:chemical-isomorphism}(b) for an illustration of 
the set of chemical graphs $G^*$ that are chemically isomorphic 
to a target chemical graph $G^\dagger$. 
Section~\ref{sec:specification} describes a way of specifying
a requirement, called a ``target specification'' such that  a prescribed 
  substructure of  graphs such as a benzene ring
  to be included in a target chemical graph
  or a partly predetermined assignment of chemical elements 
  and bond-multiplicity   to  a target graph.


\section{Specifying Target Chemical Graphs}\label{sec:specification} 

This section  presents a flexible way of specifying a topological structure
of the core and assignments of chemical elements
and bond-multiplicity of a target chemical graph. 
We define a {\em target specification} 
 $(\GC,\sco,\snc,\sab)$ with  a multigraph $\GC$ and
sets $\sco,\snc$ and $\sab$
of lower and upper bounds on several descriptors
that we describe in the following. 

\subsection{Seed Graphs}

\begin{figure}[h!] \begin{center}
\includegraphics[width=.55\columnwidth]{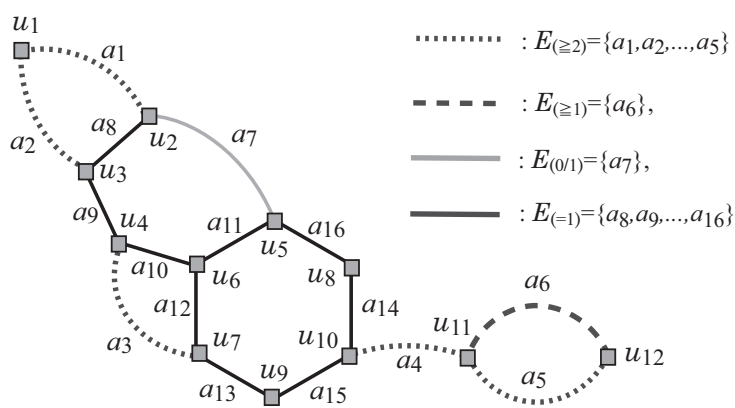}
\end{center} \caption{An illustration of a seed graph $\GC$ with 
$\Et=\{a_1,a_2,\ldots,a_5\}$, 
$\Ew=\{a_6\}$,
$\Ez=\{a_7\}$ and 
$\Eew=\{a_8,a_9,\ldots,a_{16}\}$,
where the vertices in $\VC$ are depicted with gray squares,
the edges in $\Et$ are depicted with dotted lines,
the edges in $\Ew$ are depicted with dashed lines,
the edges in $\Ez$ are depicted with gray lines and 
the edges in $\Eew$ are depicted with black solid lines.  }
\label{fig:specification_example_1} \end{figure}

A  {\em seed graph} $\GC=(\VC,\EC)$ is defined
to be a multigraph with no self-loops such that 
the edge set $\EC$ consists of four sets 
$\Et$, $\Ew$, $\Ez$ and $\Eew$, 
where each of them can be empty.
Figure~\ref{fig:specification_example_1} illustrates an example of a seed graph.
From a seed graph $\GC$, the core of a cyclic graph will be
constructed in the following way:
\begin{enumerate}
\item[-]
Each edge $e=uv\in \Et$ will be replaced
with a $u,v$-path $P_e$ of length at least 2.

\item[-] 
Each edge $e=uv\in \Ew$ will be replaced
with a $u,v$-path $P_e$ of length at least 1
(equivalently $e$ is directly used or replaced with
a $u,v$-path $P_e$ of length at least 2).

\item[-] 
Each edge $e\in \Ez$ is either used or discarded.

\item[-]
Each edge $e\in \Eew$ is always used directly. 
\end{enumerate}

\subsection{Core Specification}
The core of a target chemical graph is
constructed from a seed graph $\GC$ by a  {\em core specification} $\sco$
that consists of the following: 
\begin{enumerate}
\item[-] 
Lower and upper bound functions
$\ell_\LB, \ell_\UB:  \Et\cup\Ew \to \mathbb{Z}_+$.
For a notational convenience, set 
$\ell_\LB(e):=0$, $\ell_\UB(e):=1$, $e\in \Ez$ and
$\ell_\LB(e):=1$, $\ell_\UB(e):=1$, $e\in \Eew$. 

\item[-]
Lower and upper bounds $\cs_\LB, \cs_\UB\in \mathbb{Z}_+$ on
the core size, where we assume  
$\cs_\LB\geq |\VC|+ \sum_{e\in  \Et\cup\Ew}(\ell_\LB(e) -1)$. 

\item[-] Side constraints: 
As an option, we can specify additional linear constraints on
 the length $\ell(P_i)$ of path $P_i$, $a_i\in \EC$ such as
 $\ell(P_2)+\ell(P_3)= c$ for a constant $c$ or
$\ell(P_1)\leq \ell(P_4)+\ell(P_5)$. 
\end{enumerate}

An example of a core specification $\sco$ to the seed graph $\GC$ in 
Figure~\ref{fig:specification_example_1} is given in Table~\ref{table:core_spec}. 
 
\begin{table}[h!]\caption{Example~1 of a core specification  $\sco$. }
 \begin{center}
 \begin{tabular}{ |  c | c c c c c c |  } \hline
                        & $a_1$ &  $a_2$ &   $a_3$ &   $a_4$ &   $a_5$ &   $a_6$   \\\hline
 $\ell_\LB(a_i)$&  2 &  2 &  2 & 3 &  2 &  1 \\ \hline
 $\ell_\UB(a_i)$&  3 & 4 &  3 & 5 & 4 &  4 \\\hline
\end{tabular}
\begin{tabular}{ |  c | c |  } \hline 
$\cs_\LB=20$ & $\cs_\UB = 28$ \\\hline 
\end{tabular}
\end{center}\label{table:core_spec}
\end{table}

\begin{figure}[h!] \begin{center}
\includegraphics[width=.55\columnwidth]{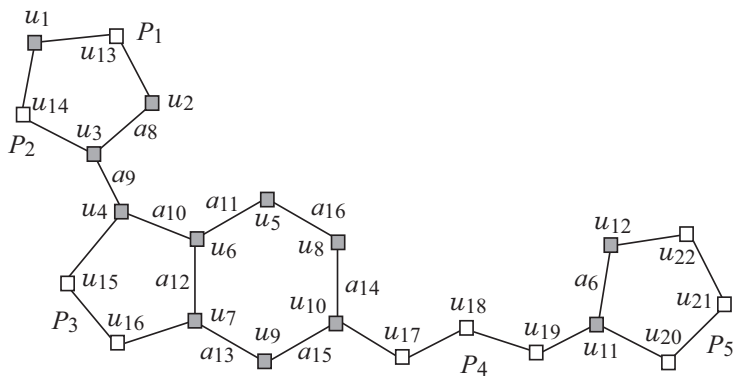}
\end{center} \caption{An illustration of a $\sco$-extension $C$ with $\cs(C)=22$, where 
the vertices in $V(C)\setminus \VC$ are depicted with white squares. }
\label{fig:specification_example_2} \end{figure}

 A {\em $\sco$-extension}  of a seed graph   $\GC$ 
is defined to be a graph $C$ such that $|V(C)|\in [\cs_\LB, \cs_\UB]$
and $C$ is obtained from
replacing each edge $e=uv\in \Et\cup\Ew$
with a $u,v$-path $P_e$ of length 
 $\ell(P_e)\in  [\ell_\LB(e), \ell_\UB(e)]$
 under specified side constraints, if any. 
 Figure~\ref{fig:specification_example_2} illustrates one of the $\sco$-extensions  of 
 the seed graph  $\GC$ in   Figure~\ref{fig:specification_example_1}
  with the core specification $\sco$ in  Table~\ref{table:core_spec}. 
The edges $a_i\in \Et$, $i\in [1,5]$ are replaced with paths 
$P_1=(u_1,u_{13},u_3)$,
 $P_2=(u_1,u_{14},u_3)$,
  $P_3=(u_4,u_{15},u_{16},u_7)$,
 $P_4=(u_{10},u_{17},u_{18},u_{19}, u_{11})$ and 
 $P_5=(u_{11},u_{20},u_{21},u_{22}, u_{12})$, respectively.
 The edge $a_6\in \Ew$ is used in the graph $C$ and
 the edge $a_7\in \Ez$ is discarded,
 where $\Eew\subseteq E(C)$. 
  

Let $\mathcal{C}(\GC,\sco)$ denote the set of all $\sco$-extensions of a seed graph   $\GC$.
We employ a graph $C\in \mathcal{C}(\GC,\sco)$ as the core
$\Cr(G)$ of a chemical graph $G$ to be inferred. 

Remember that the core of any connected cyclic graph is a simple connected
graph with minimum degree at least 2.
Possibly some $\sco$-extension of a seed graph   $\GC$ is not such a graph.
We show some sufficient condition for any $\sco$-extension to be a simple connected
graph with minimum degree at least 2. 
  Let $C_{\min} \in  \mathcal{C}(\GC,\sco)$ denote the minimum
 $\sco$-extension; i.e., $C_{\min} $ is obtained from
  the graph $(\VC, \Et\cup\Ew\cup\Eew)$ 
by replacing each edge $e\in \Et$ with
a path of the least length $\ell_\LB(e)$.
We see that if $C_{\min}$ is
   a connected graph with minimum degree at least 2 then 
any extension  $C\in \mathcal{C}(\GC,\sco)$ becomes
 a simple connected graph with minimum degree at least 2.

\subsection{Non-core Specification} 

Next we show how to specify the structure of the non-core part of a target chemical graph.
For a seed graph $\GC$,
let a  {\em non-core specification} $\snc$ consist of the following: 
\begin{enumerate}
\item[-] 
Lower and upper bounds $n_\LB,  n^*\in \mathbb{Z}_+$
on the number of vertices, where $\cs_\LB \leq n_\LB\leq n^*$.

\item[-] 
An upper bound  $\dg^\nc_{4,\UB}\in \mathbb{Z}_+$
on the number of non-core-vertices of degree 4. 
 
\item[-] Lower and upper functions 
$\ch_{\LB},\ch_{\UB}: \VC\to \mathbb{Z}_+$
and 
$\ch_{\LB},\ch_{\UB}: \Et\cup\Ew \to \mathbb{Z}_+$
on the maximum height of trees rooted at a vertex $v\in \VC$
or at an internal vertex of a path $P_e$ with $e\in \Et\cup\Ew$. 

\item[-] A branch-parameter ${\rho}\in \mathbb{Z}_+$.

\item[-] Lower and upper functions 
$\bl_\LB, \bl_\UB: \VC  \to \{0,1\}$ on the number of
leaf ${\rho}$-branches in the tree rooted at a vertex $v\in \VC$,
where $\bl_\UB(u)\leq 1$ for any vertex $u\in \VC$ 
for inferring a ${\rho}$-lean cyclic graph and 
$\bl_\UB(u)=0$ if $\ch_\UB(u)\leq {\rho}$; \\
Lower and upper functions  
$\bl_\LB,\bl_\UB: \Et\cup\Ew\to \mathbb{Z}_+$
on the number of
leaf ${\rho}$-branches in the trees rooted at internal vertices in a path $P_e$
constructed for an edge $e\in  \Et\cup\Ew$,
where 
$\bl_\UB(e)\leq  \ell_\UB(e)-1$;
and  $\ch_\LB(u)> {\rho}$ ($\ch_\UB(u)\leq {\rho}$) implies 
$\bl_\LB(e)\geq 1$ (resp., $\bl_\UB=0$).

\item[-] Side  constraints: 
As an option, we can specify additional linear constraints on
  $\ell(P_i)$ and the number $\bl(P_i)$ of leaf ${\rho}$-branches
in the trees rooted at $P_i$, $a_i\in \EC$ such as
$\bl(P_2)+\bl(P_3)\leq c$ for a constant.
\end{enumerate}
 
An example of a non-core specification $\snc$ to the seed graph $\GC$ in 
Figure~\ref{fig:specification_example_1} is given in Table~\ref{table:non-core_spec}.

\begin{table}[h!]\caption{Example~2 of a core specification  $\snc$. }
\begin{tabular}{ |  l |  } \hline
 $n_\LB=30$,  $n^* =50$. \\\hline
  branch-parameter:   ${\rho}=2$  \\\hline
\end{tabular}

\begin{tabular}{ |  c | c c c c c c   c c c c cc |  } \hline
                        & $u_1$ &  $u_2$ &   $u_3$ &   $u_4$ &   $u_5$ &   $u_6$ 
                       & $u_7$ &   $u_8$ &   $u_9$ &   $u_{10}$ &   $u_{11}$ &   $u_{12}$ \\\hline
 $\ch_\LB(u_i)$&  0 &  0 &   0 & 0 &  1 &   0
                       & 0 &   0 &  0 &   0 &  0 &  0 \\ \hline
 $\ch_\UB(u_i)$&  1 & 0 &   0 & 0 & 3 &   0
                       & 1 &   1 &  0 &   1 &  2 & 4 \\\hline
\end{tabular} 

\begin{tabular}{ |  c | c c c c c c |  } \hline
                        & $a_1$ &  $a_2$ &   $a_3$ &   $a_4$ &   $a_5$ &   $a_6$   \\\hline
 $\ch_\LB(a_i)$&  0 &  1 & 0 & 4 &  3 &  0 \\ \hline
 $\ch_\UB(a_i)$&  3 & 3 &  1 & 6 & 5 &  2 \\\hline
\end{tabular}
 
\begin{tabular}{ |  c | c c c c c c   c c c c cc |  } \hline
                        & $u_1$ &  $u_2$ &   $u_3$ &   $u_4$ &   $u_5$ &   $u_6$ 
                       & $u_7$ &   $u_8$ &   $u_9$ &   $u_{10}$ &   $u_{11}$ &   $u_{12}$ \\\hline
 $\bl_\LB(u_i)$&  0 &  0 &   0 & 0 &  0 &   0
                       & 0 &   0 &  0 &   0 &  0 &  0 \\ \hline
 $\bl_\UB(u_i)$&  1 & 1 &   1 & 1 & 1 &   0
                       & 0 &   0 &  0 &   0 &  0 &  0 \\\hline
\end{tabular} 
 
\begin{tabular}{ |  c | c c c c c c |  } \hline
                        & $a_1$ &  $a_2$ &   $a_3$ &   $a_4$ &   $a_5$ &   $a_6$   \\\hline
 $\bl_\LB(a_i)$&  0 &  0 &   0 & 1 &  1 &   0 \\ \hline
 $\bl_\UB(a_i)$&  1 & 1 &   0 & 2 & 1 &   0 \\\hline
\end{tabular}
\label{table:non-core_spec}
\end{table}

 Let $C\in \mathcal{C}(\GC,\sco)$ be a $\sco$-extension of $\GC$,
 where each edge $e=uv\in \Et\cup\Ew$ is replaced with
a $u,v$-path $P_e$ (where possibly $P_e$ is equal to $e$). 
We consider a ${\rho}$-lean cyclic graph $H$ obtained from $C$  by appending
a tree $T_v$ with at most one leaf ${\rho}$-branch at each vertex $v\in V(C)$,
where possibly $E(T_v)=\emptyset$.
We call the vertices in $C$ {\em core-vertices} of $H$
and the newly added vertices {\em non-core-vertices} of $H$. 
 For each edge $e=uv\in \Et\cup\Ew$ 
let $\F(P_e)$ denote the set of  trees $T_w$ rooted
at internal vertices $w$ of the $u,v$-path $P_e$
(where $w\neq u,v$).

We call the above ${\rho}$-lean cyclic graph $H$ obtained 
from a graph $C\in \mathcal{C}(\GC,\sco)$
a {\em $(\sco,\snc)$-extension}    of   $\GC$ 
if the following hold: 
\begin{enumerate}
\item[-]
  $n(H)\in [n_\LB, n^*]$. 
 
\item[-] 
   $\dg_4^\nc(H)\leq \dg^\nc_{4,\UB}$.  

\item[-] 
For each vertex   $v\in \VC$, the tree $T_v$ attached to $v$ satisfies 
$\h(T_v)\in [\ch_{\LB},\ch_{\UB}]$;  
For each  edge $e\in \Et\cup\Ew$,  
$\max\{\h(T)\mid T\in \F(P_e)\}\in [\ch_\LB(e),  \ch_\UB(e)]$. 

\item[-] 
Each tree $T_v$, $v\in V(C)$ contains at most one leaf ${\rho}$-branch;
i.e., $H$ is a ${\rho}$-lean graph with $\Cr(H)=C$.

\item[-] 
For each  edge $e\in \Et\cup\Ew$, 
$\sum\{\bl_{{\rho}}(T)\mid T\in \F(P_e)\}
\in  [\bl_\LB(e), \bl_\UB(e)]$.

\item[-] 
The additional linear constraints are satisfied.
\end{enumerate}

\begin{figure}[h!] \begin{center}
\includegraphics[width=.75\columnwidth]{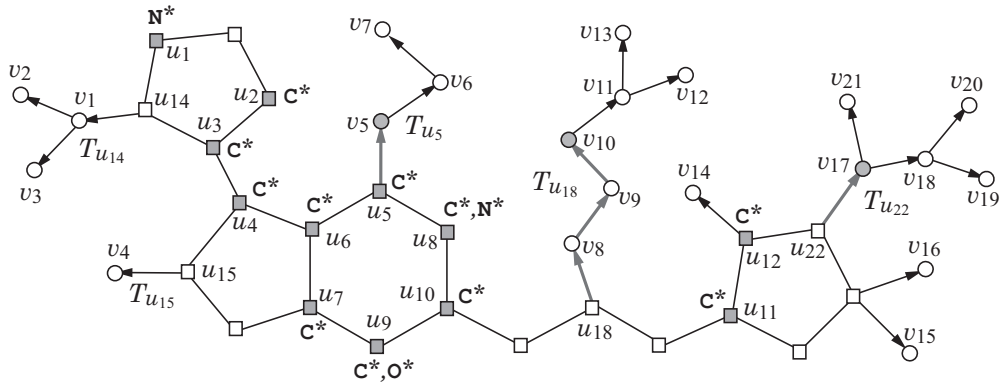}
\end{center} \caption{An illustration of 
a $(\sco,\snc)$-extension $H$ with $\Cr(H)=C$
with $n(H)=43$,  $\ch(H)=5$ and $\bl_2(H)=3$,
where the non-core-vertices  are depicted with circles, 
 the leaf 2-branches are depicted with gray circles, 
  the 2-internal edges are depicted with thick gray arrows and 
 the elements in $\Lambda^*(v)$, $v\in\VC$ are indicated with asterisk. }
\label{fig:specification_example_3} \end{figure}

 Figure~\ref{fig:specification_example_3} illustrates
  one of the $(\sco,\snc)$-extensions  of 
 the seed graph  $\GC$ in   Figure~\ref{fig:specification_example_1}
  with the specifications $\sco$ in  Table~\ref{table:core_spec}
  and $\snc$  in  Table~\ref{table:non-core_spec}. 
  
Let $\mathcal{H}(\GC,\sco,\snc)$ denote the set of all 
$(\sco,\snc)$-extensions of a seed graph $\GC$. 
We employ a graph $H\in \mathcal{H}(\GC,\sco,\snc)$
 as the underlying graph based on which we
 assign elements in $\Lambda$ and 
 bond-multiplicities to infer a chemical graph
 $G=(H,\alpha,\beta)$.

\subsection{Chemical Specification}  

A {\em chemical specification} $\sab$ consists of the following: 
\begin{enumerate}
\item[-] 
We choose a set $\Lambda$ of chemical elements 
\[ \Lambda^\co, \Lambda^\nc \subseteq \Lambda. \]
For a chemical graph $G$,
let $\na_\ta(G)$
(resp., $\na_\ta^\co(G)$ and $\na_\ta^\nc(G)$) 
 denote the number of vertices   (resp.,   core-vertices and  non-core-vertices) in $G$
 assigned   chemical element  
 $\ta\in \Lambda$   (resp.,  $\ta\in \Lambda^\co$ and $\ta\in \Lambda^\nc$).

\item[-] 
We choose sets of   symbols 
 \[  \Ldg^\co \subseteq \Ldg(\Lambda^\co,\val), ~~~
   \Ldg^\nc \subseteq \Ldg(\Lambda^\nc,\val)  \]
 such that   $i\geq 2$ for any symbol $\ta i\not\in \Ldg^\co $.  
the number of core-vertices $v$ with $\alpha(v)=\ta$ and $\deg_G(v)=i$
for  $\mu=\ta i$.
We choose sets of  edge-configurations
   \[
  \Gamma^\co\subseteq \Gamma_{<}(\Ldg^\co)\cup \Gamma_{=}(\Ldg^\co), 
~~ \Gamma^\nc\subseteq \Gamma(\Ldg^\nc),   \] 
\[ \Gamma^\inn\subseteq \Gamma^\nc, ~~~
  \Gamma^\ex\subseteq \Gamma^\nc  \]
 such that   $i,j\geq 2$ for any tuple 
 $(\ta i, \tb j ,m)\not\in \Gamma^\inn$. \\
Define $\Gamma^\co_{>}
\triangleq \{\overline{\gamma}=(\xi, \mu,m) 
\mid \gamma=(\mu,\xi,m)\in\Gamma^\co, \mu<\xi \}$.

\item[-] 
 The {\em induced} adjacency-configuration  $\ac(\gamma)$
 of  an edge-configuration $(\ta d, \tb d',m)$ is defined to be 
 the adjacency-configuration $\ac(\gamma)=(\ta, \tb, m)$.
Set the following  sets of   adjacency-configurations: 
\[  \Gac^\co:=\{\ac(\gamma)\mid \gamma\in \Gamma^\co\},
\Gacl^\co :=\{\ac(\gamma)\mid \gamma\in \Gamma^\co_{>}\}, \]
\[
  \Gac^\inn:=\{\ac(\gamma)\mid \gamma\in \Gamma^\inn\},
 \Gac^\ex :=\{\ac(\gamma)\mid \gamma\in \Gamma^\ex\}.  \]
  In a chemical specification, 
 we define the adjacency-configuration of a core-edge $uv$ 
 to be $(\ta,\tb,\beta(uv))$ with $\{\ta,\tb\}=\{\alpha(u),\alpha(v)\}$ 
 and 
 the adjacency-configuration of  a  directed non-core edge $(u,v)$
 to be $(\alpha(u),\alpha(v),\beta(uv))$. \\
Let  $\ac_\nu^\co(G)$  (resp., $\ac_\nu^\inn(G)$ and $\ac_\nu^\ex(G)$) 
denote  the number of core-edges   (resp.,  directed ${\rho}$-internal edges and
directed   ${\rho}$-external edges) in $G$ assigned  
adjacency-configuration   $\nu\in \Gac^\co$ 
  (resp.,   $\nu\in \Gac^\inn$ and  $\nu\in \Gac^\ex$).

\item[-] 
 Subsets $\Lambda^*(v)$, $v\in \VC$ of elements
 that are allowed to be assigned to vertex $v\in \VC$;  

\item[-] Lower and upper bound functions 
$\na_\LB,\na_\UB: \Lambda\to  [1,n^*]$
 and 
$\na_\LB^\typ,\na_\UB^\typ: \Lambda^\typ\to  [1,n^*]$,
$\typ\in \{\co,\nc\}$ on the number of 
   core-vertices and   non-core-vertices, respectively, 
  assigned chemical element     $\ta$. 

\item[-] Lower and upper bound functions 
$\ns_\LB,\ns_\UB: \Ldg\to  [1,n^*]$ and 
$\ns_\LB^\typ,\ns_\UB^\typ: \Ldg^\typ\to  [1,n^*]$,
$\typ\in \{\co,\nc\}$ on the number of 
    core-vertices and   non-core-vertices, respectively, 
  assigned   symbol     $\mu$.   

\item[-] Lower and upper bound functions  
$\ac_\LB^\typ,\ac_\UB^\typ: \Gac^\typ\to  \mathbb{Z}_+$,
$\typ\in \{\co,\inn,\ex\}$ on the number of 
 core-edges,  directed  ${\rho}$-internal edges and
 directed  ${\rho}$-external edges, respectively,   
assigned adjacency-configuration    $\nu$. 

\item[-] Lower and upper bound functions  
$\ec_\LB^\typ,\ec_\UB^\typ: \Gamma^\typ\to  \mathbb{Z}_+$,
$\typ\in \{\co,\inn,\ex\}$ on the number of 
 core-edges,  directed  ${\rho}$-internal edges and
 directed  ${\rho}$-external edges, respectively,   
assigned edge-configurations    $\gamma$.   

\item[-] Lower and upper functions 
$\bd_{m, \LB}, \bd_{m, \UB}:  \EC \to \mathbb{Z}_+$, $m\in [2,3]$,
where $\bd_{2, \LB}(e)+\bd_{3, \LB}(e)\leq \ell_\UB(e)$, $e\in \EC$;

\item[-] Side  constraints: 
Lower and upper bounds on the number of  some adjacency-configurations 
and edge-configurations; 
We can specify additional linear constraints on
 $\ell(P_i)$, $\bl(P_i)$ and  
 the number $\na_\ta(P_i)$ of chemical element $\ta$
  in the path $P_i$, $a_i\in \EC$
  such as
$\na_{\tt N}(P_2)+\na_{\tt N}(P_3)\leq c$ for a constant $c$
and nitrogen ${\tt N}\in \Lambda$.
\end{enumerate}
 
An example of a chemical specification $\sab$ to the seed graph $\GC$ in 
Figure~\ref{fig:specification_example_1} is given in Table~\ref{table:chemical_spec}.

For a graph $H\in \mathcal{H}(\GC,\sco,\snc)$,
let  $\alpha: V(H)\to \Lambda$ and $\beta: E(H)\to [1,3]$
be functions.
Then $G=(H,\alpha,\beta)$ is called a $(\sco,\snc,\sab)$-extension of $\GC$
if the following hold:
\begin{enumerate}
\item[-]
   $\sum_{uv\in E}\beta(uv)\leq  \val(\alpha(u))$ 
   for each vertex $u\in V(H)$;  
 $\tau(e) \in \Gamma^\co$  for each core-edge $e$; 
 $\tau(e) \in \Gamma^\inn$  for each directed ${\rho}$-internal edge; and
 $\tau(e) \in \Gamma^\ex$  for each directed ${\rho}$-external edge.
 
\item[-] 
 $\alpha(v)\in \Lambda^*(v)$ for each vertex $v\in \VC$.

\item[-]
It holds that
$\na_\ta(G)\in [\na_\LB(\ta),\na_\UB(\ta)]$, $\ta\in \Lambda$, 
$\na_\ta^\co(G)\in [\na_\LB^\co(\ta),\na_\UB^\co(\ta)]$, $\ta\in \Lambda^\co$, 
and
$\na_\ta^\nc(G)\in [\na_\LB^\nc(\ta),\na_\UB^\nc(\ta)]$, $\ta\in \Lambda^\nc$;

\item[-]
It holds that
$\ns_\mu(G)\in [\na_\LB(\mu),\na_\UB(\mu)]$, $\mu\in \Ldg$, 
$\ns_\mu^\co(G)\in [\ns_\LB^\co(\mu),\ns_\UB^\co(\mu)]$, $\mu\in \Ldg^\co$, 
and
$\ns_\mu^\nc(G)\in [\ns_\LB^\nc(\mu),\ns_\UB^\nc(\mu)]$, $\mu\in \Ldg^\nc$;

\item[-]
It holds that
$\ac_\nu^\co(G)\in [\ac_\LB^\co(\nu),\ac_\UB^\co(\nu)]$, $\nu\in \Gac^\co$, 
$\ac_\nu^\inn(G)\in [\ac_\LB^\inn(\nu),\ac_\UB^\inn(\nu)]$, $\nu\in \Gac^\inn$, 
and
$\ac_\nu^\ex(G)\in [\ac_\LB^\ex(\nu),\ac_\UB^\ex(\nu)]$, $\nu\in \Gac^\ex$;

\item[-]
It holds that
$\ec_\gamma^\co(G)\in [\ec_\LB^\co(\gamma),\ec_\UB^\co(\gamma)]$, $\gamma\in \Gamma^\co$, 
$\ec_\gamma^\inn(G)\in [\ec_\LB^\inn(\gamma),\ec_\UB^\inn(\gamma)]$, $\gamma\in \Gamma^\inn$, 
and
$\ec_\gamma^\ex(G)\in [\ec_\LB^\ex(\gamma),\ec_\UB^\ex(\gamma)]$, $\gamma\in \Gamma^\ex$;

\item[-] 
For each edge $e\in \Et\cup \Ew$, 
$|\{e\in E(P_e)\mid \beta(e)=m\}|\in [\bd_{m, \LB}(e), \bd_{m, \UB}(e)]$;

\item[-] 
The additional linear constraints are satisfied.
\end{enumerate}

\begin{figure}[h!] \begin{center}
\includegraphics[width=.75\columnwidth]{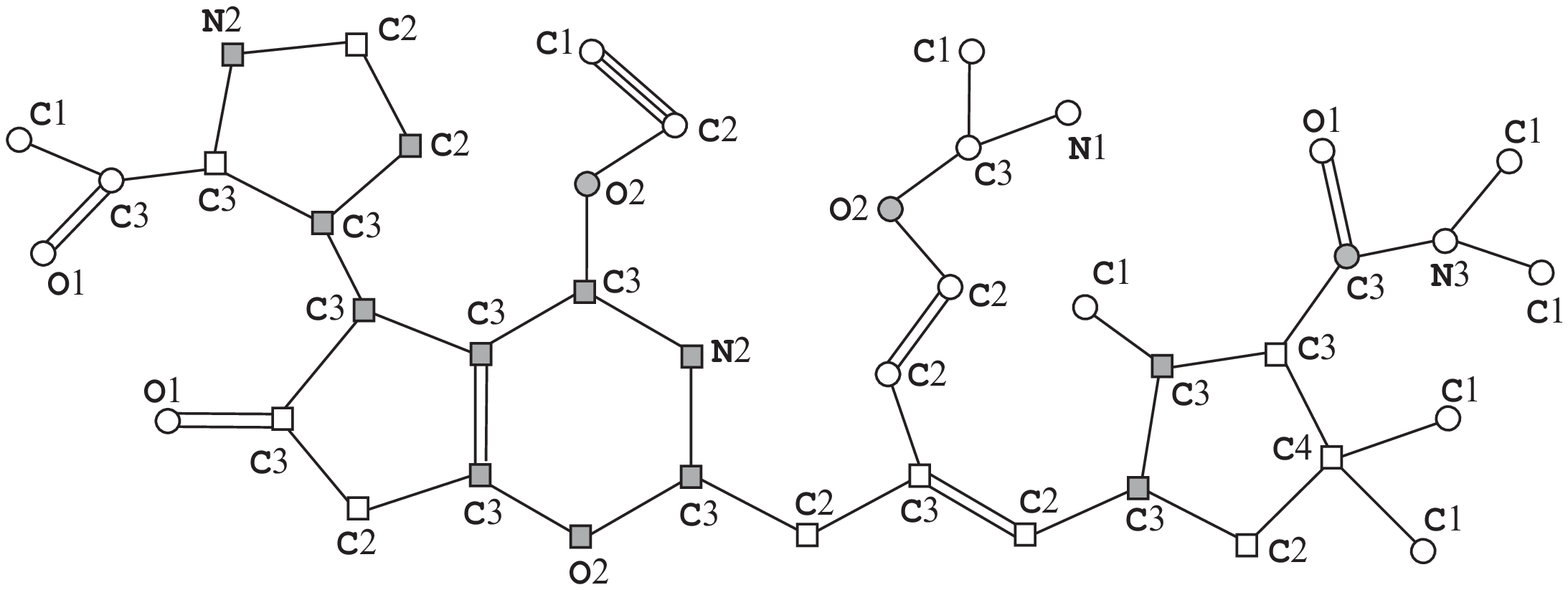}
\end{center} \caption{An illustration of a   $(\sco,\snc,\sab)$-extension 
$G=(H,\alpha,\beta)$
of the seed graph $\GC$ in Figure~\ref{fig:specification_example_1}
with the specifications $\sco,\snc$ and $\sab$in Tables~\ref{table:core_spec}, 
\ref{table:non-core_spec} and \ref{table:chemical_spec}, respectively, 
where a symbol $\mu\in \Ldg$ is depicted with a pair
of an element $\ta$ and a degree $i$ such as ${\tt C}1$. }
\label{fig:specification_example_4} \end{figure}

 Figure~\ref{fig:specification_example_4} illustrates
  one of the $(\sco,\snc,\sab)$-extensions  of 
 the seed graph  $\GC$ in   Figure~\ref{fig:specification_example_1} 
with the specifications $\sco,\snc$ and $\sab$ 
in Tables~\ref{table:core_spec}, 
\ref{table:non-core_spec} and \ref{table:chemical_spec}, respectively,

Let $\mathcal{G}(\GC,\sco,\snc,\sab)$ denote the set of all 
$(\sco,\snc,\sab)$-extensions of $\GC$.  

When a required condition on a target chemical graph to be inferred
is described with a target specification $( \sco,\snc,\sab)$
with a seed graph $\GC$, the inverse QSAR/QSPR can be formulated
as an MILP, as discussed in the next section. 

\begin{table}[h!]\caption{Example~3 of a chemical specification  $\sab$. }
\begin{tabular}{ |  c | c | c |  } \hline
  $\Lambda=\{{\tt C,N,O}\}$ & 
  $\Ldg^\co=\{    {\tt C} 2 , {\tt C } 3,   {\tt C } 4, {\tt N } 2  , {\tt O } 2   \}$ & 
  $\Ldg^\nc=\{  {\tt C } 1, {\tt C} 2 , {\tt C } 3,   {\tt C } 4, 
   {\tt N } 1, {\tt N } 3,    {\tt O } 1, {\tt O } 2\}$ \\\hline
\end{tabular}

\begin{tabular}{ |  c | l |  } \hline
  $\Gac^\co$ &
  $ \nu_1^\co=({\tt C}   , {\tt C }  , 1) ,   \nu_2 ^\co=({\tt C}   , {\tt C }  , 2) ,   
    \nu_ 3^\co=({\tt C}   , {\tt N }  , 1) ,  \nu_ 4^\co= ({\tt C}  , {\tt O }  , 1) $ \\ \hline
  $\Gacl^\co $ &
  $ \overline{\nu_ 3}^\co =({\tt N}   , {\tt C}  , 1) ,  \overline{\nu_4}^\co= ({\tt O}  , {\tt C}  , 1) $ \\ \hline
  $\Gac^\inn$ & 
  $ \nu_1 ^\inn=({\tt C}   , {\tt C } , 1),  \nu_2 ^\inn= ({\tt C}   , {\tt C }  , 2),   
  \nu_3^\inn= ({\tt C}   , {\tt O }  , 1)    $  \\\hline
  $\Gac^\ex$ & 
 $ \nu_1^\ex=({\tt C}   , {\tt C } , 1), \nu_2^\ex= ({\tt C}   , {\tt C }  , 3) ,   
  \nu_3^\ex= ({\tt C}   , {\tt N }  , 1) , \nu_4^\ex= ({\tt N}   , {\tt C}  , 1) ,  
  \nu_5^\ex=({\tt C}   , {\tt O }  , 1),  \nu_6^\ex=({\tt C}   , {\tt O }  , 2),      $  \\
   & $ 
      \nu_7^\ex=({\tt O}   , {\tt C}  , 1) $  \\\hline
\end{tabular}

\begin{tabular}{ |  c | l |  } \hline
  $\Gamma^\co$ &
  $ \gamma_1^\co= ({\tt C} 2 , {\tt C } 2, 1) ,
   \gamma_2^\co=({\tt C} 2 , {\tt C } 3, 1) ,  
   \gamma_3^\co=({\tt C} 2 , {\tt C } 3, 2) ,  
   \gamma_4^\co=({\tt C} 2 , {\tt C } 4, 1) , 
   \gamma_5^\co=({\tt C} 3 , {\tt C } 3, 1) ,  $ \\
   &
  $   
   \gamma_6^\co=({\tt C} 3 , {\tt C } 3, 2) ,
    \gamma_7^\co= ({\tt C} 3 , {\tt C } 4, 1), 
   \gamma_8^\co= ({\tt C} 2 , {\tt N } 2, 1) ,  
   \gamma_9^\co=({\tt C} 3 , {\tt N } 2, 1) ,  
   \gamma_{10}^\co=({\tt C} 3 , {\tt O } 2, 1) $ \\ \hline
  $\Gamma^\co_{>}$ &   
   $\overline{\gamma_2}^\co=({\tt C} 3 , {\tt C } 2, 1) ,  
   \overline{\gamma_3}^\co=({\tt C} 3 , {\tt C } 2, 2) ,  
   \overline{\gamma_4}^\co=({\tt C} 4 , {\tt C } 2, 1) , 
    \overline{\gamma_7}^\co= ({\tt C} 4 , {\tt C } 3, 1),   
   \overline{\gamma_8}^\co= ({\tt N} 2 , {\tt C } 2, 1) ,   $ \\
   &
  $   
   \overline{\gamma_9}^\co=({\tt N } 2, {\tt C} 3 , 1) ,  
   \overline{\gamma_{10}}^\co=({\tt O } 2, {\tt C} 3 , 1)$ \\ \hline
  $\Gamma^\inn$ & 
  $ \gamma_1^\inn=({\tt C} 2 , {\tt C } 2, 2),     
   \gamma_2^\inn=({\tt C} 3 , {\tt C } 2, 1),  
 \gamma_3^\inn=({\tt C} 3 , {\tt C } 3, 1) ,     
   \gamma_4^\inn=({\tt C} 2 , {\tt O } 2, 1) ,
   \gamma_5^\inn=({\tt C} 3 , {\tt O } 2, 1)  $  \\\hline
  $\Gamma^\ex$ & 
 $ \gamma_1^\ex=({\tt C} 3 , {\tt C } 1, 1), 
   \gamma_2^\ex=({\tt C} 2 , {\tt C } 1, 3) ,  
  \gamma_3^\ex= ({\tt C} 3 , {\tt C } 3, 1) , 
  \gamma_4^\ex=({\tt C} 4 , {\tt C } 1, 1) ,
   \gamma_5^\ex=({\tt C} 3 , {\tt N } 1, 1) , $ \\
   &
  $  
   \gamma_6^\ex=({\tt C} 3 , {\tt N } 3, 1) , 
   \gamma_7^\ex=  ({\tt C} 3 , {\tt O } 1, 2) , 
   \gamma_8^\ex=({\tt O} 2 , {\tt C } 2, 1) ,
   \gamma_9^\ex=({\tt O} 2 , {\tt C } 3, 1) ,
   \gamma_{10}^\ex=({\tt N} 3 , {\tt C } 1, 1)$  \\\hline
\end{tabular}

\begin{tabular}{ |  l|  } \hline
$\Lambda^*(u_1)=\{{\tt N}\}$, 
$\Lambda^*(u_8)=\{{\tt C, N}\}$, 
$\Lambda^*(u_9)=\{{\tt C, O}\}$, 
   $\Lambda^*(u)=\{{\tt C}\}$, $u\in \VC\setminus\{u_1,u_8,u_9\}$
   \\\hline
\end{tabular}

\begin{tabular}{ |  c | c c c   |  } \hline
                           & ${\tt C}$ &   ${\tt N}$ &     ${\tt O}$  \\\hline
 $\na_\LB(\lambda)$ &  27 &  1 &   1    \\ \hline
 $\na_\UB(\lambda) $&  37 & 4 &  8  \\\hline
\end{tabular} 
\begin{tabular}{ |  c | c c c   |  } \hline
                           & ${\tt C}$ &   ${\tt N}$ &     ${\tt O}$  \\\hline
 $\na_\LB^\co(\lambda)$ &   9 &  1 &   0       \\ \hline
 $\na_\UB^\co(\lambda) $&  23 & 4 & 5   \\\hline
\end{tabular} 
\begin{tabular}{ |  c | c c c   |  } \hline
                           & ${\tt C}$ &   ${\tt N}$ &     ${\tt O}$  \\\hline
 $\na_\LB^\nc(\lambda)$ &  9 &  1 &   2    \\ \hline
 $\na_\UB^\nc(\lambda) $&  18 & 3 &  8  \\\hline
\end{tabular}

\begin{tabular}{ |  c | c c c c c c  c c c |  } \hline
                           & ${\tt C}1$ &  ${\tt C}2$ &   ${\tt C}3$ & ${\tt C}4$ &   ${\tt N}1$ &   ${\tt N}2$ 
                           &   ${\tt N}3$ &   ${\tt O}1$ &   ${\tt O}2$  \\\hline
 $\ns_\LB(\mu)$ &  6 &  7 &   12 & 0 &  0 &   0 & 0 &  0 &   0   \\ \hline
 $\ns_\UB(\mu) $&  10 & 11 & 18 & 2 & 2 &  2   & 2 & 5 &   5  \\\hline
\end{tabular} 
 
\begin{tabular}{ |  c | c c c c c c   |  } \hline
                              & ${\tt C}2$ &  ${\tt C}3$ &   ${\tt C}4$ & ${\tt N}2$ &   ${\tt N}3$ &   ${\tt O}2$ \\\hline
 $\ns_\LB^\co(\mu)$ &  3 &  5 &   0 & 0 &  0 &   0    \\ \hline
 $\ns_\UB^\co(\mu) $&  8 & 15 & 2 & 2 & 3 &  5     \\\hline
\end{tabular} 
\begin{tabular}{ |  c | c c c c c c  c c c |  } \hline
       & ${\tt C}1$ &  ${\tt C}2$ &   ${\tt C}3$ & ${\tt C}4$ &   ${\tt N}1$ &   ${\tt N}2$ 
                           &   ${\tt N}3$ &   ${\tt O}1$ &   ${\tt O}2$  \\\hline
 $\ns_\LB^\nc(\mu)$ &  6 &  1 &   1 & 0 &  0 &   0 & 0 &  0 &   0   \\ \hline
 $\ns_\UB^\nc(\mu) $&  10 & 5 & 5 & 2 & 2 &  2   & 2 & 5 &   5  \\\hline
\end{tabular} 

\begin{tabular}{ |  c | c c c c |  } \hline
         & $\nu_1^\co$ &   $\nu_2^\co$ & $\nu_3^\co$   & $\nu_4^\co$ \\\hline
 $\ac_\LB^\co(\nu)$ &  0 &  0 & 0 & 0      \\ \hline
 $\ac_\UB^\co(\nu) $&  30 & 10 & 10 & 10  \\\hline
\end{tabular} 
\begin{tabular}{ |  c | c c c  |  } \hline
                           & $\nu_1^\inn$ &   $\nu_2^\inn$ & $\nu_3^\inn$   \\\hline
 $\ac_\LB^\inn(\nu)$ &  0 &  0  & 0      \\ \hline
 $\ac_\UB^\inn(\nu) $& 5 &   5 & 5  \\\hline
\end{tabular}  
\begin{tabular}{ |  c | c c c c c c c |  } \hline
     & $\nu_1^\ex$ &   $\nu_2^\ex$  & $\nu_3^\ex$   & $\nu_4^\ex$ & $\nu_5^\ex$ 
      & $\nu_6^\ex$ & $\nu_7^\ex$ \\\hline
 $\ac_\LB^\ex(\nu)$ &  0 & 0 &0 & 0 & 0  & 0    & 0    \\ \hline
 $\ac_\UB^\ex(\nu) $& 10 &  10 & 10 & 10 & 10 & 10 & 10 \\\hline
\end{tabular} 

\begin{tabular}{ |  c | c c c c c c c c c c |  } \hline
    & $\gamma_1^\co$ &   $\gamma_2^\co$ & $\gamma_3^\co$   & $\gamma_4^\co$ 
     & $\gamma_5^\co$
    & $\gamma_6^\co$ &   $\gamma_7^\co$ & $\gamma_8^\co$   & $\gamma_9^\co$ 
     & $\gamma_{10}^\co$                     
                            \\\hline
 $\ec_\LB^\co(\gamma)$ &  0 &  0 & 0 &  0  & 0 &  0 &  0 & 0 &  0  & 0  \\ \hline
 $\ec_\UB^\co(\gamma) $& 4 & 15 & 4 &  4  & 10 &  5 & 4 & 4 &  6 & 4   \\\hline
\end{tabular} 
\begin{tabular}{ |  c | c c c c c |  } \hline
    & $\gamma_1^\inn$ &   $\gamma_2^\inn$ & $\gamma_3^\inn$  & $\gamma_4^\inn$ 
     & $\gamma_5^\inn$                           
                             \\\hline
 $\ec_\LB^\inn(\gamma)$  &  0  & 0   & 0  & 0 & 0      \\ \hline
 $\ec_\UB^\inn(\gamma)$ &  3  &  3   & 3  & 3 & 3      \\\hline
\end{tabular} 
\begin{tabular}{ |  c | c c c c c c c c c c |  } \hline
    & $\gamma_1^\ex$ &   $\gamma_2^\ex$ & $\gamma_3^\ex$   & $\gamma_4^\ex$ 
     & $\gamma_5^\ex$
    & $\gamma_6^\ex$ &   $\gamma_7^\ex$ & $\gamma_8^\ex$   & $\gamma_9^\ex$ 
     & $\gamma_{10}^\ex$                     
                            \\\hline
 $\ec_\LB^\ex(\gamma)$ &  0 &  0 & 0 &  0 & 0  &  0 &  0 & 0 & 0 & 0 \\ \hline
 $\ec_\UB^\ex(\gamma) $&  8 &  4 & 4 &  4 & 4  &  4 &  6 & 4 & 4 &  4 \\\hline
\end{tabular}

\begin{tabular}{ |  c | c c c c c c   c c c c c c  c c c c |  } \hline
                               & $a_1$ &  $a_2$ &   $a_3$ &   $a_4$ &   $a_5$ &   $a_6$ 
                               & $a_7$ &  $a_8$ &   $a_9$ &   $a_{10}$ &   $a_{11}$ &   $a_{12}$ 
                               & $a_{13}$ &   $a_{14}$ &   $a_{15}$ &   $a_{16}$  \\\hline
 $\bd_{2, \LB}(a_i)$ &  0    &  0 &   0 & 1 &  0 &   0
                                &  0   &  0 &  0 & 0 &  0 &   1
                                &  0    &  0 &   0 & 0     \\ \hline
 $ \bd_{2, \UB}(a_i)$&  1    & 1 &   0 & 2  & 2 &   0  
                                &  0    & 0&   0 & 0 &  0 &   1
                                &  0    &  0 &   0 & 0    \\ \hline
\end{tabular} 

\begin{tabular}{ |  c | c c c c c c   c c c c c c  c c c c |  } \hline
                               & $a_1$ &  $a_2$ &   $a_3$ &   $a_4$ &   $a_5$ &   $a_6$ 
                               & $a_7$ &  $a_8$ &   $a_9$ &   $a_{10}$ &   $a_{11}$ &   $a_{12}$ 
                               & $a_{13}$ &   $a_{14}$ &   $a_{15}$ &   $a_{16}$  \\\hline
 $\bd_{3, \LB}(a_i)$ &  0    &  0 &   0 & 0 &  0 &   0
                                &  0   &  0 &  0 & 0 &  0 &   0
                                &  0    &  0 &   0 & 0    \\ \hline
 $ \bd_{3, \UB}(a_i)$&  0    & 0 &   0 & 0  & 1 &   0 
                                &  0    &  0 &   0 & 0 &  0 &   0
                                &  0    &  0 &   0 &  0     \\ \hline
\end{tabular} 
\label{table:chemical_spec}
\end{table}

\clearpage 

\subsection{Abstract Specification for Cores}\label{sec:abstract_specification} 

The framework for the inverse QSAR/QSPR
~\cite{CWZSNA20,IAWSNA20,ZCSNA20} has been applied to 
a case of chemical graphs with an abstract topological structure such as
 acyclic or monocyclic graphs 
 by Ito~et~al.~\cite{ACZSNA20} 
 and rank-2 cyclic graphs with a specified polymer topology 
with a cycle index up to 2 by  Zhu~et~al.~\cite{ZCSNA20}.

We show that such classes of cyclic graphs can be
specified with part of our target specification $(\sco,\snc,\sab)$
to a seed graph. 

In their applications~\cite{IAWSNA20,ZCSNA20}, 
 a set $\Lambda$ of chemical elements is given and 
the graph size $n(G)$, 
the core size $\cs(G)$ and the core height $\ch(G)$ of
 a target graph $G$ are required to be
 prescribed values 
$\cs^\dagger$,  $\ch^\dagger$ and $n^\dagger$. 
Then we specify the bounds on these values in $\snc$ 
so that 
$n_\LB:=n^*:=n^\dagger$;  
$\cs_\LB:=\cs_\UB:=\cs^\dagger$;   
$\ch_\LB(\typ):=\ch_\UB(\typ):=\ch^\dagger$
for some graph element $\typ\in \VC\cup \Et\cup \Ew$; and
$\ch_\LB(\typ):=0$, $\ch_\UB(\typ):=\ch^\dagger$
for the other elements $\typ\in \VC\cup \Et\cup \Ew$.
We set $\Ldg^\co$ and $\Ldg^\nc$ to be
the sets of all possible symbols in $\Lambda\times[1,4]$
and set $\Gamma^\co$, $\Gamma^\inn$ and $\Gamma^\ex$
to be the sets of  all possible edge-configurations in $\sab$.

\begin{figure}[h!] \begin{center}
\includegraphics[width=.98\columnwidth]{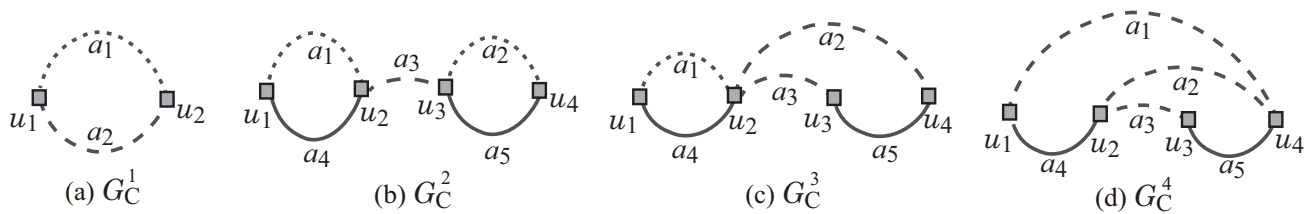}
\end{center} \caption{An illustration of seed graphs for
inferring cyclic graphs with rank at most 2: 
(a) A seed graph $\GC^1$ for monocyclic graphs;
(b) A seed graph $\GC^2$ for rank-2 cyclic graphs with the 
polymer topology $M_1\in\mathcal{PT}(2,4)$ in Figure~\ref{fig:rank_2_polymer}(d); 
(c) A seed graph $\GC^3$ for rank-2 cyclic graphs with the 
polymer topology $M_2\in\mathcal{PT}(2,4)$ in Figure~\ref{fig:rank_2_polymer}(e); 
(d) A seed graph $\GC^4$ for rank-2 cyclic graphs with the 
polymer topology $M_3\in\mathcal{PT}(2,4)$ in Figure~\ref{fig:rank_2_polymer}(f). }
\label{fig:specification_example_polymer} \end{figure}

A seed graph for inferring a chemical monocyclic graphs
can be selected as a multigraph $\GC^1$ with a vertex set  
$\VC=\{u_1,u_2\}$ and edge sets 
$\Et=\{a_1\}$ and  
$\Ew=\{a_2\}$, 
as illustrated in Figure~\ref{fig:specification_example_polymer}(a). 
We can include a linear constraint 
$\ell(a_1)\leq \ell(a_2)$ as part of the side constraint in $\sco$.
This constraints reduces the search space on an MILP. 

A seed graph for inferring a chemical rank-2 cyclic graphs
with  the  polymer topology $M_1\in\mathcal{PT}(2,4)$
 in Figure~\ref{fig:rank_2_polymer}(d)
can be selected as a multigraph $\GC^2$ with a vertex set  
$\VC=\{u_1,u_2,u_3,u_4\}$ and edge sets 
$\Et=\{a_1,a_2\}$, 
$\Ew=\{a_3\}$  and 
$\Eew=\{a_4,a_5\}$, 
as illustrated in Figure~\ref{fig:specification_example_polymer}(b). 
We can include a linear constraint 
$\ell(a_1)\leq \ell(a_2)$ as part of the side constraint in $\sco$. 

A seed graph for inferring a chemical rank-2 cyclic graphs
with  the  polymer topology $M_2\in\mathcal{PT}(2,4)$
 in Figure~\ref{fig:rank_2_polymer}(e)
can be selected as a multigraph $\GC^3$ with a vertex set  
$\VC=\{u_1,u_2,u_3,u_4\}$ and edge sets 
$\Et=\{a_1\}$, 
$\Ew=\{a_2, a_3\}$  and 
$\Eew=\{a_4,a_5\}$, 
as illustrated in Figure~\ref{fig:specification_example_polymer}(c). 
We can include a linear constraint 
$\ell(a_1)\leq \ell(a_2)+\ell(a_3)$ and $\ell(a_2)\leq \ell(a_3)$ 
in $\sco$. 

A seed graph for inferring a chemical rank-2 cyclic graphs
with  the  polymer topology $M_3\in\mathcal{PT}(2,4)$
 in Figure~\ref{fig:rank_2_polymer}(f)
can be selected as a multigraph $\GC^4$ with a vertex set  
$\VC=\{u_1,u_2,u_3,u_4\}$ and edge sets 
$\Ew=\{a_1, a_2, a_3\}$  and 
$\Eew=\{a_4,a_5\}$, 
as illustrated in Figure~\ref{fig:specification_example_polymer}(d). 
We can include a linear constraint 
$\ell(a_2)\leq \ell(a_1)+1$,
$\ell(a_2)\leq \ell(a_3)+1$ and $\ell(a_1)\leq \ell(a_3)$ in $\sco$.

 \clearpage  
  
\section{MILPs for Chemical ${\rho}$-lean Graphs} 
 \label{sec:graph_MILP}

Let $(\GC,\sco,\snc,\sab)$ be a target specification,
where ${\rho}$ denotes the branch-parameter in $\snc$. 
In this section, we formulate
to an  MILP  $\mathcal{M}(x,g;\mathcal{C}_2)$  in Stage~4 
for inferring a chemical ${\rho}$-lean cyclic graph  $G\in \mathcal{G}(\GC,\sco,\snc,\sab)$.

\subsection{Scheme Graphs}\label{sec:scheme_graph}
 
 Recall that we treat the underlying  graph 
of a chemical cyclic graph as a mixed graph to define our descriptors, 
where
core-edges are undirected edges and non-core-edges are directed.  
To formulate an MILP that infers a chemical cyclic graph, 
we further assign a  direction of each core-edge  
 so that constraints on a function $\tau:E\to \Gamma$ can be described 
 notationally simpler. 

Our method first gives directions to the edges in a given seed graph
such that  
$\VC=\{u_1,u_2,\ldots, u_p\}$, 
$\EC=\{a_1,a_2,\ldots, a_q\}$ and each edge $a_i\in \EC$
is a directed edge $a_i=(u_j,u_h)$ with $j<h$.
Let $\widetilde{\VC}$  denote the set of vertices $u\in \VC$ such that 
$\bl_\UB(u)=1$ and $\widetilde{\tC}=|\widetilde{\VC}|$. 
Our method first arranges the order of verices in $\VC$
so that  
\[ \mbox{ 
$\bl_\UB(u_i)=1$, $i\in [1,\widetilde{\tC}]$ and 
$\bl_\UB(u_i)=0$, $i\in[\widetilde{\tC}+1, \tC]$. }\]

Next our method adds some more vertices and edges to the resulting digraph $\GC$
to construct a digraph, called 
 a {\em scheme graph} $\mathrm{SG}=(\mathcal{V},\mathcal{E})$  
 so that 
    any  ${\rho}$-lean graph  $H\in \mathcal{H}(\GC,\sco,\snc)$  
    (i.e., any $(\sco,\snc)$-extension $H$ of $\GC$) 
  can be chosen as a subgraph of the scheme graph $\mathrm{SG}$.  

To construct a scheme graph, our method first computes 
some integers that determine the size of each building block in  $\mathrm{SG}$.

For a given specification $(\sco,\snc)$, 
define 
\[ \begin{array}{ll}
\displaystyle{ 
 \widetilde{\ch}_\LB \triangleq  \sum_{v\in\VC}\ch_\LB(v) +
   \sum_{e\in\Ew\cup\Et}\ch_\LB(e) } &    \\
 %
%
\displaystyle{ 
 \bl^*_\LB \triangleq \sum_{v\in\VC}\bl_\LB(v) +
\sum_{e\in\Ew\cup\Et} \bl_\LB(e), } & 
\displaystyle{ 
 \bl^*_\UB \triangleq \sum_{v\in\VC} \bl_\UB(v)  +
\sum_{e\in\Ew\cup\Et} \bl_\UB(e) , }
\end{array} \]
\[
\displaystyle{ 
\ell^*_{\inl} \triangleq   \sum_{v\in\VC} \max\{\ch_\UB(v)-{\rho}, 0\}
 + \sum_{e\in\Ew\cup\Et}\bl_\UB(e)\cdot\max\{\ch_\UB(e)-{\rho}, 0\},  }\] 
\[ \beta^*_i \triangleq |\Et(u_i)|+|\Ew(u_i)|
+\sum_{e\in \Eew(u_i)}(1+\bd_{2,\LB}(e)+2\bd_{3,\LB}(e))
+\bl_\LB(u_i), ~  u_i\in\VC, \]
\[ \Delta_i := 4-\beta^*_i,  ~ u_i\in \VC. \]

Define integers that determine the size of a scheme graph $\mathrm{SG}$ as follows.
\begin{equation}\label{eq:SG_parameter}
\begin{array}{l}
  \tC:=|\VC|,  \widetilde{\tC}:=|\{u\in \VC\mid \bl_\UB(u)=1\}|,  \mC:=|\EC|, \\
 \tT:=\cs_\UB-|\VC|, \\
\tF:= \min \bigl[ n^*-\cs_\LB 
  - \max\{\widetilde{\ch}_\LB, {\rho}\cdot \bl^*_\LB\},
            \ell^*_{\inl} \bigr], \\
 \mbox{$\dmax:=3$ if $\dg^\nc_{4,\UB}=0$; 
              $\dmax:=4$ if $\dg^\nc_{4,\UB}\geq 1$, } \\
  \nC(i):=\Delta_i \cdot ((\dmax\!-\!1)^{{\rho}}-1)/(\dmax-2), \\
 \nT:=2((\dmax\!-\!1)^{{\rho}}-1)/(\dmax-2), \\ 
\nF:=(\dmax\!-\!1)((\dmax\!-\!1)^{{\rho}}-1)/(\dmax-2), \\
\displaystyle{ 
\mUB^\co
     :=\min \{ n^*+\mathrm{r}(\GC), 
       \mC+ \!\!\!\! \sum_{e\in\Et\cup\Ew} \!\!\!\!  \ell_\UB(e) \} , }\\
\displaystyle{ 
\mUB^\nc
     :=  \min\{n^*, \tF+ \!\! \sum_{i\in[1,\tC]}\nC(i) +\nT\cdot\tT+\nF\cdot\tF -1\}, }\\ 
\mUB 
    :=\min \{n^*+\mathrm{r}(\GC), \mUB^\co+\mUB^\nc\},  
\end{array}  \end{equation}
where
$\nC(i)$ is the  number of ``edges''  in the rooted  tree $T(\Delta_i, \dmax-1,   {\rho})$, 
$\nT$ is the  number of ``edges''  in the rooted  tree $T(2, \dmax-1,   {\rho})$ and 
     $\nF$  is the  number of ``edges''  in the rooted  tree $T(\dmax-1, \dmax-1,   {\rho})$.
Recall that any core-vertex is allowed to be of degree 4.

\begin{figure}[h!] \begin{center}
\includegraphics[width=.98\columnwidth]{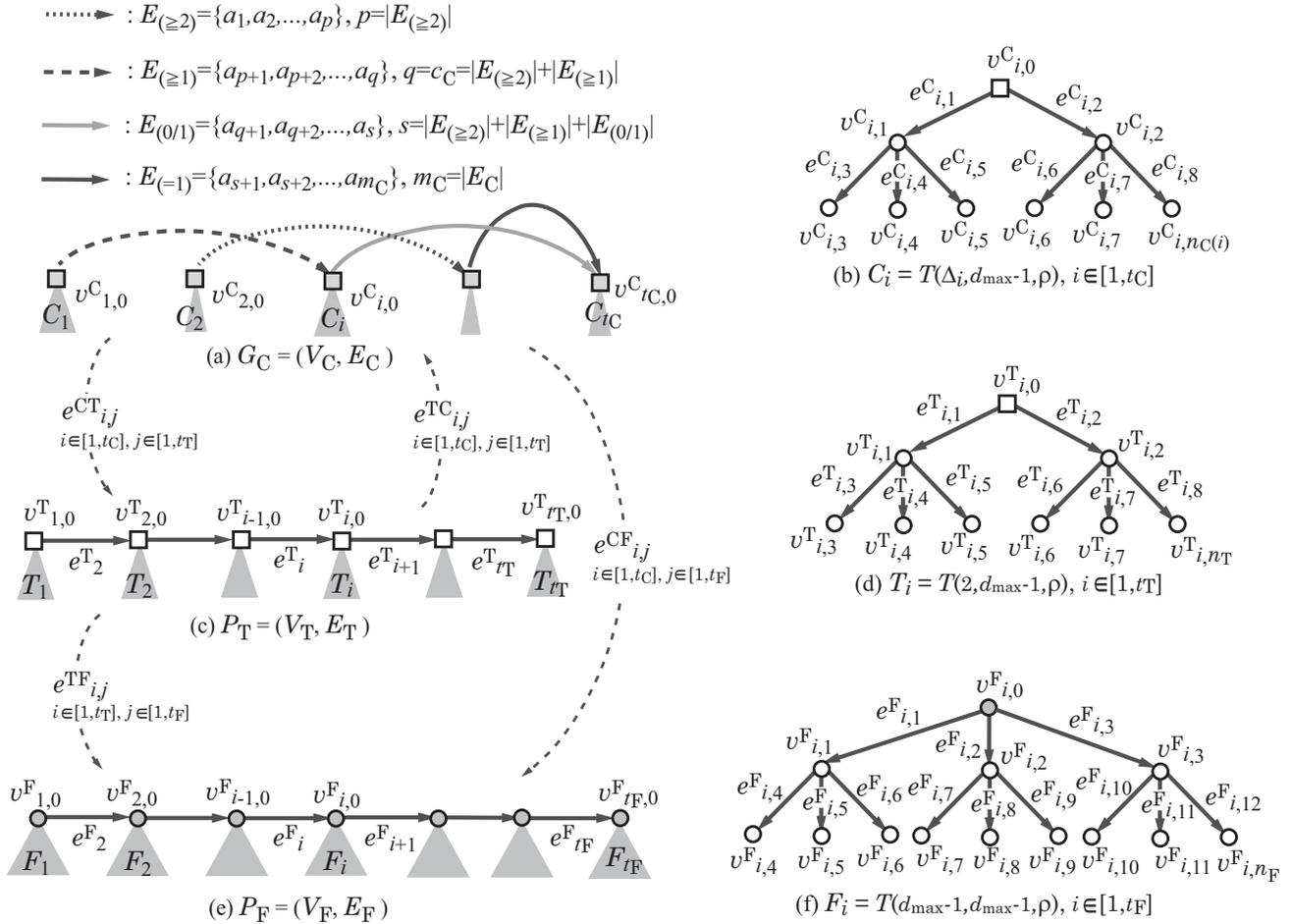}
\end{center}
\caption{An illustration of the structure of a scheme graph  $\mathrm{SG}$,
where vertices depicted with squares represent core-vertices,
vertices depicted with gray circles represent ${\rho}$-internal-vertices,
and 
vertices depicted with white circles represent ${\rho}$-external-vertices:
(a) A seed graph  $\GC=(\VC, \EC=\Et\cup\Ew\cup\Ez\cup\Eew)$;
(b) A tree $C_i$, $i\in  [1,\tC]$ rooted at a core-vertex $\vC_{i,0}\in\VC$;   
(c)  A path $\PT=(\VT, \ET)$    of length $\tT-1$; 
(d)  A  tree $T_i$, $i\in [1,\tT]$ rooted at a core-vertex $\vT_{i,0}\in\VT$; 
(e)   A path $\PF=(\VF, \EF)$  of length $\tF-1$; 
(f) A rooted tree $F_i$, $i\in  [1,\tF]$ rooted at a 
${\rho}$-internal vertex $\vF_{i,0}\in\VF$.  }
\label{fig:scheme_graph}  \end{figure}

Formally the scheme graph $\mathrm{SG}=(\mathcal{V},\mathcal{E})$
is defined  with a vertex set $\mathcal{V}= \VC\cup \VT\cup \VF\cup 
 \VC^\ex\cup \VT^\ex\cup \VF^\ex$ and  
 an edge set 
$\mathcal{E}=\EC\cup \ET\cup \EF\cup
   \ECT\cup \ETC\cup   \ECF\cup   \ETF \cup 
 \EC^\ex\cup \ET^\ex\cup \EF^\ex$ 
 that consist of the following sets.

\bigskip\noindent 
{\bf Construction of  the core $\Cr(H)$ of a $(\sco,\snc)$-extension $H$ of $\GC$:}  
Denote the vertex set $\VC$ and the edge set $\EC$ in the seed graph $\GC$
 by  $\VC=\{\vC_{i,0}\mid i\in [1,\tC]\}$
 and $\EC=\{a_i\mid i\in [1,\mC]\}$, respectively, 
 where $\VC$ is always included in  $\Cr(H)$. 
For including additional core-vertices  in  $\Cr(H)$,
introduce  a path  
$\PT =(\VT=\{\vT_{1,0},\vT_{2,0},\ldots,$ $\vT_{\tT,0}\},
            \ET=\{\eT_1, \eT_2,\ldots, \eT_{\tT}\})$   of length $\tT-1$
and a set $\ECT$ (resp., $\ETC$)
   of directed edges $\eCT_{i,j}=(\vC_{i,0},\vT_{j,0})$ 
    (resp., $\eTC_{i,j} =(\vT_{j,0}, \vC_{i,0})$)  
    $i\in [1,\tC]$, $j\in [1,\tT]$.
In  $\Cr(H)$, an edge $a_k=(\vC_{i,0},\vC_{i',0})\in \Et\cup \Ew$ is allowed 
to be replaced with a path $P_k$ 
from core-vertex $\vC_{i,0}$ to  core-vertex $\vC_{i',0}$
that visits a set of consecutive vertices $\vT_{j,0},\vT_{j+1,0},\ldots, \vT_{j+p,0}\in \VT$
and edge $\eTC_{i,j}=(\vC_{i,0},\vT_{j,0})\in \ECT$,
then edges $\eT_{j+1},\eT_{j+2},\ldots, \eT_{j+p}\in \ET$
and finally edge  $\eTC_{i',j+p}=(\vT_{j+p,0}, \vC_{i',0})\in \ETC$. 
The  vertices in $\VT$ in the path will be core-vertices in  $\Cr(H)$.

\bigskip\noindent 
{\bf Construction of paths with ${\rho}$-internal edges
in a $(\sco,\snc)$-extension $H$ of $\GC$:}  
Introduce a path
$\PF =(\VF=\{\vF_{1,0},\vF_{2,0},\ldots,$ $\vF_{\tF,0}\},$
            $\EF=\{\eF_1, \eF_2,\ldots, \eF_{\tF}\})$  
              of length $\tF-1$,               
 a set $\ECF$  of directed edges $\eCF_{i,j}=(\vC_{i,0}, \vF_{j,0})$, $i\in [1,\tC]$, 
 $j\in [1,\tF]$, and
  a set $\ETF$  of directed edges $\eTF_{i,j}=(\vT_{i,0}, \vF_{j,0})$, $i\in [1,\tT]$, $j\in [1,\tF]$. 
In $H$, a  path  $P$ with ${\rho}$-internal edges that starts from a core-vertex
$\vC_{i,0}\in \VC$ (resp.,   $\vT_{i,0}\in \VT$) 
visits a set of consecutive vertices $\vF_{j,0},\vF_{j+1,0},\ldots, \vF_{j+p,0}\in \VF$
and edge  $\eCF_{i,h}=(\vC_{i,0},\vF_{j,0})\in \ECF$
(resp.,   $\eTF_{i,j}=(\vT_{i,0},\vF_{j,0})\in \ETF$) and edges 
$\eF_{j+1},\eF_{j+2},\ldots, \eF_{j+p}\in \EF$. 
In $H$, the edges and the vertices (except for $\vC_{i,0}$) in the path $P$
 are regarded as ${\rho}$-internal edges and 
 ${\rho}$-internal vertices, respectively.  

\bigskip\noindent 
{\bf Construction of  ${\rho}$-fringe-trees
in a $(\sco,\snc)$-extension $H$ of $\GC$:}  
In $H$, the root of a  ${\rho}$-fringe-tree can be any vertex 
in $\VC\cup\VT\cup \VF$.
Let $\mathrm{X}\in \{\mathrm{C,T,F}\}$. 
Introduce a rooted tree $X_i$, $i\in [1,\tX]$ 
at each vertex $\vX_{i,0}$,
where each $C_i$ is  
   isomorphic to  $T(\dmax-2,\dmax-1,{\rho})$,    
 each  $T_i$ is   
   isomorphic to  $T(2,\dmax-1,{\rho})$ and    
 each  $F_i$  is  
   isomorphic to  $T(\dmax-1,\dmax-1,{\rho})$.
The $j$-th vertex (resp., edge) in each rooted tree $X_i$ is denoted
by $\vX_{i,j}$ (resp.,  $\eX_{i,j}$) 
See Figure~\ref{fig:scheme_graph}.
Let $\VX^\ex$ and $\EX^\ex$ denote the 
set of non-root vertices $\vX_{i,j}$ and the set of edges $\eX_{i,j}$
over all rooted trees  $X_i$, $i\in[1,\tX]$. 
In $H$, a ${\rho}$-fringe-tree is selected 
as a subtree of $X_i$, $i\in [1,\tX]$ 
with root $\vX_{i,0}$.


We see that the scheme graph    $\mathrm{SG}=(\mathcal{V},\mathcal{E})$
for a specification $(\GC,\sco, \snc, \sab)$ satisfies the following. 
\[ |\mathcal{V}|= O( (n^*+\cs_\UB) (\dmax-1)^{{\rho}}  ),  ~~~
 |\mathcal{E}|=   O(|\EC|+|\mathcal{V}|  +n^*\cdot \cs_\UB  ).  \]  
 
\subsection{Formulating an MILP for Choosing 
a Chemical Graph from a Scheme Graph}\label{sec:sMILP_Th2}
 
 Let  $K$ denote the dimension of a feature vector $x=f(G)$ used in
constructing a prediction function $\psi$ over a set of chemical graphs $G$.    
Note that  sets of chemical symbols
  and edge-configuration in Stages~4 and 5
  can be subsets of those used in constructing a prediction function $\psi$ 
  in Stage~3.  
Based on the above scheme graph $\mathrm{SG}$,
we obtain an MILP formulation that satisfies the following result. 

\begin{theorem} \label{Th2} 
Let  $(\sco,\snc,\sab)$ be a target specification and
 $|\Gamma|=|\Ldg^\co|+  |\Ldg^\nc|+ |\Gamma^\co|
  + |\Gamma^\inn|+ |\Gamma^\co|$  for sets of chemical symbols
  and edge-configuration in $\sab$. 
Then there is an MILP $\mathcal{M}(x,g;\mathcal{C}_2)$  
that consists of variable vectors 
$x\in  \mathbb{R}^K$ and 
$g\in \mathbb{R}^q$ for an integer 
 $q= O( 
 \cs_\UB(|\EC|+n^*)+ (|\EC|+|\mathcal{V}|)|\Gamma|    ) $ 
 and a set $\mathcal{C}_2$ of  
$O([\cs_\UB(|\EC|+n^*)+ |\mathcal{V}|]|\Gamma|)$ 
 constraints on $x$ and $g$ such that:  
   $(x^*,g^*)$ is feasible to  $\mathcal{M}(x,g;  \mathcal{C}_2)$
  if and only if  $g^*$ forms a  chemical ${\rho}$-lean  graph 
  $G\in \mathcal{G}(\GC,\sco,\snc,\sab)$   such that  $f(G)=x^*$. 
\end{theorem}
   
 Note that  our MILP requires only  $O(n^* )$ variables and constraints 
 when the branch-parameter ${\rho}$, 
 integers $|\EC|$,   $\cs_\UB$  and $|\Gamma|$ are constant.   

We explain the basic idea of our MILP  in Theorem~\ref{Th2}. 
The MILP mainly consists of the following three types of constraints.
\begin{enumerate}
\item[C1.] 
Constraints for selecting a ${\rho}$-lean  graph  $H\in  \mathcal{H}(\GC,\sco,\snc)$
 as a subgraph of the scheme graph $\mathrm{SG}$; 

\item[C2.] 
Constraints for assigning chemical elements to vertices and multiplicity to edges
to determine a chemical graph $G=(H,\alpha,\beta)$;  and

\item[C3.] 
Constraints for computing descriptors
 from the selected chemical graph $G$.
\end{enumerate} 

In the constraints of C1, more formally we prepare the following.  \\
{Variables: }\\ 
 a binary variable $\vX(i,j)\in\{0,1\}$ for each vertex 
$\vX_{i,j}\in \VX$,
$\mathrm{X}\in \{\mathrm{C,T,F}\}$ so that
 $\vX(i,j)=1$ $\Leftrightarrow$ vertex $\vX_{i,j}$ is used in a  graph $H$
 selected from $\mathrm{SG}$; \\
  a binary variable  $\eX(i)\in\{0,1\}$ (resp.,  $\eC(i)\in\{0,1\}$)
for each edge 
 $\eX_{i}\in \ET \cup \EF$ (resp.,  $\eC_i=a_i\in \Et\cup\Ew\cup\Ez$) 
 so that
 $\eX(i)=1$   $\Leftrightarrow$ edge $\eX_{i}$  
  is used in a  graph $H$  selected from $\mathrm{SG}$.  
To save the number of variables in our MILP formulation, we do not prepare
a binary variable  $\eX(i,j)\in\{0,1\}$ 
 for any edge $\eX_{i,j}\in  \ECT\cup \ETC\cup \ECF\cup \ETC$,
 where we represent a choice of edges in these sets 
 by a set of $O(n^*|\EC|)$ variables (see \cite{AN20} for the details); \\ 
{Constraints:  } \\
 linear constraints so that  
each ${\rho}$-fringe-tree of a  graph $H$
from $\mathrm{SG}$ is selected a subtree of some of the rooted trees
$C_i$, $i\in [1,\tC]$, $T_i$, $i\in [1,\tT]$ and $F_i$, $i\in [1,\tF]$; \\ 
  linear constraints such that  
 each edge $\eC_i=a_i\in \Eew$ is always used as a core-edge in $H$
 and  
 each edge $\eC_i=a_i\in \Ez$ is used as a core-edge in $H$ if necessary;  \\
  linear constraints such that 
 for each edge $a_k=(\vC_i,\vC_{i'})\in \Et$,
 vertex $\vC_i\in \VC$ is connected to vertex $\vC_{i'}\in \VC$ in $H$
by  a path $P_k$  that passes through some core-vertices in $\VT$ and edges
 $\eCT_{i ,j}, \eT_{j+1}, \eT_{j+2},\ldots, \eT_{j+p}, \eTC_{i', j+p}$ for some
 integers $j$ and $p$;  \\
  linear constraints such that 
 for each edge $a_k=(\vC_i,\vC_{i'})\in \Ew$,
 either the edge $a_k$ is  used as a core-edge in $H$ or 
 vertex $\vC_i\in \VC$ is connected to vertex $\vC_{i'}\in \VC$ in $H$
by  a path $P_k$   as in the case of  edges in $\Et$; \\
linear constraints for selecting  
 a path $P$ with   ${\rho}$-internal edges
 $\eCF_{i ,j}$ (or $\eTF_{i,j}$), $\eF_{j+1}, \eF_{j+2},\ldots, \eF_{j+p}$ for some
 integers $j$ and $p$.  
 
Based on these, we include constraints with some more additional variables
so that a selected subgraph $H$ is a connected  graph
and satisfies the core specification $\sco$ and the non-core specification $\snc$.  
See  constraints (\ref{eq:co_first}) to  (\ref{eq:co_last})  in Appendix~\ref{sec:co}
for choosing core-edges from the path $\PT$.
See  constraints (\ref{eq:int_first}) to  (\ref{eq:int_last})  in Appendix~\ref{sec:int}
for choosing internal ${\rho}$-internal vertices/edges  from the path $\PF$.
See 
constraints (\ref{eq:ex_first}) to  (\ref{eq:ex_last})  in Appendix~\ref{sec:ex}
for choosing internal ${\rho}$-external vertices/edges  from the trees
$C_i$, $T_i$ and $F_i$.  

In the constraints of C2,  we prepare an integer variable 
$\aX(i,j)$ for each vertex $\vX_{i,j}\in \mathcal{V}$,
$\mathrm{X}\in \{\mathrm{C,T,F}\}$
 in the scheme graph
that represents the chemical element $\alpha(\vX_{i,j})\in \Lambda$
if $\vX_{i,j}$ is in a selected graph $H$ (or $\alpha(\vX_{i,j})=0$ otherwise);  
  integer variables  $\bC: \EC\to [0,3]$, $\bT: \ET\to [0,3]$ and 
$\bF: \EF\to [0,3]$ that  represent  the bond-multiplicity of  edges in 
$\EC \cup \ET\cup \EF$; 
and   integer variables  $\beta^+,\beta^-:\Et\cup\Ew \to[0,3]$ 
 and  $\beta^\inn: \widetilde{\VC}\cup\VT \to[0,3]$ that
represent  the bond-multiplicity of edges in 
 $\ECT\cup \ETC\cup \ECF\cup \ETF$. 
This determines a chemical graph $G=(H,\alpha,\beta)$. 
Also we include constraints for a selected chemical graph
$G$ to satisfy the valence condition at each vertex $v$ with 
the edge-configurations   $\tau(e)$ of  
the edges incident to $v$
and the chemical specification $\sab$. 
See constraints (\ref{eq:beta_first})  to (\ref{eq:beta_last}) 
 in  Appendix~\ref{sec:beta}  
for assigning multiplicity; and 
 constraints (\ref{eq:alpha_first}) to  (\ref{eq:alpha_last}) 
 in  Appendix~\ref{sec:alpha}  
   for assigning chemical clements and satisfying valence condition.

In the constraints of C3,  we introduce a variable for each descriptor
and constraints with some more variables to compute the value of 
each descriptor in $f(G)$ for a selected chemical graph $G$.
See  constraints (\ref{eq:Deg_first})  to (\ref{eq:Deg_last}) 
 in  Appendix~\ref{sec:Deg}  
 for descriptor of the number of specified degree; 
 constraints (\ref{eq:BDbond_first}) to  (\ref{eq:BDbond_last}) in  Appendix~\ref{sec:BDbond}  
  for lower and upper bounds on the number of bonds in 
  a chemical specification $\sab$; 
 constraints (\ref{eq:AC_first}) to (\ref{eq:AC_last}) in  Appendix~\ref{sec:AC}  
 for descriptor of the number of adjacency-configurations; 
  constraints (\ref{eq:CS_first}) to (\ref{eq:CS_last}) in  Appendix~\ref{sec:CS}  
  for descriptor of the number of chemical symbols;  
 and 
 constraints (\ref{eq:EC_first}) to (\ref{eq:EC_last}) in  Appendix~\ref{sec:EC}  
 for descriptor of the number of edge-configurations. 
 When we use adjacency-configuration in a feature vector $f(G)$
 instead of edge-configuration, we do not need to include 
 the constraints in Appendix~\ref{sec:EC}.


\section{A New Graph Search Algorithm}\label{sec:graph_search}

This section designs a new algorithm for generating ${\rho}$-lean cyclic
graphs $G$ that have the same feature vector $f(G^\dagger)$
of a given chemical ${\rho}$-lean graph $G^\dagger$.

\subsection{The New Aspects} 
We design a new algorithm for generating cyclic chemical
graphs based on the following aspects: 
\begin{enumerate}
\item[(a)] Treat the non-core components of a cyclc graphs with a certain limited structure  
that frequently appears among chemical compounds registered 
in the chemical data base;  
\item[(b)]  Instead of manipulating target  graphs directly,
first compute the frequency vectors $\f(G')$
(some types of feature vectors)   of subtrees $G'$ of all target   graphs
and then construct  
a limited number of target graphs $G$ from the process of computing the vectors; and  
\item[(c)]  First construct a chemical graph $G^\dagger\in \G$
 with $\f(G^\dagger)=x^*$  by solving an MILP in Stage~4
and restrict ourselves to a family of  chemical graphs
$G^*\in\G$ that have a common structure 
with the initial chemical graph $G^\dagger$.
\end{enumerate}
 
In (a), we choose a small branch-parameter ${\rho}$ such as ${\rho}=2$ 
and treat chemical ${\rho}$-lean cyclic graphs $G$ such that 
each $2$-fringe-tree in $G$ satisfies the  size constraint (\ref{eq:fringe-size}).  
  
We design a method in (b) by extending
the dynamic programming  algorithm for generating acyclic chemical graphs
proposed by Azam et al.~\cite{AZSSSZNA20}. 
The first phase of the algorithm 
computes some compressed forms of all substructures of target objects   before
the second phase realizes a final object based on the computation process of
the first phase. 

The idea of (c) is first introduced in this paper.
Informally speaking, we first decompose the chemical graph $G^\dagger$ into
a collection of chemical subtrees $T^\dagger_1,T^\dagger_2,\ldots,T^\dagger_p$ 
and then compute vectors $x^*_1,x^*_2,\ldots,x^*_p$ such that
$\f(T^\dagger_i)=x^*_i$, $i\in [1,p]$  and
any collection of other chemical trees $T^\dagger_i$ with $\f(T^\dagger_i)=x^*_i$ always 
gives rise to a target chemical graph $G^*\in G$. 
Thus this allows us to generate chemical trees $T^\dagger_i$ with $\f(T^\dagger_i)=x^*_i$
for each index $i\in [1,p]$ {\em independently} 
before we combine them to get a chemical graph $G^\dagger\in G$ in Stage~5.

In the following, we describe a new algorithm that
for a given ${\rho}$-lean chemical graph
 $G^\dagger=(H^\dagger,\alpha^\dagger,\beta^\dagger)$,
 generates  chemical ${\rho}$-lean cyclic graphs $G^*=(H ,\alpha ,\beta )$
 such that 
 \[\mbox{ $\f(G^*)=\f(G^\dagger)$ and 
 the core $\Cr(H)$ is isomorphic to the core $\Cr(H^\dagger)$, }\] 
 where $G^*$ may not be isomorphic to $G^\dagger$
 and the elements in $\Lambda$ may not correspond
 between the two cores; i.e., possibly 
 $\alpha(v)\neq \alpha^\dagger(\phi(v))$ for some core-vertex $v$ of $H$
 in the graph-isomorphism $\phi$ between $\Cr(H)$ and $\Cr(H^\dagger)$.

  In this section, we describe our new algorithm in a general setting 
  where a branch-parameter is any integer ${\rho}\geq 1$
  and a chemical graph $G$ to be inferred is any chemical 
   ${\rho}$-lean cyclic graph. 

\subsection{Nc-trees and C-trees}\label{sec:nc-_c-trees}

\begin{figure}[!ht] \begin{center}
\includegraphics[width=.98\columnwidth]{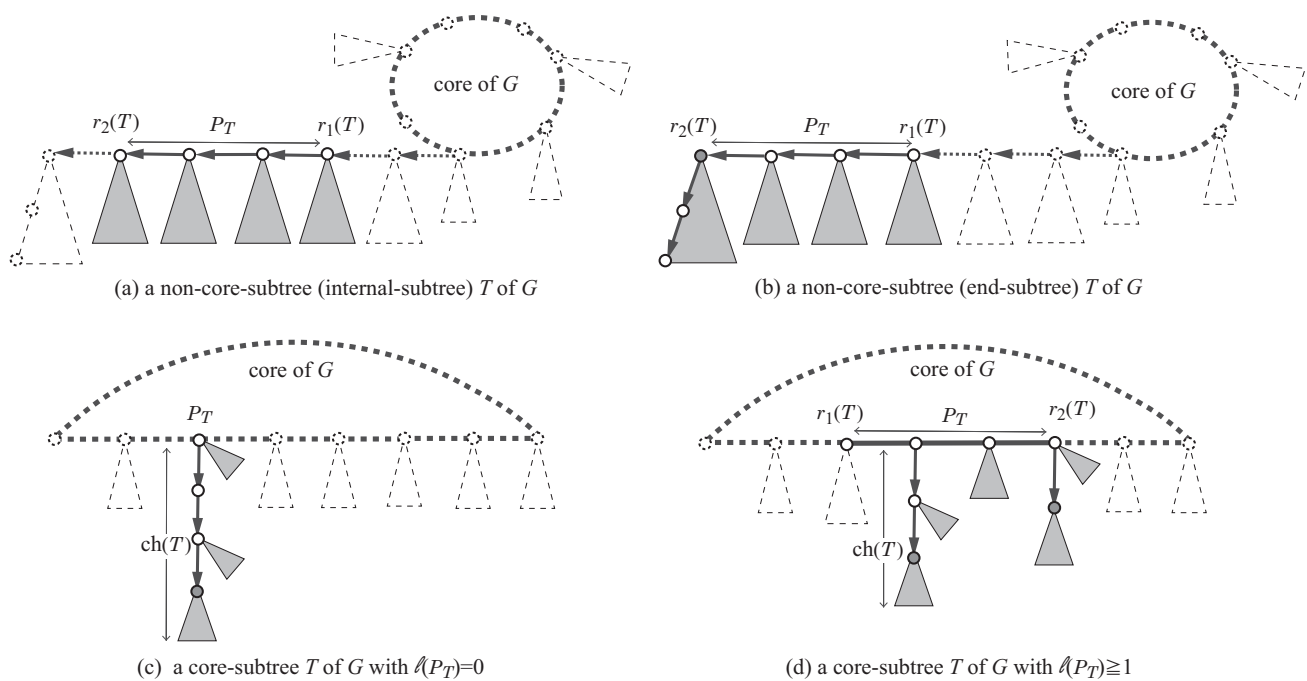}
\end{center}
\caption{An illustration of subtrees of a chemical ${\rho}$-lean cyclic graph $G$,
where  thick lines depict the cycle of the core of $G$,
 gray circles depict leaf ${\rho}$-branches in $G$
 and arrows depict non-core directed edges: 
(a)~A non-core-subtree (internal-subtree) $T$ of $G$ represented by
an nc-tree (a chemical bi-rooted tree);
(b)~A non-core-subtree (end-subtree) $T$ of $G$ represented by
an nc-tree (a chemical bi-rooted tree);
(c) ~A  core-subtree  $T$ of $G$ with $\ell(P_T)=0$  represented by
a c-tree (a chemical rooted tree);
(d)~A  core-subtree  $T$ of $G$ with $\ell(P_T)\geq 1$ represented by
a c-tree (a chemical bi-rooted tree).
}
\label{fig:subtrees_cyclic} 
\end{figure}

\noindent
{\bf Nc-trees  } 
Let ${\rho}$ be a branch-parameter.
  and   $H$ be a    ${\rho}$-lean cyclic graph.  
 We have introduced core-subtrees  in Section~\ref{sec:preliminary}.
 We define ``non-core-subtrees'' as follows depending on 
  branch-parameter ${\rho}$.
 Let $T$ be a  connected subgraph of $H$.
 We call $T$ a {\em non-core-subtree} of $H$
    if $T$ consists of  
  a path $P_{\inn}$ of a ${\rho}$-pendent-tree  of $H$   and 
  the ${\rho}$-fringe-trees rooted at vertices   in $P_{\inn}$.
 We call a  non-core-subtree $T$ of $H$
an {\em internal-subtree} (resp., an {\em end-subtree}) of $H$
if neither (resp., one) of the two end-vertices of $P_{\inn}$ 
is a leaf ${\rho}$-branch of $H$, as illustrated in Figure~\ref{fig:subtrees_cyclic}(a)
(resp., in Figure~\ref{fig:subtrees_cyclic}(b)).
   
To represent a non-core-subtree of a  ${\rho}$-lean cyclic graph $H$,
 we introduce ``nc-trees.''
We define an {\em nc-tree}   to be 
 a chemical bi-rooted tree $T$  such that
each rooted tree $T'\in  \F(T)$ has a height at most ${\rho}$.
For an nc-tree $T$, define \\
~~ $\widetilde{V}_{\co}(T)\triangleq  \emptyset$
(resp., $\widetilde{E}_{\co}(T)\triangleq \emptyset$);  \\
~~ $\widetilde{V}_{\inn}(T)\triangleq V(P_T)$ 
(resp., $\widetilde{E}_{\inn}(T)\triangleq E(P_T)$) for the backbone path $P_T$ of $T$;\\ 
~~ $\widetilde{V}_{\ex}(T)\triangleq V(T)\setminus \widetilde{V}_{\inn}(T)$ 
(resp.,  
$\widetilde{E}_{\ex}(T)\triangleq E(T)\setminus \widetilde{E}_{\inn}(T)$). \\ 
Define the number $\bc_{\rho}(T)$ of ${\rho}$-branch-core-vertices in $T$ 
to be  $\bc_{\rho}(T)=0$ and the core height $\ch(T)$ of $T$ to be $0$.  

\bigskip
\noindent
{\bf C-trees  } 
To represent a core-subtree   of a  ${\rho}$-lean cyclic graph $H$,
 we introduce ``c-trees.''
For a branch-parameter ${\rho}$,
we call a bi-rooted tree {\em ${\rho}$-lean}
   if each rooted tree $T'\in \F(T)$ contains at most one
${\rho}$-branch; i.e., there is no non-leaf ${\rho}$-branch
and no two ${\rho}$-branch-pendent-trees meet at the same vertex
in $P_T$.  
A {\em c-tree} is defined to be a  chemical  ${\rho}$-lean bi-rooted tree $T$.
See Figure~\ref{fig:subtrees_cyclic}(c) and (d) 
for illustrations  of  c-trees $T$ with $\ell(P_T)=0$ and $\ell(P_T)\geq 1$,
respectively. 
For a c-tree $T$, define \\
~~ $\widetilde{V}_{\co}(T)\triangleq V(P_T)$ 
(resp., $\widetilde{E}_{\co}(T)\triangleq E(P_T)$)
 for the backbone path $P_T$ of $T$;    \\
~~ $\widetilde{V}_{\inn}(T)$ (resp.,  $\widetilde{E}_{\inn}(T)$) to be 
the set of ${\rho}$-internal vertices (resp.,   ${\rho}$-internal vertices)  \\
~~~~  in the rooted trees $T'\in  \F(T)$;  \\
~~ $\widetilde{V}_{\ex}(T)\triangleq V(T)\setminus
 (\widetilde{V}_{\co}(T)\cup \widetilde{V}_{\inn}(T))$ 
(resp.,  
$\widetilde{E}_{\ex}(T)\triangleq E(T)\setminus 
 (\widetilde{E}_{\co}(T)\cup \widetilde{E}_{\inn}(T))$).\\
Define the number $\bc_{\rho}(T)$ of ${\rho}$-branch-core-vertices in $T$ 
to be the number of rooted trees in $T'\in  \F(T)$ 
with $\h(T')> {\rho}$. 
Define the core height $\ch(T) \triangleq \h(T)$ for the bi-rooted tree $T$.  
Note that  
  $\widetilde{V}_{\ex}(T)$ (resp., $\widetilde{E}_{\ex}(T)$) is 
the set of ${\rho}$-external vertices (resp.,   ${\rho}$-external vertices)   
in the rooted trees in  $\F(T)$.

\begin{figure}[!ht] \begin{center}
\includegraphics[scale=0.49]{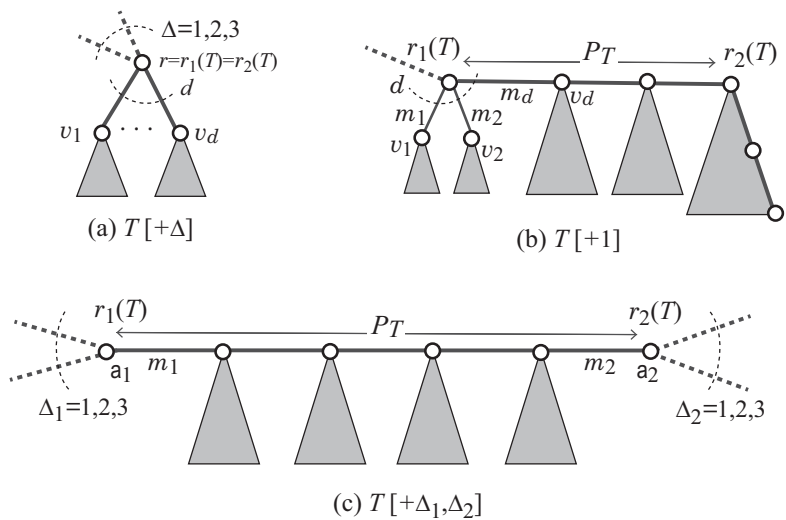}
\end{center}
\caption{An illustration of fictitious  trees:   
	(a) $T[+\Delta]$ of a  rooted nc- or c-tree $T$; 
		(b)  $T[+1]$ of a bi-rooted nc-tree $T$;  
		(c)  $T[+\Delta_1,\Delta_2]$ of a bi-rooted c-tree $T$. 	}
\label{fig:fictitious_trees} \end{figure}

\bigskip \noindent
{\bf Fictitious  Trees }  
For  an nc-tree or a c-tree $T$ and  an integer $\Delta\in[1,3]$, 
let   $T[+\Delta]$  denote  a fictitious chemical graph obtained from $T$
by regarding the degree of terminal $r_1(T)$ as $\deg_T(r_1(T))+\Delta$.
Figure~\ref{fig:fictitious_trees}(a) and (b) 
illustrate  fictitious trees $T[+\Delta]$ in the case of  $r_1(T)=r_2(T)$
and $T[+1]$ in the case of $\Delta=1$ and $r_1(T)\neq r_2(T)$.

For a c-tree $T$ with $r_1(T)\neq r_2(T)$ and integers $\Delta_1,\Delta_2\in [1,3]$, 
let    $T[+\Delta_1,\Delta_2]$ denote  a fictitious chemical graph obtained from $T$
by regarding the degree of  terminal $r_i(T)$, $i=1,2$ as
$\deg_T(r_i(T))+\Delta_i$. 
Figure~\ref{fig:fictitious_trees}(c) illustrates a fictitious bi-rooted c-tree
  $T[+\Delta_1,\Delta_2]$.

\subsection{Frequency Vectors}\label{sec:frequency_vector}
 
For a finite set $A$ of elements, let 
$\mathbb{Z}_+^A$ denote the set of functions
$\w:A\to \mathbb{Z}_+$.
A function $\w\in \mathbb{Z}_+^A$ is called
a {\em non-negative integer vector} (or a vector) on $A$
and the value $\x(a)$ for an element $a\in A$ is called
the {\em entry} of $\x$ for $a\in A$. 
For a vector $\w\in \mathbb{Z}_+^A$ and  an element $a\in A$,
let $\w+\1_{a}$ (resp., $\w-\1_{a}$)
denote the vector  $\w'$ 
such that  $\w'(a)=\w(a)+1$ (resp., $\w'(a)=\w(a)-1$)
 and  $\w'(b)=\w(b)$
 for the other elements   $b\in A\setminus\{a\}$.  
For a vector $\w\in \mathbb{Z}_+^A$ and a subset $B\subseteq A$,
let $\w_{[B]}$ denote the {\em projection} of $\w$ to $B$;
i.e., $\w_{[B]}\in \mathbb{Z}_+^B$ such that 
$\w_{[B]}(b)=\w(b)$, $b\in B$.
  
To introduce a ``frequency vector'' of 
 a subgraph of  a chemical cyclic graph, we define
  sets of symbols that correspond to some descriptors of a chemical cyclic graph.
Let $\Gamma^\co$, $\Gamma^\inn$ and $\Gamma^\ex$ be 
  sets
of edge-configurations in Section~\ref{sec:chemical_model}.
We define a vector whose entry is the frequency of
an edge-configuration in sets $\Gamma^\typ$, $\typ\in\{\co, \inn, \ex\}$
or the number of ${\rho}$-branch-core-vertices.
We use  a symbol $\tbc$ to denote the number of ${\rho}$-branch-core-vertices
in our frequency vector. 
To distinguish edge-configurations from different sets among 
three sets $\Gamma^\typ$, $\typ\in\{\co, \inn, \ex\}$,
we use $\gamma^\typ$ to denote the entry of 
an edge-configuration $\gamma\in \Gamma^\typ$, $\typ\in\{\co, \inn, \ex\}$.
We denote by $\langle \Gamma^\typ \rangle$ the set of
entries $\gamma^\typ$, $\gamma\in  \Gamma^\typ$, $\typ\in\{\co, \inn, \ex\}$.
Define the set of all entries of a frequency vector to be  
\[   
\displaystyle{
\Sigma\triangleq \{\tbc\} \cup 
 \bigcup_{\typ\in\{\co, \inn, \ex \}}  \langle \Gamma^\typ \rangle  .
}   \]

Given an nc-tree or c-tree $G$ or a chemical ${\rho}$-lean cyclic graph $G$,
define the frequency vector $\f(G)$,  
to be a vector $\w\in \mathbb{Z}_+^\Sigma$ that consists of the following entries:  
\begin{itemize}
\item[-] 
$\w(\gamma^{\co})=|\{uv\in \widetilde{E}_{\co} \mid 
(\alpha(u),\deg_G(u),\alpha(v),\deg_G(v))\in 
\{(\ta, i,\tb, j,m), (\tb, j,\ta, i,m)\}, \beta(uv)=m\}|$,
$\gamma=(\ta i,\tb j,m)^{\co}\in \langle \Gamma^\co \rangle$; 
\item[-]  
$\w(\gamma^\typ)=|\{(u,v)\in \widetilde{E}_\typ \mid 
 \alpha(u)=\ta, \deg_G(u)=i, \alpha(v)=\tb, \deg_G(v)=j, \beta(uv)=m\}|$,
$\gamma=(\ta i,\tb j, m)^\typ\in\langle \Gamma^\typ \rangle$,
$\typ\in\{\inn, \ex\}$; 
\item[-]
  $\w(\tbc)=\bc_{\rho}(G)$.
\end{itemize} 
Note that 
any other descriptors of a chemical ${\rho}$-lean cyclic graph $G\in \G(\x^*)$
except for the core height can be determined by the entries of 
the frequency vector $\w=\f(G)$.
For example, the vector 
$\z\in \mathbb{Z}_+^{\{\dg1,\dg2,\dg3,\dg4\}}$ 
with the numbers $\dg i$ of core-vertices of degree $i\in [1,4]$
is given by 
\[\displaystyle{
\z= \frac{1}{2}\Bigl [ \sum_{\gamma^\co=(\ta i, \ta j, m)^\co
    \in \langle \Gamma^\co \rangle}
 \!\!\!\! \w(\gamma^\co)\cdot  (\1_{\dg i} \!+\! \1_{\dg j}) 
 - \sum_{ i_v=\deg_G(v): v\in V_B}
 (\Delta_v\! -\! 2)\1_{\dg i_v} \Bigr ] } \]
 and  the vector 
$\z'\in \mathbb{Z}_+^{\Ldg}$ 
with the numbers of symbols $\mu\in \Ldg$ of core-vertices  
is given by 
\[\displaystyle{
\z'= \frac{1}{2}\Bigl [ \sum_{\gamma^\co=(\mu, \xi, m)^\co
\in\langle \Gamma^\co \rangle}
 \!\!\!\! \w(\gamma^\co )\cdot  (\1_{\mu } \!+\! \1_{\xi }) 
 - \sum_{\mu_v=\ta_v \deg_G(v)   : v\in V_B}
 (\Delta_v\! -\! 2)\1_{ \mu_v   } \Bigr ]. } \] 
Similarly  the vector 
$\z''\in \mathbb{Z}_+^{\Ldg}$ 
the numbers of symbols $\mu\in \Ldg$ of non-core-vertices  
is given by 
\[\displaystyle{
\z'= 
\sum_{\gamma^\inn=(\mu, \xi, m)^\inn\in \langle \Gamma^\inn \rangle}
 \!\!\!\! \w(\gamma^\inn )\cdot  \1_{\xi } 
 +
\sum_{\gamma^\ex=(\mu, \xi, m)^\ex\in \langle \Gamma^\ex \rangle}
 \!\!\!\! \w(\gamma^\ex )\cdot  \1_{\xi }. } \] 
 
For an  nc-tree or c-tree $T$,
the frequency  vector  $\f(T[+\Delta])$ of a fictitious tree $T[+\Delta]$
is  defined as follows: 
Let  $r=r_1(T)$,  $d=\deg_T(r)$, $N_T(r)=\{v_1,v_2,\ldots,v_d\}$,  
$\gamma_j=(\mu_j=\ta_j d,\xi_j, m_j)=\tau((r,v_i))\in \Gamma$,  $j\in [1,d]$. 
Let    
$\widetilde{\gamma}_j =(\ta_j \{d\!+\!\Delta\}, \xi_j,m_j)$,  $j\in [1,d]$.
Set  $\typ:=\inn$   if $T$ is an nc-tree,   and  $\typ:=\co$    if $T$ is a c-tree.
When  $r_1(T)=r_2(T)$, 
\[ \f(T[+\Delta]) =  \f(T)  
+ \sum_{j\in [1,d]} ( \1_{ \widetilde{\gamma}_j ^\typ} -\1_{ \gamma_j^\typ }).\]
Let $r_1(T)\neq r_2(T)$, and $v_d$ belong to $P_T$.
When $T$ is an nc-tree, 
\[ \f(T[+\Delta]) =  \f(T)  
+ \sum_{j\in [1,d-\! 1]} ( \1_{ \widetilde{\gamma}_j ^\ex} -\1_{ \gamma_j^\ex}) 
+   \1_{ \widetilde{\gamma}_d ^\inn } -\1_{ \gamma_d^\inn } .  \]
 
The frequency vector  
$\f(T[+\Delta_1,\Delta_2])$ of a fictitious tree $T[+\Delta_1,\Delta_2]$
for a bi-rooted  c-tree $T$ with $r_1(T)\neq r_2(T)$
is  defined as follows: 
For each $i=1,2$, 
let  $r_i=r_i(T)$, $\ta_i=\alpha(r_i)$, $d_i=\deg_T(r_i)$, 
$m_i=\beta(e_i)$ of the unique edge incident to $r_i$  and 
$\gamma_i=(\mu_i=\ta_i d_i, \xi_i,m_i)=\tau(e_i)\in \Gamma$,  $i=1,2$. 
  Let  
 $\widetilde{\gamma}_i=(\ta_i\{d_i\!+\!\Delta_i\}, \xi_i,m_i)$, $i=1,2$.  
Then 
\[ \f(T[+\Delta_1,\Delta_2]) =  \f(T) 
 + \sum_{i=1,2}( \1_{ \widetilde{\gamma}_i^\co } -\1_{ \gamma_i^\co }  ) . \]

\subsection{A Chemical Graph Isomorphism}
  
For a chemical ${\rho}$-lean  cyclic graph  for  a branch-parameter ${\rho}\geq 1$,
we choose  a  path-partition 
$\Pt=\{P_1,P_2,\ldots,P_p\}$ of  the core $\Cr(H)=(V^\co, E^\co)$, 
where  $|\Pt|\leq |E^\co|$.
Let $V_B$ denote the set of all end-vertices of paths $P\in\Pt$,
where  $V_{\rho}^{\bc}\subseteq V_B\subseteq V^\co$.

Define the {\em base-graph} $G_B=(V_B,E_B)$ 
of $H$ by $\Pt$ to be  the multigraph obtained
from $H$ replacing each path $P_j\in\Pt$
with a single edge $e_j$ joining the end-vertices of $P_i$,
where $E_B=\{e_1,e_2,\ldots,e_p\}$.  
We call a vertex in $V_B$ and an edge in $E_B$
a {\em base-vertex} and a {\em base-edge}, respectively. 
For a notational convenience in distinguishing the two end-vertices 
$u$ and $v$ of a base-edge $e=uv\in E_B$, 
we regard each base edge $e=uv$ 
as a directed edge $e=(u,v)$.
For each base-edge $e\in E_B$, let $P_e$ denote the path $P_j\in E_B$
that is replaced by edge $e=e_j$.  

We define the ``components'' of $G$ by $\Pt$ as follows.

\bigskip\noindent
{\bf Vertex-components }
For each base-vertex $v\in V_B$,
define   the component at vertex $v$ 
(or the {\em $v$-component}) $T_v$ of $G$ to be  
 the chemical core-subtree rooted at $v$ in $G$; i.e.,
  $T_v$ consists of all pendent-trees rooted at $v$.
We regard $T_v$ as a c-tree  rooted at the core-vertex $v$ of $G$
and define the {\em code} $\code_v(T_v)$ of $T_v$ to be a tuple
$(\ta_v,d_v,m_v,  \Delta_v, \x_v)$ such that 
\[\mbox{
$\ta_v=\alpha(v)$, 
$d_v= \deg_{G}(v)-\deg_{G_B}(v)$, 
$m_v=\sum_{vv'\in E(T_v)}\beta(vv')$, }\]
\[\mbox{ 
$\Delta_v= \deg_{G_B}(v)$ and  
$\x_v=\f(T_v[+\Delta_v])$.
 }\]

\bigskip\noindent
{\bf Edge-components }
For each base-edge $e=e_j=(u,v)\in E_B$,
define   the component at edge $e$ 
(or the {\em $e$-component}) $T_e$  of $G$ to be  
  the chemical core-subtree of $G$ that consists of
the core-path  $P_j\in\Pt$   and
 all pendant-trees of $G$ rooted at internal vertices of path $P_j$.
 We  regard   $T_e$ as a bi-rooted c-tree
with $r_1(T_e)=u$ and $r_2(T_e)=v$ for the base-edge $e=uv$ 
and define the {\em code} $\code_e(T_e)$ of $T_e$ to be a tuple
$(\ta_u^e,  m_u^e,   \ta_v^e,  m_v^e,                  
\Delta_u^e,\Delta_v^e,  \x_e)$ such that 
\[ \mbox{ 
$\ta_u^e=\alpha(u)$, $\ta_v^e=\alpha(v)$,  
$\Delta_u^e=\deg_{G_B}(u)-1$, 
$\Delta_v^e=\deg_{G_B}(v)-1$,  }\] 
\[ \mbox{  
  $\x_e=\f(T_e[+\Delta_u^e, \Delta_v^e] )$,
   }\] 
\[ \mbox{ 
$m_u^e=\beta(uu')$ and $m_v^e=\beta(vv')$ for
the edges $uu',vv'\in E(P_j)$ incident to $u$ and $v$. }\]
Observe that  
 \[  \f(G)= \sum_{v\in V_B} \x_v  +  \sum_{e\in E_B} \x_e. \]

We introduce a specification $\sch$ as a set
of functions $\ch_\LB:V_B\cup E_B \to \mathbb{Z}_+$.

We call two chemical graphs {\em $(\Pt,\sch)$-isomorphic}
if they consist of vertex and edge components with the same codes and heights;
i.e., two chemical ${\rho}$-lean  cyclic graphs 
  $G^i=(H^i,\alpha^i,\beta^i)$, $i=1,2$  are 
   $(\Pt,\sch)$-isomorphic if the following hold: 
\begin{enumerate}
\item[-]  
  $\Cr(H_1)$ and  $\Cr(H_2)$ are graph-isomorphic,
  where we assume that   $\Cr(H_1)=\Cr(H_2)=(V^\co, E^\co)$
  and $G_B=(V_B,E_B)$ denotes the  base-graph 
of both graphs $H_1$ and $H_2$ by $\Pt$; 
\item[-] 
For the $v$-components $T^i_v$ of $G^i$, $i=1,2$ at each base-vertex $v\in V_B$, 
\[ \code_v(T^1_v) = \code_v(T^2_v) \mbox{ and } 
  \h(T^2_v), \h(T^1_v)\in [\ch_\LB(v),\ch_\UB(v)]  ;  
\]
\item[-] For the $e$-components $T^i_e$ of $G^i$, $i=1,2$ 
at each base-edge $e\in E_B$, 
\[ 
 \code_e(T^1_e) = \code_e(T^2_e) \mbox{ and } 
  \h(T^2_v), \h(T^1_v)\in [\ch_\LB(v),\ch_\UB(v)]  ;  
\]
\end{enumerate}
See Section~\ref{sec:preliminary} for the definition of
height $\h(T)$ of a bi-rooted tree $T$. 

The   $(\Pt,\sch)$-isomorphism also implies that 
  $\f(G^1)=\f(G^2)$,  
  $n(G^1)=n(G^2)$,   $\cs(G^1)=\cs(G^2)$,
$\bl_{\rho}(G^1)=\bc_{\rho}(G^1)=\bl_{\rho}(G^2)=\bc_{\rho}(G^2)$ and
$|\ch(G^1) - \ch(G^2)|\leq 
\max_{\typ\in V_B\cup E_B}(\ch_\UB(\typ)- \ch_\LB(\typ) )$.

\subsection{Computing Isomorphic Chemical Graphs from a Given Chemical Graph $\G^\dagger$}

Now we assume that  a chemical ${\rho}$-lean  cyclic graph  $G^\dagger$ for
  a branch-parameter ${\rho}\geq \ch(G^\dagger)$ is available
  in such case where a target chemical graph $G^\dagger$ 
  is constructed by solving an MILP in Stage~4.
  Let $T_v^\dagger$ (resp., $T_e^\dagger$) denote
  the $v$-component (resp.,   the $e$-component) of $G^\dagger$.

\bigskip\noindent
{\bf Target $v$-components }
Let $\h^*_v$ denote the height $h(T^\dagger_v)$ of the $v$-component of $G^\dagger$.
For each base-vertex $v\in V_B$, 
fix a code $(\ta_v,d_v,m_v, \Delta_v, \x^*_v):=\code_v(T^\dagger_v)$
and call a  rooted c-tree  $T$  a {\em target $v$-component}
if   \[\mbox{
   $\code_v(T)=(\ta_v,d_v,m_v, \Delta_v, \x_v^*)$
    and $\h(T)\in [\ch_\LB(v), \ch_\UB(v)]$,  }\]
where the condition on $\h(T)$ is equivalent to
$\h(T)=\h^*_v$ when $\x^*_v(\tbc)=1$, 
since  $G^\dagger$  is a ${\rho}$-lean  cyclic graph
and 
the set of ${\rho}$-internal edges in any target component forms
a single path of length $\h^*_v$ from the root to a unique leaf ${\rho}$-branch.

\bigskip\noindent
{\bf Target $e$-components }
 For each base-edge $e=(u,v)=e_j\in E_B$,  
fix a code $(\ta_u^e,  m_u^e,   \ta_v^e,  m_v^e,$
$\Delta_u^e,\Delta_v^e,  \x_e^*) :=\code_e(T^\dagger_e)$
and call a  bi-rooted c-tree  $T$  a {\em target $e$-component}
if   \[\mbox{
   $\code_e(T)=(\ta_u^e,  m_u^e,   \ta_v^e,  m_v^e,$
$\Delta_u^e,\Delta_v^e,  \x_e^*)$
    and $\h(T)\in [\ch_\LB(e), \ch_\UB(e)]$.  }\]
Let $\T_e$ denote the set of all target components of  
a base-edge $e\in E_B$.

  Given a collection of target $v$-components $T_v\in \T_v$, $v\in V_B$ and
  target $e$-components $T_e\in \T_e$, $e\in E_B$,
there is a chemical ${\rho}$-lean cyclic graph $G^*$ that is $(\Pt,\sch)$-isomorphic to
the original chemical  graph $G^\dagger$. 
 Such a graph $G^*$  can be obtained from $G_B$ 
by replacing each base-edge $e\in E_B$ with $T_e$  
and attaching $T_v$ at each base-vertex $v\in V_B$.

From this observation, our aim is now to generate
some number of target $v$-components for each base-vertex $v$
and   target $e$-components for each base-edge $e$.
In the following, we denote   
$\ta_u^e$, $\ta_v^e$, $\Delta_u^e$,  $\Delta_v^e$, 
$m_u^e$ and $m_v^e$ for each base-edge $e=(u,v)\in E_B$ 
 by  $\ta_1^e$, $\ta_2^e$, $\Delta_1^e$,  $\Delta_2^e$, 
$m_1^e$ and $m_2^e$, respectively for a notational simplicity. 
For each base-edge $e=e_j\in E_B$, let
\[ \delta_1^e:=\lfloor (\ell(P_j)-1)/2 \rfloor \mbox{ and } 
    \delta_2^e:=\lceil (\ell(P_j)-1)/2 \rceil .\] 

\subsubsection{Dynamic Programming Algorithm on Frequency Vectors}
We start with describing a sketch of our new algorithm for generating
graphs $G^*$ in Stage~5 before we present some technical details
of the algorithm in the following sections.

We start with enumerating chemical rooted trees
with height at most ${\rho}$, which can be a ${\rho}$-fringe-tree 
of a target component.
Next we extend each of the rooted tree to
an nc-tree $T$ and then to a c-tree $T$ under a constraint
that the frequency vector of $T$ does not exceed
a given vector $\x=\x^*_v$, $v\in V_B$ or
$\x=\x^*_e$, $e\in E_B$.

For a vector  $\x\in \mathbb{Z}_+^\Sigma$,
we formulate the following sets of nc-trees and c-trees
and of their frequency vectors:
\begin{enumerate}
\item[(i)]
  $\T_{\inl}^{(0)}(\ta, d, m;\x)$, 
$\ta\in \Lambda$, $d\in [0,\min\{\val(\ta),\dmax\}-2]$,
  $m\in[d,\val(\ta)-2]$:
 the set of   rooted nc-trees $T$ with a root $r$ such that
 \[\mbox{
$\alpha(r)=\ta$, $\deg_T(r)=d$, $\beta_T(r)=m$,   
 $\h(T)\leq {\rho}$ and $\f(T[+2])\leq  \x $;  }\]  
 Let 
  $\W_{\inl}^{(0)}(\ta, d, m;\x)$ denote the set of the frequency vectors
  $\w=\f(T[+2])$ for all nc-trees $T\in \T_{\inl}^{(0)}(\ta, d, m;\x)$;
\item[(ii)]  $\T_{\en}^{(0)}(\ta, d, m;\x)$, 
$\ta\in \Lambda$,  $d\in [1,\min\{\val(\ta),\dmax\}-1]$, 
 $m\in[d,\val(\ta)-1]$:
 the set of rooted nc-trees $T$ with a root $r$ such that
 \[\mbox{
 $\alpha(r)=\ta$, $\deg_T(r)=d$, $\beta_T(r)=m$,
   $\h(T)= {\rho}$ and  $\f(T[+1])\leq  \x $;  }\]   
 Let 
  $\W_{\en}^{(0)}(\ta, d, m;\x)$ denote the set of the frequency vectors
  $\w=\f(T[+1])$ for all nc-trees $T\in \T_{\en}^{(0)}(\ta, d, m;\x)$;
\item[(iii)] 
 $\T_{\en}^{(h)}(\ta , d , m ;\x  )$, $\x=\x_\typ^*$, 
  $\ta \in \Lambda$, 
 $d \in [1,\min\{\val(\ta),\dmax\}-1]$, $m \in [d ,\val(\ta )-1]$,  
 $h\in[ 1,  \ch_\UB(\typ)]$:
  the  set of bi-rooted nc-trees  $T$ such that 
\[\mbox{
 $\alpha(r_1(T))=\ta $, $\deg_T(r_1(T))=d $,  $\beta(r_1(T))=m $,
  $\ell(P_T)=h$,   $\f(T[+1])\leq \x$, }\]
\[\mbox{
  $\h(T')\leq {\rho}$ for all trees $T'\in \F(T)$ and }\]
\[\mbox{
  $\h(T'')={\rho}$ for the tree $T''\in \F(T)$ rooted at terminal $r_2(T)$;  }\]  
Let 
 $\W_{\en}^{(h)}(\ta , d , m  )$ denote the set of all frequency vectors
   $\w=\f(T[+1])$  for  all bi-rooted nc-trees  $T\in \T_{\en}^{(h)}(\ta , d , m  )$;
\item[(iv)]    $\T_{\co+\Delta}^{(0)}(\ta, d, m, h;\x)$, $\x=\x_\typ^*$,  
$\ta\in \Lambda$, $\Delta\in [2,3]$,
 $d\in [0,\min\{\val(\ta),\dmax\}-\Delta]$, 
 $m\in[d,\val(\ta)- \Delta]$, $h\in [0,\ch_\UB(\typ)]$:
 the set of rooted  c-trees  $T$ with a root $r$ such that
 \[\mbox{
$\alpha(r)=\ta$, $\deg_T(r)=d$, $\beta_T(r)=m$,
   $\h(T)= h$   and   $\f(T[+\Delta])\leq  \x $;  }\]  
   Let $\W_{\co+\Delta}^{(0)}(\ta, d, m, h;\x)$ denote
   the set of the frequency vectors $\w=\f(T[+\Delta])$
   for all c-trees $T\in \T_{\co+\Delta}^{(0)}(\ta, d, m, h;\x)$; 
\item[(v)] 
   $\T_{\co+1,\Delta}^{(q)}(\ta ,d ,m ,\tb,1,m',h; \x)$, $\x=\x_e^*$,
  $\ta,\tb  \in \Lambda$,  $\Delta\in [2,3]$,
  $d \in [1,\min\{\val(\ta),\dmax\}-1]$, 
  $m \in [d ,\val(\ta )-1]$, $m' \in [1 ,\val(\tb )-\Delta]$,
    $h\in [0, \ch_\UB(e)]$,
  $q\in[1, \ell(P_e) ]$:  
   the  set of bi-rooted c-trees $T$ such that 
\[\mbox{ $\alpha(r_1(T))=\ta$,   $\deg_T(r_2(T))=1$,   $\beta(r_1(T))=m$,  }\]
\[\mbox{  $\alpha(r_2(T))=\tb$,   $\deg_T(r_1(T))=d $,   $\beta(r_2(T))=m'$,  }\]
\[\mbox{   $\ell(P_T)=q$, $\h(T) =h$
 and   $\f(T[+1,\Delta])\leq \x$;   }\]
Let    $\W_{\co+1,\Delta}^{(q)}(\ta ,d ,m ,\tb,1,m',h;\x)$ 
   denote the set of the frequency vectors $\w=\f(T[+1,\Delta])$ 
for all bi-rooted c-trees 
$T\in \T_{\co+1,\Delta}^{(q)}(\ta ,d ,m ,\tb,1,m',h;\x)$.
\end{enumerate}

 Note that $\w(\tbc)=0$ for any vector $\w$ in the above set in (i)-(iii).         

Our algorithm consists of six steps. 
Step~1 computes the sets of trees and vectors in (i), (ii) and (iii) with $h\leq {\rho}$,
where each tree in these sets is of height at most ${\rho}$.
Note that the frequency vectors of some two trees in a tree set $\T$ in
the above can be identical.
In fact, the size $|\T|$ of a set $\T$ of trees can be considerably larger than
that $|\W|$ of the set $\W$ of their frequency vectors.
We mainly maintain a whole vector set $\W$, and
for each vector $\w\in \W$, we store 
at least one tree $T_{\w}\in \T$ such that $\w$ is the frequency
vector of a fictitious tree of $T_{\w}$, where we call such a tree $T_{\w}$
a {\em sample tree} of the vector $\w$.
With this idea, Steps~2-5 compute only vector sets $\W$ in
(iii) with $h>{\rho}$, (iv) and (v).
In each of these steps, we compute a set $\Sb$ of sample trees
of the vectors in each vector set $\W$. 
 The last step constructs at least one target component 
 for each base-vertex or base-edge, and then
 combines them to obtain a graph $G^*$ to be inferred in Stage~5.
 
 We derive recursive formula that hold among the above sets.
 Based on this, we compute the vector sets in (iii) in Step~2,
 those in (iv) in Step~3 and  those in (v) in Step~4.
 During these steps, we can find a target $v$-component
 for each base-vertex $v\in V_B$.
 For each base-edge $e=(u_1,u_2)\in E_B$,
 Step~5 compare vectors $\w_1$ and $\w_2$,
 where $\w_i$ is the frequency vector of a c-tree $T_i$ that is 
 extended from the end-vertex $u_i$, 
 to examine whether $T_1$ and $T_2$ give rise to a target $e$-component.
 
\subsubsection{Step~1: Enumeration of ${\rho}$-fringe-trees}
 \label{sec:fringe-tree}

 Step~1   computes the following sets in (i)-(iv).  
 \begin{enumerate}
\item[(i)]  
For each  base-vertex $v\in V_B$ such that
 $\ch_\LB(v)\leq {\rho}$ and $\x^*_v(\tbc)=0$, 
  compute the set 
$\T_{\co+\Delta_v}^{(0)}(\ta_v, d_v, m_v, h ;\x^*_v)$, 
$h\in [\ch_\LB(v),  \min\{{\rho}, \ch_\UB(v)\} ]$ of rooted c-trees. 
Note that 
every c-tree in the set $\T_{\co+\Delta_v}^{(0)}(\ta_v, d_v, m_v, h ;\x^*_v)$
with   $h\in [\ch_\LB(v),   \min\{{\rho}, \ch_\UB(v)\} ]$ 
is a target $v$-component in $\T_v$; \\
Set $\Sb(v)$ be a subset of $\bigcup_{h\in [\ch_\LB(v),   \min\{{\rho}, \ch_\UB(v)\} ]}
\T_{\co+\Delta_v}^{(0)}(\ta_v, d_v, m_v, h ;\x^*_v)$; 

 \item[(ii)]  
 For each  base-vertex $v\in V_B$ such that ${\rho}< \ch_\UB(v)$  and $\x^*_v(\tbc)=1$
 and   integers   $m\in[d_v-1, \val(\ta_v)- \Delta_v-1]$ and $p\in [0,k]$,
 compute the sets 
$\T_{\co+(\Delta_v+1)}^{(0)}(\ta_v, d_v-1, m, p;\x^*_v)$ of rooted c-trees
and 
$\W_{\co+(\Delta_v+1)}^{(0)}(\ta_v, d_v-1, m, p ;\x^*_v)$
of   their frequency vectors; \\  
For each vector $\w\in \W_{\co+(\Delta_v+1)}^{(0)}(\ta_v, d_v-1, m, p;\x^*_v)$,  
choose some number of sample trees $T_{\w}$,
and store them in a set $\Sb_{\co+(\Delta_v+1)}^{(0)}(\ta_v, d_v-1, m, p;\x^*_v)$;

\item[(iii)] 
 For each  base-vertex $v\in V_B$ such that ${\rho}< \ch_\UB(v)$  and $\x^*_v(\tbc)=1$, 
 each possible tuple $(\ta,d,m)$,     compute 
the sets  $\T_{\inl}^{(0)}(\ta, d, m; \x^*_v)$ and  $\T_{\en}^{(0)}(\ta, d, m; \x^*_v)$ 
of rooted nc-trees and 
the sets  $\W_{\inl}^{(0)}(\ta, d, m; \x^*_v)$ and $\W_{\en}^{(0)}(\ta, d, m; \x^*_v)$  
 of  their frequency vectors; \\  
For each vector $\w\in \W_{\inl}^{(0)}(\ta, d, m; \x^*_v)$ 
(resp.,  $\w\in \W_{\en}^{(0)}(\ta, d, m; \x^*_v)$), 
choose
some number of sample trees $T_{\w}\in \T_{\co+\Delta}^{(0)}(\ta, d, m, h ; \x^*_e)$,
and store them in a set $\Sb_{\inl}^{(0)}(\ta, d, m; \x^*_v)$ 
(resp.,  $\w\in \W_{\en}^{(0)}(\ta, d, m; \x^*_v)$);

\item[(iv)]  
 For each  base-edge $e\in E_B$ and each possible tuple $(\ta,d,m)$, 
   compute 
the sets  $\T_{\inl}^{(0)}(\ta, d, m; \x^*_e)$ and  
  $\T_{\en}^{(0)}(\ta, d, m; \x^*_e)$ 
of rooted nc-trees and   
 $\T_{\co+\Delta}^{(0)}(\ta, d, m, h ; \x^*_e)$, 
 $h\in [0, \min\{{\rho},  \ch_\UB(e)\}]$
 of rooted c-trees
and 
the sets  $\W_{\inl}^{(0)}(\ta, d, m; \x^*_e)$, $\W_{\en}^{(0)}(\ta, d, m; \x^*_e)$  
and    $\W_{\co+\Delta}^{(0)}(\ta, d, m, h ; \x^*_e)$ of  their frequency vectors; \\  
For each vector 
 $\w\in \W_{\inl}^{(0)}(\ta, d, m; \x^*_e)$
 (resp., $\w\in \W_{\en}^{(0)}(\ta, d, m; \x^*_e)$  and
 $\w\in \W_{\co+\Delta}^{(0)}(\ta, d, m, h ; \x^*_e)$), 
choose some number of sample trees $T_{\w}$ 
and store them in a set 
 $\Sb_{\inl}^{(0)}(\ta, d, m; \x^*_e)$
 (resp., $\Sb_{\en}^{(0)}(\ta, d, m; \x^*_e)$  and
 $\Sb_{\co+\Delta}^{(0)}(\ta, d, m, h ; \x^*_e)$), 
\end{enumerate}

To compute the above sets of trees and vectors, 
we enumerate all possible trees with height at most 2
under the size constraint (\ref{eq:fringe-size})
by a branch-and-bound procedure.

\subsubsection{Step~2: Generation of Frequency Vectors of End-subtrees}
 \label{sec:internal-subtree}

\begin{figure}[!ht] \begin{center}
\includegraphics[scale=0.50]{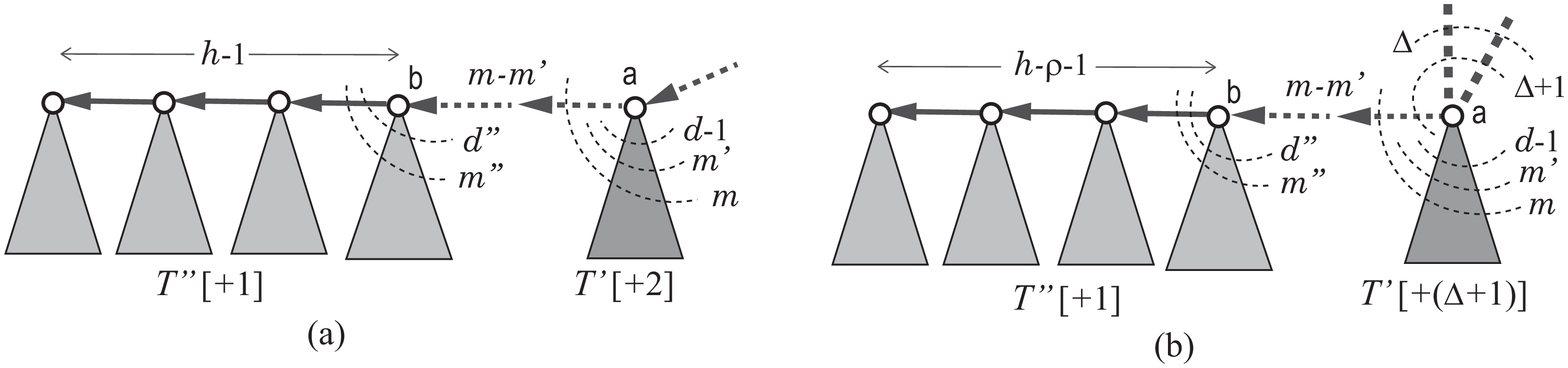}
\end{center}
\caption{ (a) An illustration of   computing  a vector
 $\w\in \W_{\en}^{(h)}(\ta,d,m;\x^*_\typ)$
      from the frequency vectors 
 $\w'=\f(T'[+2])\in \W_{\inl}^{(0)}(\ta , d -1, m';\x^*_\typ)$
    of a bi-rooted nc-tree $T'$   and 
 $\w'' =\f(T''[+1])  \in \W_{\en}^{(h-1)}(\tb, d'', m'';\x^*_\typ)$ 
  of an nc-tree $T''$;  
(b)  An illustration of computing a vector 
 $\w\in \W_{\co+\Delta}^{(0)}(\ta,d,m,h;\x^*_\typ)$
      from the frequency vectors 
$\w'=\f(T'[+(\Delta+1)])
\in \W_{\co+(\Delta+1)}^{(0)}(\ta, d-\! 1, m', p;\x^*_\typ)$, $p\in[0,{\rho}]$
  of a c-tree $T'$  and 
$\w'' =\f(T''[+1])  \in \W_{\en}^{(h-{\rho}-1)}(\tb, d'', m'';\x^*_\typ)$.   	}
\label{fig:extend_nc-trees} \end{figure}

For each   base-vertex $\typ=v\in V_B$ or each base-edge $\typ=e\in E_B$
such that ${\rho}<\ch_\UB(\typ)$ and each possible tuple $(\ta,d,m)$, 
 Step~2  computes  the  set 
 $\W_{\en}^{(h)}(\ta , d , m ;\x^*_\typ  )$ 
in the ascending order of $h=1,2,\ldots, \ch_\UB(\typ) -{\rho}-1$.  
Observe that  each vector $\w\in \W_{\en}^{(h)}(\ta , d , m;\x^*_\typ)$ 
is obtained as $\w = \w'  + \w''  + \1_{\gamma^\inn}$
from a combination of vectors  
$\w'\in \W_{\inl}^{(0)}(\ta , d -1, m' ;\x^*_\typ)$ 
 and 
$\w''  \in \W_{\en}^{(h-1)}(\tb, d'', m'' ;\x^*_\typ)$ 
such that  
\[\begin{array}{l}  
  m' \leq \val(\ta ) - 2,  ~~    1 \leq m-m'  \leq \val(b) - m'' ,   \\
  \w' + \w''  + \1_{\gamma^\inn}   \leq \x^*_\typ   
 \mbox{ for } \gamma = (\ta \{d\!+\!1\},  \tb \{d''\!+\!1\}, m -m' ) \in \Gamma. 
\end{array}\]  
Figure~\ref{fig:extend_nc-trees}(a) illustrates this process of computing
a vector $\w \in \W_{\en}^{(h)}(\ta , d , m  ;\x^*_\typ)$.

For each vector $\w\in \W_{\en}^{(h)}(\ta , d , m  ;\x^*_\typ)$
obtained  from a combination  $\w'\in \W_{\inl}^{(0)}(\ta , d -1, m';\x^*_\typ)$ 
 and  
$\w'' \in \W_{\en}^{(h-1)}(\tb, d'', m'';\x^*_\typ)$, 
we construct at least one sample nc-tree $T_{\w}$ from their sample nc-trees
$T_{\w'}\in \Sb_{\inl}^{(0)}(\ta , d -1, m';\x^*_\typ)$  and
 $T_{\w''}\in \Sb_{\en}^{(h-1)}(\tb, d'', m'';\x^*_\typ)$
 and store them in a set $\Sb_{\en}^{(h)}(\ta , d , m  ;\x^*_\typ)$. 
 
\subsubsection{Step~3: Generation of Frequency Vectors of Rooted Core-subtrees} 
 
For each  base-vertex $\typ=v\in V_B$ or each base-edge $\typ=e\in E_B$
such that ${\rho}<\ch_\UB(\typ)$ and
 each possible tuple $(\ta,d,m,h)$ with $h\in[{\rho}+1,\ch_\UB(\typ)]$, 
 Step~3  computes  the  set   
 $\W_{\co+\Delta}^{(0)}(\ta, d, m, h ; \x^*_\typ  )$.
Observe that  each vector $\w \in \W_{\co+\Delta}^{(0)}(\ta,d,m,h; \x^*_\typ )$
 is obtained as $\w = \w'  + \w''  + \1_{\gamma^\inn} +\1_{\tbc}$
  from a combination of vectors  
$\w' \in \W_{\co+(\Delta+1)}^{(0)}(\ta, d-\! 1, m', p; \x^*_\typ )$, $p\in [0,k]$ 
 and 
$\w'' \in   \W_{\en}^{(h-{\rho}-1)}(\tb, d'', m''; \x^*_\typ )$
such that 
\[\begin{array}{l}
  m' \leq \val(\ta ) - \Delta-1,  ~~    1 \leq m-m'  \leq \val(b) - m'' ,   \\
  \w' + \w''  + \1_{\gamma^\inn} +\1_{\tbc}   \leq \x^*_\typ   
 \mbox{ for  }  
  \gamma = (\ta \{d\!+\!\Delta\}, \tb\{d''\!+\!1\}, m- m' ) \in \Gamma,
\end{array}\] 
where $\w'(\tbc)= \w''(\tbc)=0$. 
Figure~\ref{fig:extend_nc-trees}(b) illustrates this process of computing
a vector $\w \in \W_{\co+\Delta}^{(0)}(\ta,d,m,h;\x^*_\typ)$.
 
For each vector $\w \in \W_{\co+\Delta}^{(0)}(\ta,d,m,h; \x^*_\typ )$    obtained
from a combination of vectors \\
 $\w' \in \bigcup_{p\in [0,k]}\W_{\co+(\Delta+1)}^{(0)}(\ta,d-\! 1, m'', p; \x^*_\typ )$ 
 and $\w'' \in \W_{\en}^{(h-{\rho}-1)}(\tb, d',m'; \x^*_\typ )$,
we construct at least one sample c-tree $T_{\w}$ from their sample nc-trees
$T_{\w'}\in \bigcup_{p\in [0,k]}\Sb_{\co}^{(0)}(\ta,d-\! 1, m'', p; \x^*_\typ )$  and
 $T_{\w''}\in \Sb_{\en}^{(h)}(\tb, d',m'; \x^*_\typ )$
 and store them in a set $\Sb_{\co+\Delta}^{(0)}(\ta,d,m,h; \x^*_\typ )$. 
  
 For each base-vertex $v\in V_B$ with ${\rho}<\ch_\UB(v)$ and $\x^*_v(\tbc)=1$, 
  all sample  c-trees
   $T_{\w}\in \Sb_{\co+\Delta_v}^{(0)}(\ta_v,d_v,m_v, \h^*_v ; \x^*_v )$  
are target $v$-components  in $\T_v$,
and we set $\Sb(v):= \Sb_{\co+\Delta_v}^{(0)}(\ta_v,d_v,m_v, \h^*_v; \x^*_v )$.

\begin{figure}[!ht] \begin{center}
\includegraphics[scale=0.47]{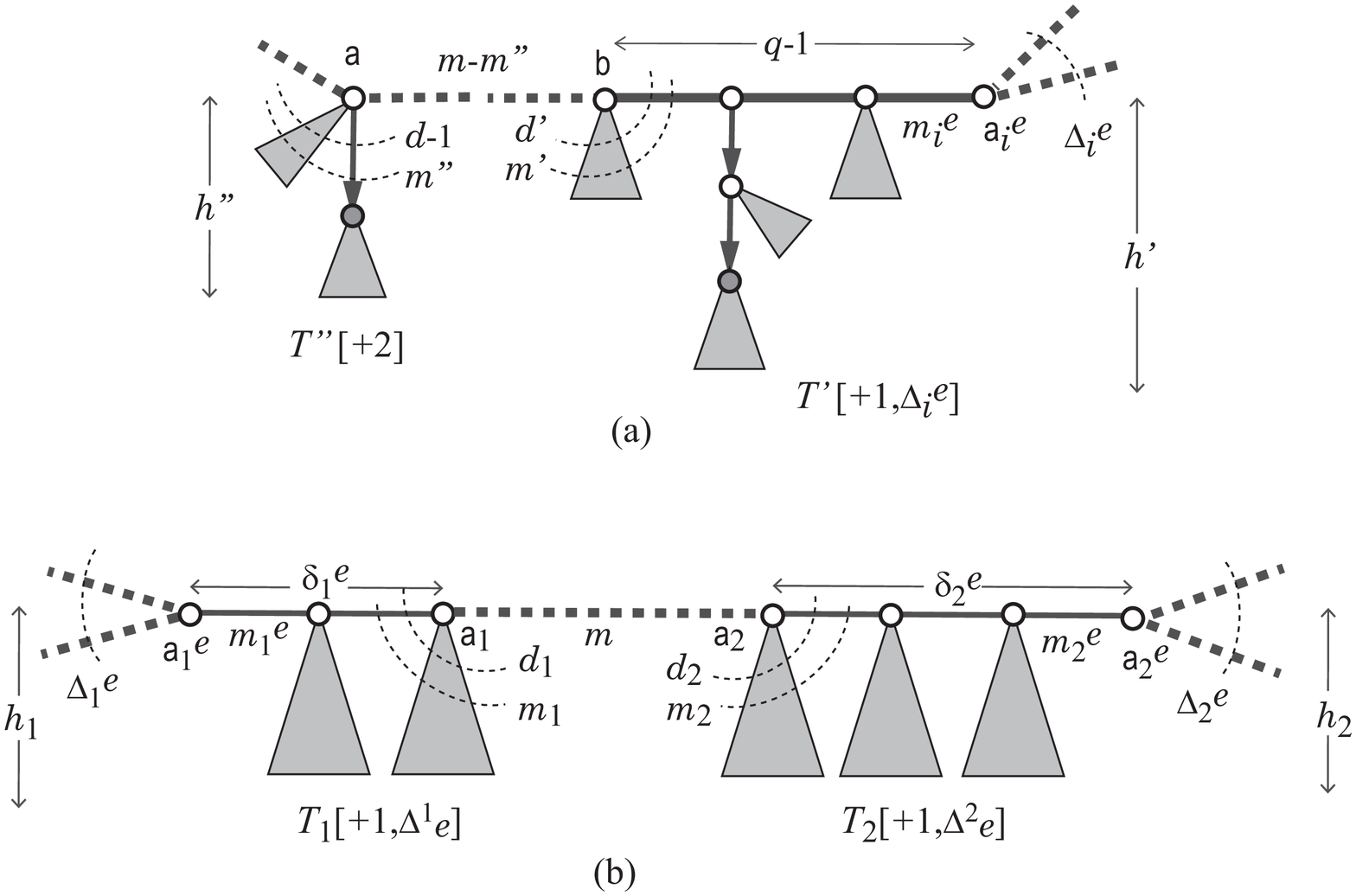}
\end{center}
\caption{
(a) An illustration of computing a vector
 $\w\in \W_{\co+1,\Delta^e_i}^{(q)}(\ta ,d ,m ,\ta^e_i,1,m^e_i,h;\x^*_e)$
 for a base-edge $e\in E_B$ 
 from the frequency vectors 
$\w' = \f(T''[+1,\Delta^e_i])
   \in \W_{\co+1,\Delta^e_i}^{(q-1)}(\tb, d', m',  \ta^e_i,1,m^e_i, h';\x^*_e)$
  of a c-tree $T'$   and 
$\w''= \f(T''[+2])\in \W_{\co+2}^{(0)}(\ta , d -1, m'', h'';\x^*_e)$  
 of a c-tree $T''$;  
 (b) An illustration of  computing a feasible vector pair  $(\w^1,\w^2)$ with 
 $\w^i=\f(T_i[+1,\Delta_i^e])
  \in \W_{\co+1,\Delta_i^e}^{(\delta_i^e)}(\ta_i,d_i,m_i,\ta_i^e,1,m_i^e,h_i;\x^*_e)$,
  of   c-trees $T_i$  $i=1,2$  for a base-edge $e\in E_B$. }
\label{fig:extend_c-trees} \end{figure}

\subsubsection{Step~4: Generation of Frequency Vectors of Bi-rooted Core-subtrees}

For each   base-edge $e\in E_B$, each index $i=1,2$  and
 each possible tuple $(\ta,d,m,h)$ with $h\in [\ch_\LB(e), \ch_\UB(e)]$, 
 Step~4  computes  the  set   
  $\W_{\co+1,\Delta^e_i}^{(q)}(\ta ,d ,m ,\ta^e_i,1,m^e_i, h ; x^*_e)$
  in the ascending order $q=1,2,\ldots,  \delta^e_i $.
Observe that  each vector 
 $\w\in \W_{\co+1,\Delta^e_i}^{(q)}(\ta ,d ,m ,\ta^e_i,1,m^e_i, h ; x^*_e)$,
 is obtained as  $\w  = \w' + \w''  + \1_{\gamma^\co}$ 
 from a combination of vectors   
$\w'  \in \W_{\co+1,\Delta^e_i}^{(q-1)}(\tb, d', m', \ta^e_i,1,m^e_i,h'; x^*_e)$,
   and 
$\w''\in \W_{\co+2}^{(0)}(\ta , d -1, m'', h''; x^*_e)$ 
such that 
\[\begin{array}{l} 
  h=\max\{h',h''\}  \in [\ch_\LB(e), \ch_\UB(e)],  ~ 
  m'' \leq \val(\ta_1) - 2, ~   1\leq m - m''  \leq \val(b) - m' ,   \\
  \w' + \w''  + \1_{\gamma^\co}   \le \x^*_e   
 \mbox{ for  }  
 \gamma = (\ta  d , \tb \{d'\!+\!1\}, m -m'' ) \in \Gamma_{\leq}. 
\end{array}\]
Figure~\ref{fig:extend_c-trees}(a)  illustrates this process of computing
a vector $\w\in \W_{\co+1,\Delta^e_i}^{(q)}(\ta ,d ,m ,\ta^e_i,1,m^e_i,h; x^*_e)$,

For each vector 
$\w\in \W_{\co+1,\Delta^e_i}^{(q)}(\ta ,d ,m ,\ta^e_i,1,m^e_i,h; x^*_e)$ 
obtained from a combination 
   $\w'  \in \W_{\co+1,\Delta^e_i}^{(q-1)}(\tb,$ $ d', m', \ta^e_i,1,m^e_i,h'; x^*_e)$ 
   and 
$\w''\in \W_{\co+2}^{(0)}(\ta , d -1, m'', h''; x^*_e)$,   
we construct at least one sample c-tree $T_{\w}$ from their sample nc-trees
$T_{\w'}\in \Sb_{\co+1,\Delta^e_i}^{(q-1)}(\tb, d', m', \ta^e_i,1,m^e_i,h'; x^*_e)$  and
 $T_{\w''}\in \Sb_{\co+2}^{(0)}(\ta , d -1, m'', h''; x^*_e)$
 and store them in a set
  $\Sb_{\co+1,\Delta^e_i}^{(q)}(\ta ,d ,m ,\ta^e_i,1,m^e_i,h; x^*_e)$.

\subsubsection{Step~5: Enumeration of Feasible  Vector Pairs}
\label{sec:feasible_pair}

For each edge $e\in E_B$, a {\em feasible vector pair}  
 is defined to be a pair of vectors
$\w^i   \in \W_{\co+1,\Delta_i^e}^{(\delta_i^e)}
(\ta_i,d_i,$ $m_i,\ta_i^e,1,m_i^e,h_i ; \x^*_e)$, $i=1,2$  that satisfies 
\[ \max\{h_1,h_2\}\in [\ch_\LB(e), \ch_\UB(e)] \mbox{ and } 
\x^*_e = \w^1  +\w^2 + \1_{\gamma}       \]
for an edge-configuration 
$\gamma = (\ta_1 \{d_1\!+\!1\}, \ta_2 \{d_2\!+\!1\}, m)\in \Gamma_{\leq}$
with an  integer $m \in [1, \min\{3, \val(\ta_1) - m_1, \val(\ta_2) - m_2\} ]$.
The second equality is equivalent with 
a condition that $\w^1$ is equal to the vector $ \x^*_e -\w^2-\1_{\gamma}   $,
which we call  the {\em $ \gamma $-complement}
of $\w^2$, and denote it by  $\overline{\w^2}$.
Figure~\ref{fig:extend_c-trees}(b)  illustrates this process of computing
a feasible vector pair $(\w^1,\w^2)$.

For each edge $e\in E_B$,  
 Step~5  enumerates the set $\W_\pair(e)$ of all feasible vector pairs $(\w^1,\w^2)$.
To efficiently search for a feasible pair of vectors
in two sets  $\W_{\co+1,\Delta_i^e}^{(\delta_i^e)}
(\ta_i,d_i,m_i,\ta_i^e,1,m_i^e,h_i ; \x^*_e)$,   $i = 1,2$
with $\max\{h_1,h_2\}\in [\ch_\LB(e), \ch_\UB(e)]$, 
we first compute the 
$ \gamma $-complement vector $\overline{\w}$  
of each vector
 $\w\in \W_{\co+1,\Delta_2^e}^{(\delta_2^e)}
(\ta_2,d_2,m_2,\ta_2^e,1,m_2^e,h_2)$ 
for each   an edge-configuration 
$\gamma = (\ta_1 \{d_1+1\}, \ta_2 \{d_2+1\}, m)\in \Gamma$
   with  
$m \in [1, \min\{3, \val(\ta_1) - m_1, \val(\ta_2) - m_2\} ]$,
and denote by  $\overline{\W_{\co}^{(\delta_2)}}$ the set of the resulting 
$\gamma$-complement vectors.
Observe that $(\w^1,\w^2)$ is a feasible vector pair if and only if
 $\w_1=\overline{\w_2}$.
To find such pairs,  we merge the sets
 $\W_{\co+1,\Delta_1^e}^{(\delta_1^e)}
(\ta_1,d_1,m_1,\ta_1^e,1,m_1^e,h_1;\x^*_e)$  
 and $\overline{\W_{\co}^{(\delta_2)}}$
into a sorted list $L_{\gamma }$.
Then each feasible vector pair $(\w^1,\w^2)$ appears
as a consecutive pair of vectors $\w_1$ and $\overline{\w_2}$
in the list $L_{\gamma, \mu}$.

\subsubsection{Step~6: Construction of Target Chemical Graphs}
 \label{sec:construct}
 
The task of Step~6 is to  construct for each feasible vector pair  
$(\w^1,\w^2)\in \W_\pair(e)$, 
construct  at least one target $e$-component  $T_{(\w_1, \w_2)}\in \T_e$ by combining
the sample c-trees $T_i=T_{\w^i}\in 
\Sb_{\co+1,\Delta_i^e}^{(\delta_i^e)}
(\ta_i,d_i,m_i,\ta_i^e,1,m_i^e,h_i ; \x^*_e)$,   $i = 1,2$ with
an edge $r_1(T_1)r_1(T_2)$ with a bond-multiplicity $m$,
and store these target $e$-components  $T_{(\w_1, \w_2)}$ in
 a set $\Sb(e)$.
Figure~\ref{fig:extend_c-trees}(b)  illustrates
two sample c-trees $T_i$, $i=1,2$ to be combined 
with a new edge $e=r_1(T_1)r_1(T_2)$. 

For each base-vertex  $v\in B$ and each base-edge $e\in E_B$,
a set $\Sb(v)$ of target $v$-components and a set $\Sb(e)$ 
of target $e$-components
have be constructed. 
Let $\mathcal{C}$ be a collection obtained 
by choosing a target $v$-component $T_v\in \Sb(v)$ 
for each base-vertex  $v\in B$ and 
and a target $e$-component $T_e\in \Sb(e)$ for each base-edge $e\in E_B$.
Then a chemical graph $G^*(C)$ obtained by assembling these components
is a target chemical graph to be inferred in Stage~5.
The number of chemical graphs $G^*(C)$ in this manner is
\[ \prod_{v\in V_B}|\Sb(v)|\times\prod_{e\in E_B}|\Sb(e)| , \]
where we ignore a possible automorphism over the resulting graphs.

For a base-edge $e \in E_B$ with
a relatively large instance size $\delta^e_2$, 
the number $|\W_\pair(e)|$ 
of feasible vector pairs in Step~5 still can be very large.
 In fact, the size $|\W|$ of a vector set $\W$ to be computed in Steps~2 to 4
 can also be considerably large during an execution of the algorithm.
 For such a case, 
   we impose a time limitation on the running time for computing $\W$ 
 and a memory limitation on the number of vectors stored in a vector set $\W$.
With these limitations,  
 we can compute only a limited subset $\widehat{\W}$ 
 of each vector set $\W$  in Steps~2 to 4.
 Even with such a subset $\widehat{\W}$,
 we still can find a large size of a subset $\widehat{\W}_\pair(e)$
  of $\W_\pair(e)$ in Step~5.  

Our algorithm also can deliver a lower bound on the number 
$|\T_\typ(\x^*_\typ)|$, $\typ\in V_B\cup E_B$ of
all target components   in the following way.
In Step~1, we also compute the number $t(\w)$ of rooted trees $T\in \T^{(0)}$
 in (i)-(iii).
In Steps~2, 3  and 4, when a vector $\w$ is constructed from
two vectors $\w'$ and $\w''$, we iteratively compute the number $t(\w)$
of all trees $T$ such that $\w$ is the frequency vector of 
a fictitious tree of $T$  by 
$t(\w):=t(\w')\times t(\w'')$.  
In Step~5,  when a feasible vector pair $(\w^1,\w^2)\in \W_\pair(e)$ is obtained
for a base-edge $e\in E_B$,
we know that the number of the corresponding target $e$-components 
is $t(\w^1)\times t(\w^2)$. 
Possibly we compute a subset $\widehat{\W}_\pair(e)$
  of $\W_\pair$ in Step~4.
Then $(1/2)\sum_{(\w^1,\w^2)\in\widehat{\W}_\pair}t(\w^1)\times t(\w^2)$
gives a lower bound on the number $|\T_e|$ of target $e$-components,
where we divided by 2 since 
an axially symmetric target $e$-component can correspond
to   two vector pairs in $\W_\pair(e)$.
A lower bound on the  number $|\T_v|$ of target $v$-components
for a base-vertex $v\in V_B$ can be obtained in a similar way.

\subsection{Choosing a Path-partition}
 \label{sec:sigma-extension} 

This section describes how to apply our new algorithm for generating
chemical isomers in Stage~4 after we obtain a chemical ${\rho}$-lean cyclic 
graph $G^\dagger$.

Let $(\sco,\snc,\sab)$ be a specification of substructures
and $G^\dagger\in \G(\sco,\snc,\sab)$ be an $(\sco,\snc,\sab)$-extension,
where we assume that the minimum $\sco$-extension $C_{\min}\in\mathcal{C}(\sco)$
is a simple connected graph with the minimum degree at least 2.
Let $C^\dagger=(V^\co,E^\co)$ denote the core of $\Cr(G^\dagger)$
and 
 $E_0$ denote the set of edges $e\in \Ez$ that are removed in the construction
 of  $C^\dagger$ from the seed graph $\GC$.

To generate chemical ${\rho}$-lean cyclic graphs $G^*\in \G(\sco,\snc,\sab)$ 
by our new algorithm,  we first choose a path-partition 
$\Pt=\{P_1,P_2,\ldots,P_p\}$ of the core $C^\dagger$.
Recall that  the base-graph $G_B=(V_B,E_B=\{e_1,e_2,\ldots,e_p\})$
is  determined by the partition $\Pt$
so that each edge $e_j\in E_B$ directly joins the end-vertices  of each path $P_j\in \Pt$
and  $V_B$ is the set of end-vertices of paths in $\Pt$.
We choose a path-partition 
$\Pt=\{P_1,P_2,\ldots,P_p\}$ so that  the next condition is satisfied.
\[ V_B=\VC  , ~  E_B=\EC \setminus E_0; \]
  i.e., each edge $e_i\in E_B$ corresponds to an edge $a_i\in \EC \setminus E_0$.
   
 We next set a specification $\sch$ to be a set of the
 lower and upper bound functions $\ch_\LB,\ch_\UB:\VC\cup\EC\to\mathbb{Z}_+$
 for the set $V_B=\VC$  of vertices and the set $E_B=\EC \setminus E_0$ 
   of edges.
 Observe that any $(\Pt,\sch)$-isomer $G^*$ of $\G^\dagger$ is a  $(\sco,\snc,\sab)$-extension,
 since the assignment of elements of $\Lambda$ to the base-vertices in $V_B$
 remains unchanged among all $(\Pt,\sch)$-isomer  of $\G^\dagger$. 
 The converse is not true in general; i.e., there may be a $(\sco,\snc,\sab)$-extension $G$
 that is not a $(\Pt,\sch)$-isomer of $\G^\dagger$.

 In Stage~5, we run our new algorithm for generating $(\Pt,\sch)$-isomers $G^*$ of $\G^\dagger$.

 We remark that when 
 lower and upper bound functions $\ch_\LB$ and $\ch_\UB$ in 
 a core specification $\sco$ are uniform overall vertices or edges,
 we can choose a path-partition $\Pt$ in a more flexible way.
 In this case, we can apply our algorithm if a path-partition $\Pt$ satisfies
 \[ \VC^*\cup  V_3^\co\subseteq V_B  \] 
 for  the set  $V_{3}^\co$ of core-vertices
 $v\in V^\co$ of degree at least 3, i.e.,  $\deg_{C^\dagger}(v)\geq 3$.
We can choose a path in $\Pt$ so that it ends with a core-vertex
$v\in V^\co\setminus (\VC^*\cup  V_3)$ of degree 2, i.e., 
 $\deg_{C^\dagger}(v) =2$.

\subsection{A Possible Extension to the General Graphs}
 \label{sec:possible_extension} 

When a given chemical graph $G^\dagger$ is not a ${\rho}$-lean cyclic 
graph, we can extend the definition of the chemical graph isomorphism
in a flexible way.
Suppose that  $G^\dagger$ is an acyclic graph.
We first choose a vertex $r$ as the root of tree $G^\dagger$,
where $r$ is not necessarily a graph-theoretically designated vertex
such as a center or a centroid. 
For a branch-parameter ${\rho}$ such as ${\rho}=2$,
find the set $V_\br$ of all ${\rho}$-branches of the tree.
We set $V_B:=V_\br\cup\{r\}$ to be a set of base-vertices and
$\Pt=\{P_1,\ldots,P_p\}$ denote the collection of paths 
with end-vertices of base-vertices  in $V_B$ 
and no internal vertices from $V_B$. 
For each path $P_{u,v}\in \Pt$ between two base-vartices $u,v\in V_B$,
prepare a base-edge $e=uv$ and let $E_B$ denote the set of 
the resulting base-edges.
Then the base-graph $G_B=(V_B,E_B)$ in this case is a tree. 
Based on $G_B$, we can define the vertex and edge components of $G^\dagger$
in a similar way, and can generate target $v$-components $T_v$, $v\in V_B$
 and target $e$-components $T_e$, $e\in E_B$,
 each of them independently, 
by our algorithm with a slight modification
or the algorithm for trees $T$ with $\bl_{\rho}(T)=2$ due to
 Azam~et~al.~\cite{AZSSSZNA20}.

Now consider the case where $G^\dagger$ is cyclic but not ${\rho}$-lean.
In this case,  some tree $T$ rooted at a core-vertex may contain
more than one leaf ${\rho}$-branch. 
Let $\T_{\br}$ denote the set of all these rooted trees,
and let   $V_\br$ denote  the set of all ${\rho}$-branches of the trees in $\T_\br$.
Let $V^\co_{\br}$ denote the set of core-vertices $v$ at which
some tree $T_v\in \T_{\br}$ is rooted.
We find a path-partition $\Pt^\co$ of the core $\Cr(G^\dagger)$
so that each vertex in  $V^\co_{\br}$ is used as an end-vertex 
of some path $P\in \Pt^\co$.
For the trees in $\T_\br$, we find a path-partition $\Pt^\nc$
with paths between two ${\rho}$-branches in $V_\br$
in an analogous way of the above tree case. 
Finally set $\Pt:=\Pt^\co\cup \Pt^\nc$ and define the base-graph
$G_B=(V_B,E_B)$ based on $\Pt$.
We see that target $v$-components $T_v$, $v\in V_B$
 and target $e$-components $T_e$, $e\in E_B$ can be generated
by our algorithm with a slight modification.

     
  
\section{Concluding Remarks}\label{sec:conclude}

In this paper, we employed the new mechanism of
utilizing a target chemical graph $G^\dagger$ obtained in Stage~4
of the framework for inverse QSAR/QSPR  
to generate a  larger number of target graphs $G^*$ in Stage~5.
We showed that a family of  graphs $G^*$ 
 that are chemically isomorphic to $G^\dagger$ 
can be obtained by the dynamic programming algorithm
 designed in Section~\ref{sec:graph_search}.
Based on the new mechanism of Stage~5, we proposed
a target specification on a seed graph as 
a flexible way of specifying a family of target chemical graphs.
With this specification, we can realize requirements
on partial topological substructure of the core of graphs
and partial assignment of chemical elements and bond-multiplicity
within the framework for inverse QSAR/QSPR  
by ANNs and MILPs. 

The current topological specification proposed in this paper
does not allow to fix part of the non-core structure of a graph.
We remark that it is not technically difficult to extend the MILP
formulation in  Section~\ref{sec:graph_MILP} 
and the algorithm for computing chemical isomers in 
Section~\ref{sec:graph_search}
so that a more general specification for such a case can be handled.

 \appendix
  
 

\section{All Constraints in an MILP Formulation for Chemical  
 Cyclic Graphs}\label{sec:full_milp}
 
We define a standard encoding of a finite set $A$ of elements
to be a bijection $\sigma: A \to [1, |A|]$, 
where we denote by $[A]$   the set $[1, |A|]$ of integers
and by $[{\tt e}]$ the encoded element $\sigma({\tt e})$.
Let $\epsilon$ denote {\em null}, a fictitious chemical element 
that does not belong to any set of chemical elements,
chemical symbols, adjacency-configurations and
edge-configurations in the following formulation.
Given a finite set $A$, let $A_\epsilon$ denote the set $A\cup\{\epsilon\}$
and define a standard encoding of $A_\epsilon$
  to be a bijection $\sigma: A \to [0, |A|]$ such that
$\sigma(\epsilon)=0$, 
where we denote by $[A_\epsilon]$   the set $[0, |A|]$ of integers
and by $[{\tt e}]$ the encoded element $\sigma({\tt e})$,
where $[\epsilon]=0$.

We choose a branch-parameter ${\rho}$ and 
subsets
 $\Lambda^\co, \Lambda^\nc \subseteq \Lambda$ of chemical elements, 
subsets  $\Ldg^\co\subseteq \Lambda^\co\times[2,4]$
and $ \Ldg^\nc\subseteq \Lambda^\nc\times[1,4]$ of chemical symbols,  
subsets
$\Gamma^\co\subseteq \Gamma_{<}(\Ldg^\co)\cup \Gamma_{=}(\Ldg^\co)$
and 
$\Gamma^\inn,\Gamma^\ex\subseteq \Gamma(\Ldg^\nc)$ of
edge-configurations.

 \bigskip 
 Let $(\sco,\snc,\sab)$ be a specification, and
 let $G$ be a chemical ${\rho}$-lean graph in $\G(\sco,\snc,\sab)$. 

 \subsection{Selecting  Core-vertices and Core-edges} 
\label{sec:co}
 
Recall that  
\[ \begin{array}{ll}
   \Eew = \{e\in \EC\mid \ell_\LB(e)=\ell_\UB(e)=1 \}; &
   \Ez =\{e\in \EC\mid \ell_\LB(e)=0, \ell_\UB(e)=1 \}; \\
  \Ew=\{e\in \EC\mid \ell_\LB(e)=1,  \ell_\UB(e)\geq 2 \}; &
  \Et= \{e\in \EC\mid \ell_\LB(e)\geq 2 \}; \end{array} \]
\begin{enumerate}
\item[-]
Every edge $a_i\in \Eew$ is  included in  $G$;

\item[-]
Each edge $a_i\in \Ez$ is   included in $G$ if necessary;
 
\item[-]
For each edge  $a_i  \in \Et$, edge $a_i$ is not included in $G$
and instead a path 
\[P_i=(\vC_{\tailC(i)}, \vT_{j-1,0},\vT_{j,0},\ldots,
    \vT_{j+t,0}, \vC_{\hdC(i)})\]
     of length at least 2
  from vertex $\vC_{\tailC(i),0}$ to vertex $\vC_{\hdC(i),0}$ 
  visiting some core-vertices in $\VT$ is constructed in $G$; and  
 
\item[-]
For each edge $a_i  \in \Ew$, either  edge $a_i$   is directly used in $G$ or
the above path $P_i$ of length at least 2   is constructed in $G$.  
 \end{enumerate}
 
Let  $\tC\triangleq |\VC|$ and denote $\VC$ by 
$\{\vC_{i,0}\mid i\in [1,\tC]\}$.
Regard the seed graph $\GC$ as a digraph such that
each edge $a_i$ with end-vertices $\vC_{j,0}$ and $\vC_{j',0}$
is directed from  $\vC_{j,0}$ to $\vC_{j',0}$ when $j<j'$.
 For each directed edge $a_i  \in \EC $,
 let $\hdC(i)$ and $\tailC(i)$ denote the head and tail of $\eC(i)$;
 i.e., $a_i=(\vC_{\tailC(i),0}, \vC_{\hdC(i),0})$. 
  
Assume that $\EC=\{a_i\mid i\in[1,\mC]\}$,
$\Et=\{a_k\mid k\in[1,p]\}$,
$\Et=\{a_k\mid k\in[p+1,q]\}$,
$\Ez=\{a_i\mid i\in[q+1,t]\}$ and 
$\Eew=\{a_i\mid i\in[t+1,\mC]\}$ 
for integers $p,q$ and $t$. 
Define 
 \[ \kC \triangleq  |\Et\cup \Ew| , ~~ \widetilde{\kC} \triangleq  |\Et| .\]
To control construction of such a path $P_i$
 for each edge  $a_k\in  \Et\cup \Ew $,
we regard the index $k\in [1,\kC]$ of each edge $a_k\in  \Et\cup \Ew$
as the ``color'' of the edge.
To introduce necessary linear constraints 
that can construct such a path $P_k$ properly   in our MILP,
we assign the color $k$ to the vertices $\vT_{j-1,0},\vT_{j,0},\ldots,$ 
$\vT_{j+t,0}$ in $\VT$
when the above path  $P_k$ is used in $G$.
 
For each index $s\in [1,\tC]$, let  
 $\Eew^+(s)$ (resp., $\Eew^-(s)$) denote the set of 
 edges $e\in \Eew$ such that 
the tail (resp., head) of $a_i$ is $\vC_{s,0}$.
Similarly for 
$\Ez^+(s)$,  $\Ez^-(s)$, $\Ew^+(s)$,  $\Ew^-(s)$,
$\Et^+(s)$ and $\Et^-(s)$.

Let $\Iew$ denote the set of indices $i$ of edges $a_i\in \Eew$.
Similarly for 
$\Iz$, $\Iw$, $\It$, 
$\Iew^+(s)$,  $\Iew^-(s)$,
$\Iz^+(s)$,  $\Iz^-(s)$, 
$\Iw^+(s)$,  $\Iw^-(s)$,
$\It^+(s)$ and $\It^-(s)$.
Note that $[1, \kC]=\It\cup \Iw$ and 
$[\widetilde{\kC}+1,\mC]=\Iw\cup \Iz\cup\Iew$.

\bigskip 
\noindent
{\bf constants: } \\ 
 ~~ $\tC=|\VC|$, $\widetilde{\kC}=  |\Et|$, $\kC= |\Et\cup \Ew|$,
      $\tT=\cs_\UB-|\VC|$, $\mC=|\EC|$, \\
~~~~~  where 
      $a_i\in \EC\setminus (\Et\cup \Ew)$, $i\in [\kC+1,\mC]$; \\
       
\noindent
{\bf constants for core specification $\sco$: } \\ 
~~ $\cs_\LB, \cs_\UB \in [2,n^* ]$; lower and upper bounds on $\cs(G)$; \\ 
~~ $\ell_\LB(k), \ell_\UB(k)\in [1, \tT]$, $k\in [1,\kC]$: 
lower and upper bounds on the length of path $P_k$;  \\
       
\noindent
{\bf constants for core specification $\snc$: } \\ 
~~ $\bl_\LB(i),\bl_\UB(i)\in [0,1]$, $i\in [1,\tT]$: 
lower and upper bounds on $\bl_{\rho}(T_i)$ of the tree rooted \\
~~~~~ at a vertex $\vC_{i,0}$; \\

\noindent
{\bf variables: } \\
~~ $\eC(i)\in[0,1]$,  $i\in [1, \mC]$: 
$\eC(i)$ represents edge $a_i\in \EC$, $i\in [1,\mC]$ \\
~~~~~~~ 
 ($\eC(i)=1$, $i\in \Iew$;  $\eC(i)=0$, $i\in \It$)    \\
~~~~~~~  
  ($\eC(i)=1$ $\Leftrightarrow $   edge $a_i$ is  used in  $G$);   \\
~~ $\vT(i,0)\in[0,1]$,   $i\in [1,\tT]$: \\
~~~~~~  $\vT(i,0)=1$ $\Leftrightarrow $ vertex $\vT_{i,0}$ is used in  $G$;   \\
~~ $\eT(i)\in[0,1]$, $i\in [1,\tT+1]$:  $\eT(i)$ represents edge 
$\eT_{i}=(\vT_{i-1,0}, \vT_{i,0})\in \ET$,  \\
~~~~~~ 
where $\eT_{1}$ and $\eT_{\tT+1}$ are fictitious edges
  ($\eT(i)=1$ $\Leftrightarrow $   edge $\eT_{i}$ is  used in  $G$);   \\
~~ $\chiT(i)\in [0,\kC]$, $i\in [1,\tT]$: $\chiT(i)$ represents
 the color assigned to core-vertex $\vT_{i,0}$ \\
~~~~~~ 
  ($\chiT(i)=k>0$
   $\Leftrightarrow $  vertex $\vT_{i,0}$ is  assigned color $k$; \\
   ~~~~~~    
   $\chiT(i)=0$ means that vertex $\vT_{i,0}$ is not used in $G$);   \\
~~ $\clrT(k)\in [\ell_\LB(k)-1, \ell_\UB(k)-1]$, $k\in [1,\kC]$
$\clrT(0)\in [0, \tT]$: the number of vertices 
$\vT_{i,0}\in \VT$ \\
~~~~~~ with color $c$ 
(the range $[\ell_\LB(c)-1, \ell_\UB(c)-1]$
 expresses a constraint in $\sco$);\\
~~ $\dclrT(k)\in [0,1]$,   $k\in [0,\kC]$:
      $\dclrT(k)=1$    $\Leftrightarrow $ $\chiT(i)=k$ 
      for some $i\in [1,\tT]$;\\ 
~~   $\chiT(i,k)\in[0,1]$,  $i\in [1,\tT]$, $k\in [0,\kC]$  
  ($\chiT(i,k)=1$    $\Leftrightarrow $ $\chiT(i)=k$); \\  
~~  $\tldgC^+(i)\in [0,4]$, $i\in [1,\tC]$: 
the out-degree of vertex $\vC_{i,0}$ with the used edges $\eC$ in $\EC$; \\
~~  $\tldgC^-(i)\in [0,4]$, $i\in [1,\tC]$: 
the in-degree of vertex $\vC_{i,0}$  with the used edges $\eC$ in $\EC$; \\

\noindent
{\bf variables for specification $\sco$: } \\
 ~~ $\cs\in [\cs_\LB, \cs_\UB]$: the core size
   (the range $[\cs_\LB, \cs_\UB]$ 
   expresses a constraint in $\sco$);\\
  
\noindent
{\bf constraints: }   
\begin{align} 
  \eC(i)=1,  ~~~  i\in \Iew,       &&  \label{eq:co_first}  \\
  \eC(i)=0,  ~~ \clrT(i)\geq 1,   ~~~  i\in \It,     &&   \label{eq:co_first} \\
  \eC(i)+ \clrT(i)\geq 1,  ~~~~~  \clrT(i)\leq \tT\cdot (1-\eC(i) ), 
~~~  i\in \Iw,    &&    \label{eq:co1c} 
\end{align}   
  
\begin{align}  
\sum_{ c\in \Iw^-(i)\cup \Iz^-(i)\cup \Iew^-(i) }\!\!\!\!\!\! \eC(c) 
 = \tldgC^-(i),  ~~ 
\sum_{ c\in \Iw^+(i)\cup \Iz^+(i)\cup \Iew^+(i) }\!\!\!\!\!\! \eC(c) 
 = \tldgC^+(i),  &&   i\in [1,\tC],   \label{eq:co_5}
\end{align}   

\begin{align} 
\chiT(i,0)=1 -\vT(i,0), ~~~
\sum_{k\in [0,\kC]} \chiT(i,k)=1,  ~~~ 
\sum_{k\in [0,\kC]}k\cdot \chiT(i,k)=\chiT(i),  && i\in[1,\tT],  \label{eq:co2} 
\end{align}   

\begin{align}  
\sum_{i\in[1,\tT]} \chiT(i,k)=\clrT(k), ~~
\tT\cdot \dclrT(k)\geq  \sum_{i\in [1,\tT]} \chiT(i,k)
\geq \dclrT(k), &&  k\in [0,\kC],    \label{eq:co3}   
\end{align}     
 
\begin{align}  
\vT(i-1,0)\geq \vT(i,0), && \notag \\
 \kC\cdot (\vT(i-1,0)-\eT(i )) \geq \chiT(i-1)-\chiT(i )
  \geq \vT(i-1,0) - \eT(i ), && i\in[2,\tT], \label{eq:co6} 
 \end{align}

\begin{align}  
 \tC +\sum_{i\in [1,\tT]} \vT(i, 0) =\cs.  && 
  \label{eq:co_last} 
 \end{align}

  \newpage
\subsection{Constraints for Including Internal Vertices and Edges} 
\label{sec:int}

Define the set of colors for the vertex set 
$\{u_i\mid i\in [1,\widetilde{\tC}] \}\cup \VT$
 to be $[1,\cF]$ with 
\[ \cF \triangleq \widetilde{\tC} + \tT =|\{u_i\mid i\in[1,\widetilde{\tC}]\}\cup \VT|. \]
Let each core-vertex   $\vC_{i,0}$, $i\in[1,\widetilde{\tC}]$ 
(resp., $\vT_{i,0}\in \VT$)
  correspond to 
a color $i\in [1,\cF]$ (resp., $i+\widetilde{\tC} \in [1,\cF]$).
Let  $\tailF(i):=i$ for each color $i\in [1,\widetilde{\tC} ]$  and 
$\tailF(i):=i - \tC$ for each color $i\in [\widetilde{\tC} +1,\cF]$.
When a path $P=(u, \vF_{j,0}, \vF_{j+1,0},\ldots, \vF_{j+t,0})$ 
from a vertex $u\in \VC\cup \VT$ 
  is used in $G$, we assign the color $i\in [1,\cF]$ of the vertex $u$
to the vertices $\vF_{j,0}, \vF_{j+1,0},\ldots, \vF_{j+t,0}$ in $\VF$.

\bigskip 
\noindent
{\bf constants: } \\ 
~~ $\cF$: the maximum number of different colors 
assigned to the vertices in $\VF$; \\
\noindent
{\bf constants for non-core specification $\snc$: } \\ 
~~ $\bl_\LB(i) \in [0,1]$,  $i\in [1, \widetilde{\tC}]$: 
a lower   bound  on the number of leaf ${\rho}$-branches \\
~~~~~~ in the tree rooted at $u_i\in \VC$; \\
~~ $\bl_\LB(k),\bl_\UB(k)\in [0,\ell_\UB(k)-1]$, 
 $k\in[1,\kC]=\It\cup\Iw$: 
lower and upper bounds on the sum \\
~~~~~  of  $\bl_{\rho}(T)$ of the trees rooted at internal vertices of a path $P_k$
 for an edge $a_k\in \Ew\cup \Et$; \\ 

\noindent
{\bf variables: } \\\ 
~~ $\bl_G\in [0, \max[ \max\{\bl_\UB(v)\mid v\in \VC\},
  \max\{\bl_\UB(e)\mid e\in \EC\} ] $:     $\bl_{\rho}(G)$; \\
~~ $\vF(i,0)\in[0,1]$,   $i\in [1,\tF]$: \\
~~~~~~  $\vF(i,0)=1$ $\Leftrightarrow $ vertex $\vF_{i,0}$ is used in  $G$;   \\
~~ $\eF(i)\in[0,1]$, $i\in [1,\tF+1]$:  $\eF(i)$ represents edge 
$\eF_{i}=\vF_{i-1,0} \vF_{t,0}$,  \\
~~~~~~ 
where $\eF_{1}$ and $\eF_{\tF+1}$ are fictitious edges
  ($\eF(i)=1$ $\Leftrightarrow $   edge $\eF_{i}$ is  used in  $G$);   \\
~~ $\chiF(i)\in [0,\cF]$, $i\in [1,\tF]$: $\chiF(i)$ represents
 the color assigned to non-core-vertex $\vF_{i,0}$ \\
~~~~~~ 
  ($\chiF(i)=c$ $\Leftrightarrow $  vertex $\vF_{i,0}$ is  assigned color $c$);   \\
~~ $\clrF(c)\in [0, \tF]$, $c\in [0,\cF]$: the number of vertices $\vF_{i,0}$
 with color $c$;\\ 
~~ $\dclrF(c)\in [\bl_\LB(c), 1]$,  $c\in [1, \widetilde{\tC}]$:
      $\dclrF(c)=1$    $\Leftrightarrow $ $\chiF(i)=c$ for some $i\in [1,\tF]$; \\
~~ $\dclrF(c)\in[0,1]$,  $c\in [\widetilde{\tC}+1,\cF]$:
      $\dclrF(c)=1$    $\Leftrightarrow $ $\chiF(i)=c$ for some $i\in [1,\tF]$; \\
~~   $\chiF(i,c)\in[0,1]$,
 $i\in [1,\tF]$, $c\in [0,\cF]$:  
   $\chiF(i,c)=1$    $\Leftrightarrow $ $\chiF(i)=c$; \\  
~~ $\sigma(c)\in[0,1]$, $c\in [0,\cF]$: 
    $\sigma(c)=1$ $\Leftrightarrow$ the ${\rho}$-pendent-tree rooted at 
    vertex $\vC_{c,0}$ $(c\leq \widetilde{\tC})$ \\
~~~~~~  or $\vT_{c-\widetilde{\tC}, 0}$ $(c>\widetilde{\tC})$ has the core height $\ch_G$.\\
~~ $\sX(i)\in[0,1]$, $i\in[1,\tX]$, $\mathrm{X}\in\{\mathrm{C,T,F}\}$:
    $\sX(i)=1$ $\Leftrightarrow$ the subtree of $X_i$ rooted  \\
~~~~~~  at vertex $\vX_{i,0}$
     has the core height $\ch_G$.\\  

\noindent
{\bf variables for chemical specification $\snc$: } \\
~~ $\bl(k,i)\in [0,1]$, $k\in[1,\kC]= \It\cup\Iw$,  $i\in[1,\tT]$: 
    $\bl(k,i)=1$ $\Leftrightarrow$ path $P_k$ contains vertex $\vT_{i,0}$ \\
~~~~~~ as an internal vertex
    and the ${\rho}$-fringe-tree rooted at $\vT_{i,0}$ contains a leaf ${\rho}$-branch;\\

\noindent
{\bf constraints: }   
\begin{align} 
\chiF(i,0)=1 -\vF(i,0), ~~~
\sum_{c\in [0,\cF]} \chiF(i,c)=1,  ~~~ 
\sum_{c\in [0,\cF]}c\cdot \chiF(i,c)=\chiF(i),  &&  i\in[1,\tF],  \label{eq:int_first} 
\end{align}

\begin{align}  
\sum_{i\in[1,\tF]} \chiF(i,c)=\clrF(c), ~~~ \tF\cdot \dclrF(c)\geq
\sum_{i\in [1,\tF]} \chiF(i,c)\geq \dclrF(c), &&  c\in [0,\cF],    \label{eq:int3}   
\end{align}   
 
\begin{align}  
 \eF(1)=\eF(\tF+1)=0,  && \label{eq:int4} 
 \end{align}   
 
\begin{align}  
\vF(i-1,0)\geq \vF(i,0), && \notag \\
 \cF\cdot (\vF(i-1,0)-\eF(i)) \geq \chiF(i-1)-\chiF(i) 
 \geq \vF(i-1,0)- \eF(i), && i\in[2,\tF], \label{eq:int6} 
 \end{align}

\begin{align}  
\sum_{c\in [1,\cF]} \dclrF(c) =\bl_G,   ~~ 
    \sum_{c\in [1,\cF]}\sigma(c)
    +     \sum_{ i\in[1,\tX], \mathrm{X}\in\{\mathrm{C,T,F}\}} \sX(i) =1, &&    \label{eq:int9} 
 \end{align}


\begin{align}  
 \bl(k,i)\geq  \dclrF(\widetilde{\tC} + i)+\chiT(i,k)-1 , ~~~
 ~~~~~   k \in[1,\kC],   i\in[1,\tT], &&    
 \end{align}   
 
\begin{align}  
 \sum_{k \in[1,\kC],  i\in[1,\tT]} \bl(k,i)
 \leq \sum_{i\in[1,\tT]}\dclrF( \widetilde{\tC} +i),   &&    
  \label{eq:int12} 
 \end{align}   
  
  \begin{align}  
 \bl_\LB(k)\leq  \sum_{ i\in[1,\tT]} \bl(k,i) \leq  \bl_\UB(k) , ~~~~~~
     k \in[1,\kC]. &&    
       \label{eq:int_last} 
 \end{align}

  \newpage
\subsection{Constraints for Including Fringe-trees} 
\label{sec:ex}

We set $\dmax=3$ if $\dg^\nc_{4,\UB}=0$,
and $\dmax=4$ if $\dg^\nc_{4,\UB}\geq 1$.

\begin{figure}[h!] \begin{center}
\includegraphics[width=.75\columnwidth]{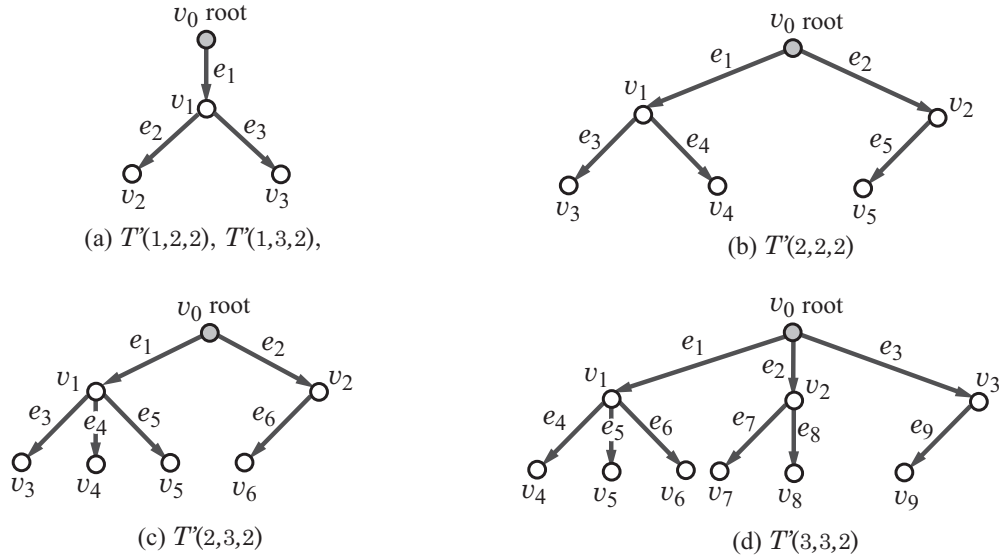}
\end{center} \caption{An illustration of reduced trees $T'$ of rooted trees $T(a,b,c)$:
(a) A reduced tree for $T(1,2,2)=T(\dmax-2,\dmax-1,{\rho})$ 
with $\dmax=3$ and ${\rho}=2$ and for $T(1,3,2)=T(\dmax-3,\dmax-1,{\rho})$ with $\dmax=4$ and ${\rho}=2$; 
(b) A reduced tree for 
$T(2,2,2)=T(\dmax-1,\dmax-1,{\rho})$ with $\dmax=3$ and ${\rho}=2$; 
(c)  A reduced tree for  
$T(2,3,2)=T(\dmax-2,\dmax-1,{\rho})$ with $\dmax=4$ and ${\rho}=2$; 
(d)  A reduced tree for 
$T(3,3,2)=T(\dmax-1,\dmax-1,{\rho})$ with $\dmax=4$ and ${\rho}=2$. }
\label{fig:regular-tree_reduced} \end{figure}

Since we use the proper sets  $\Pprc(1,2,2)$, $\Pprc(2,2,2)$, $\Pprc(2,3,2)$
and $\Pprc(3,3,2)$ of  ordered index pairs   in Section~\ref{sec:preliminary},
we observe that some vertices in the trees  $T(1,2,2)$, $T(2,2,2)$, 
$T(1,3,2)$, $T(2,3,2)$
and $T(3,3,2)$  will not be used in any choice of subtrees.
In this MILP formulation, we use the reduce trees $T'$ illustrated in
Figure~\ref{fig:regular-tree_reduced}(a)-(d).

We introduce some notations of rooted trees in the scheme graph $\mathrm{SG}$.\\
~~  For each rooted  tree $X_i$ such that $C_i=T(\Delta_i,\dmax-1,{\rho})$, 
   $T_i=T(\dmax-2,\dmax-1,{\rho})$ and    $F_i=T(\dmax-1,\dmax-1,{\rho})$, 
     $\mathrm{X}\in\{\mathrm{C,T,F}\}$,\\ 
~~~~~~  $\CldX(j)$ denotes  the set  of the indices  $h$ 
  of children $\vX_{i,h}$ of a vertex $\vX_{i,j}$;\\
~~~~~~  $\prtX(j)$ denotes the index $h$ of  the parent  $\vX_{i,h}$ 
    of a non-root vertex $\vX_{i,j}$;\\
~~~~~~ $\DsnX(p)$ denotes the set  of indices $j$ of a vertex $\vX_{i,j}$ whose
depth is $p$;  \\ 
~~~~~~ $\PprcX$ is a proper set 
$\Pprc(d_\mathrm{X},\dmax-1,{\rho})$ of index pairs, \\ 
~~~~~~ 
where $d_\mathrm{C}=\Delta_i$, 
  $d_\mathrm{T}=2$ and $d_\mathrm{F}=\dmax-1$; \\
~~~~~~ $\jX_p$ denotes the index $j\in \DsnX(p)$ of the vertex $\vX_{i,j}$
  with depth $p$  in the leftmost path\\
~~~~~~~~~~~ from the root  (where assume that the height of a rooted tree \\
~~~~~~~~~~~  satisfying $\PprcX$ is 
    given by the leftmost path  from the root);  \\

We assume that every rooted tree 
$T\in \Tprc(\dmax-2,\dmax-1,{\rho})$ 
(resp., $T\in \Tprc(\dmax-1,\dmax-1,{\rho})$) satisfies
a property that the leftmost path (or the path that visits 
children with the smallest index) from the root is 
alway is the height $\h(T)$.

To express the condition that
a ${\rho}$-fringe-tree is chosen from a rooted tree $C_i$, $T_i$  or  $F_i$, 
we introduce the following set of variables and constraints.  
  
\bigskip
\noindent 
{\bf constants for non-core specification $\snc$: } \\ 
~~ $n_\LB, n^*\geq \cs_\LB$: lower and upper bounds on $n(G)$; \\
~~ $\ch_{\LB}(i),\ch_{\UB}(i)\in [0,n^* ]$, $i\in [1,\tT]$: 
lower and upper bounds on $\h(T_i)$ of the tree rooted \\
~~~~~ at a vertex $\vC_{i,0}$; \\
~~ $\ch_{\LB}(k),\ch_{\UB}(k)\in [0,n^* ]$, $k \in[1,\kC]= \It\cup\Iw$: 
lower and upper bounds on the maximum  \\
~~~~~  height $\h(T)$ of the tree $T\in \F(P_k)$ of rooted at 
an internal vertex of a path $P_k$  \\
~~~~~  for an edge $a_k\in \Ew\cup \Et$; \\

\noindent
{\bf variables: } \\ 
~~ $n_G\in [n_\LB, n^*]$: $n(G)$;\\
~~ $\ch_G\in [0, \max[ \max\{\ch_\UB(v)\mid v\in \VC\},
                                    \max\{\ch_\UB(e)\mid e\in \EC\} ] $:     $\ch(G)$; \\
~~ $\vT(i,j)\in[0,1]$,   $i\in [1,\tT]$, $j\in [1,\nT]$: \\
~~~~~~  $\vT(i,j)=1$ $\Leftrightarrow $ vertex $\vT_{i,j}$ is used in  $G$;   \\
~~~~~~ $\vT(i,j)=1$ and $j\geq 1$
 $\Leftrightarrow $ edge $\eT_{i,j}$ is used in  $G$;   \\
~~ $\vX(i,j)\in[0,1]$,   $i\in [1,\tC]$, $j\in [0,\nC(i)]$
 (resp.,  $i\in [1,\tF]$, $j\in [0,\nF]$), 
  $\mathrm{X=C}$ (resp., $\mathrm{X=F}$): \\
~~~~~~  $\vX(i,j)=1$ $\Leftrightarrow $ vertex $\vX_{i,j}$ is used in  $G$;   \\
~~~~~~ $\vX(i,j)=1$ and $j\geq 1$
 $\Leftrightarrow $ edge $\eX_{i,j}$ is used in  $G$;   \\
 ~~ $\hX(i)\in [0,{\rho}]$, $i\in [1,\tX]$,
$\mathrm{X}\in \{\mathrm{C,T,F}\}$: the height of tree $X_i$; \\

\noindent
{\bf variables for chemical specification $\snc$: } \\
~~ $\sigma(k,i)\in[0,1]$, $k \in[1,\kC]=\It\cup\Iw,  i\in [1,\tT]$: 
    $\sigma(k,i)=1$ $\Leftrightarrow$ a tree rooted at  a vertex $\vT_{i,0}$  \\
~~~~~~  with color $k$  has the largest height among such trees;\\

\bigskip
\noindent
{\bf constraints: } 
\begin{align}    
\vC(i,0)=1, &&  i\in [1,\tC],   \label{eq:ex_first}  
\end{align}

\begin{align}    
   \vF(i,\jF_{{\rho}}) \geq \vF(i,0) - \eF(i+1),
      && i\in [1,\tF]~(\eF(\tF+1)=0), \label{eq:ex3}  
\end{align}   
 
\begin{align}   
\vX(i,j)\geq \vX(i,h), && i\in [1,\tX],  (j,h)\in \PprcX,  
\mathrm{X}\in \{\mathrm{C,T,F}\},  
\label{eq:ex4}   
\end{align}   
\begin{align}    
 \sum_{p\in [1,{\rho}]} \vX(i,\jX_p) = \hX(i) ,
 &&   i\in[1,\tX], \mathrm{X}\in \{\mathrm{C,T,F}\}, \label{eq:ex5}  
\end{align}   

\begin{align}    
\sum_{j\in [0,\nC(i)]}\vX(i,j)
\leq 2+ 2 \sum_{h\in \CldC(0)}\vC(i,h),  && i\in [1,\tC], 
  \label{eq:ex6}  
\end{align}   

\begin{align}    
\sum_{j\in [0,\nX]}\vX(i,j)
\leq 2+ 2 \sum_{h\in \CldX(0)}\vX(i,h),  && i\in [1,\tX], 
\mathrm{X}\in \{\mathrm{T,F}\},   \label{eq:ex6b}  
\end{align}   
    
\begin{align}  
\ch_G\geq  \clrF(c) +{\rho} \geq \ch_G -n^*\cdot(1-\sigma(c)), 
  && c\in [1,\cF],   \label{eq:int10} 
 \end{align}   
    
\begin{align}  
\ch_G\geq  \hX(i)  \geq \ch_G -n^*\cdot(1-\sX(i)),   &&
 i\in [1,\tX],  \mathrm{X}\in\{\mathrm{C,T,F}\} \label{eq:int10} 
 \end{align}   

\begin{align}    
   \sum_{  i\in [1,\tC], j\in [0,\nC(i)]}   \vC(i,j) 
  +    \sum \limits_{\substack{  i\in [1,\tX], j\in [0,\nX], \\
               \mathrm{X}\in \{\mathrm{T,F}\}  }}   \vX(i,j) = n_G, ~~
 && 
  \label{eq:ex2} 
\end{align}   

\begin{align}  
\hC(i)    \geq \ch_\LB(i)- n^*  \dclrF(i),  ~~
\clrF(i)+{\rho} \geq \ch_\LB(i) , ~~~~~~~~~~~~~~   &&\notag \\
\hC(i)          \leq \ch_\UB(i) ,  ~~
\clrF(i)+{\rho} \leq \ch_\UB(i)+ n^*  (1-\dclrF(i)),  ~ 
           i\in [1,\widetilde{\tC}],    &&  
            \label{eq:int14} 
 \end{align}   
 
\begin{align}  
 \ch_\LB(i) \leq  \hC(i)   \leq  \ch_\UB(i) ,  ~~ 
           i\in [\widetilde{\tC}+1,\tC],    &&  
             \label{eq:int14} 
 \end{align}   
 
\begin{align}   
 \hT(i)    \leq \ch_\UB(k)+ n^*  (\dclrF( \widetilde{\tC}+ i)+1-\chiT(i,k)),  &&\notag \\
\clrF(\widetilde{\tC}+i)+{\rho}
 \leq \ch_\UB(k)+ n^*  (2-\dclrF( \widetilde{\tC}+ i)-\chiT(i,k)),   && \notag\\
    k \in[1,\kC],  i\in [1,\tT],  &&   
    \label{eq:int15} 
 \end{align}   
 
\begin{align}   
 \sum_{i\in[1,\tT]}\sigma(k,i) =\dclrT(k),  ~~~~    k \in[1,\kC],   &&  
 \label{eq:int16} 
 \end{align}   
 
\begin{align}  
 \chiT(i,k)\geq \sigma(k,i), && \notag\\
 \hT(i)    \geq \ch_\LB(k) - n^*  (\dclrF( \widetilde{\tC}+ i)+1-\sigma(k,i) ),
  && \notag\\ 
\clrF(\widetilde{\tC}+i)+{\rho}
 \geq \ch_\LB(k) - n^*  (2-\dclrF( \widetilde{\tC}+ i)-\sigma(k,i)),  && 
   k \in[1,\kC],  i\in [1,\tT]. ~~~~~  
    \label{eq:ex_last} 
 \end{align}

  \newpage
\subsection{Descriptor for the  Number of Specified Degree} 
\label{sec:Deg}

We include constraints to compute descriptors $\dg_i^\co(G)$
and $\dg_i^\nc(G)$, $i\in[1,4]$. \\

\bigskip
\noindent
{\bf constants for non-core specification $\snc$: } \\ 
~~ $\dg^\nc_{4,\UB}\in [0, n^* - \cs_\LB]$: 
   an upper bound on $\dg_4^\nc(G)$. \\

\noindent
{\bf variables: } \\
~~    $\degX(i,j)\in [0,4]$,  $i\in [1,\tX]$, $j\in [0,\nX]$
 ($\nC\!=\! \nC(i)$),  
 $\mathrm{X}\in \{\mathrm{C,T,F}\}$: \\
~~~~~~ the degree $\mathrm{deg}_G(\vX_{i,j})$ of vertex $\vX_{i,j}$ in $G$;\\
~~ $\degCT(i)\in [0,4]$,  $i\in [1, \tC]$: the number of edges
from vertex $\vC_{i,0}$ to vertices $\vT_{j,0}$, $j\in [1,\tT]$; \\  
~~ $\degTC(i)\in [0,4]$,  $i\in [1, \tC]$: the number of edges
from  vertices $\vT_{j,0}$, $j\in [1,\tT]$ to vertex $\vC_{i,0}$; \\  
~~ $\degCex(i) \in [0,3]$, $i\in [1, \tC]$: the number of children
 of vertex $\vC_{i,0}$ in the ${\rho}$-fringe-tree $C_i$;\\
 ~~    $\ddgX(i,j,d)\in[0,1]$,  $i\in [1,\tX]$, $j\in [0,\nX]$ 
 ($\nC\!=\! \nC(i)$),   
 $d\in [1,4]$,  $\mathrm{X}\in \{\mathrm{C,T,F}\}$: \\
~~~~~~       $\ddgX(i,j,d)=1$ $\Leftrightarrow$   $\degX(i,j)=d$; \\
~~   $\dg(d)\in[\dg_\LB(d),\dg_\UB(d)]$,  $d \in[1,4]$:
    the number  of vertices $v$ with $\deg_G(v)=d$; \\ 
~~  $\dg^{\co}(d),\dgC(d),\dgT(\dg) \in[0,\cs_\UB]$,
 $d \in[1,4]$:     the number of core-vertices \\
~~~~~~~  $v\in V(G)$ 
    (resp., $v\in V(G)\cap \VC$ and $v\in V(G)\cap \VT$)
    with $\deg_G(v)=d$; \\ 
 ~~  $\dg^{\nc}(d),\dg^\inn(d),\dg^\ex(d) \in [0,n^* -\cs_\LB]$,$d \in[1,4]$:
   the number of non-core-vertices $v\in V(G)$ \\
   ~~~~~~~    (resp., ${\rho}$-internal vertices $v\in V(G)\cap \VF$ 
   and  ${\rho}$-external-vertices $v\in V(G)$)
    with $\deg_G(v)=d$; \\ 
~~ $\dgXp(d)  \in [0,n^* -\cs_\LB]$,
    $d \in[1,4],p\in[1,{\rho}]$,   $\mathrm{X}\in \{\mathrm{C,T,F}\}$: \\
~~~~~~~     the number of  ${\rho}$-external-vertices $v\in V(G)\cap V(X_i)$ 
    with depth $p\in[1,{\rho}]$ and  $\deg_G(v)=d$; \\

\noindent
{\bf constraints: } 
\begin{align}   
 \vX(i,j) +\sum_{h\in \CldX(j)}\vX(i,h) = \degX(i,j), &&  
 i\in [1,\tX], j\in [1,\nX] ~(\nC\!=\! \nC(i)),   
 \mathrm{X}\in \{\mathrm{C,T,F}\}, \label{eq:Deg_first}  
\end{align}   

\begin{align}   
\sum_{   k\in \It^+(i)\cup \Iw^+(i)} \dclrT(k) = \degCT(i), ~~
 \sum_{   k\in \It^-(i)\cup \Iw^-(i)} \dclrT(k) = \degTC(i), &&    \notag \\
   \sum_{h\in \CldC(0)}\vC(i,h) = \degCex(i),  
    &&    i\in [1, \tC],     \label{eq:Deg1}  
\end{align}   

\begin{align}   
\tldgC^-(i)+\tldgC^+(i)   + \degCT(i)  + \degTC(i) 
  + \dclrF(i) + \degCex(i)      = \degC(i,0),  
    &&    i\in [1, \widetilde{\tC}],     \label{eq:Deg2}  
\end{align}   

\begin{align}      
\tldgC^-(i)+\tldgC^+(i)  + \degCT(i)  + \degTC(i)   
    + \degCex(i)  = \degC(i,0),  
     &&   i\in [\widetilde{\tC}+1,\tC],     \label{eq:Deg2b}  
\end{align}   

\begin{align}   
  2\vT(i,0)   + \dclrF(\widetilde{\tC}+i) 
  + \sum_{h\in \CldT(0)}\vT(i,h)
 =\degT(i,0),   \hspace{2cm}  \notag \\
  i\in [1,\tT]~(\eT(1)=\eT(\tT+1)=0), \label{eq:Deg3}  
\end{align}

\begin{align}
   \vF(i,0) +\eF(i+1)
  +\sum_{h\in \CldF(0)}\vF(i,h) = \degF(i,0),  \hspace{2cm}  && \notag \\
  i\in [1,\tF] ~(\eF(1)=\eF(\tF+1)=0), && \label{eq:Deg4}  
\end{align} 

\begin{align}   
\sum_{d\in [0,4]}\ddgX(i,j,d)=1, ~
\sum_{d\in [1,4]}d\cdot\ddgX(i,j,d)=\degX(i,j), \hspace{2cm} \notag \\
 i\in [1,\tX], j\in [0,\nX] ~(\nC\!=\! \nC(i)),      
 \mathrm{X}\in \{\mathrm{T, C, F}\}, \label{eq:Deg5}  
\end{align}   

\begin{align}   
\sum_{ i\in [1,\tC],  j\in \DsnX(p)} \ddgX(i,j,d)   
   = \dgXp(d), && d \in[1,4], p\in[1,{\rho}], \mathrm{X}\in \{\mathrm{C,T,F}\},  \label{eq:Deg7}  
\end{align} 

\begin{align}   
\sum_{ i\in [1,\tC]} \ddgC(i,0,d)=\dgC(d),  ~
\sum_{ i\in [1,\tT]} \ddgT(i,0,d)=\dgT(d),  ~
 \sum_{ i\in [1,\tF] }  \ddgF(i,0,d) =\dg^\inn(d),   && \notag\\   
   \dg^\inn(d)+ \sum_{p\in[1,{\rho}], \mathrm{X}\in \{\mathrm{C,T,F}\}}
    \dgXp(d)= \dg^\nc(d), ~
\dgC(d)+\dgT(d)= \dg^\co(d),     
    && d\in [1,4],  \label{eq:Deg7}  
\end{align} 

\begin{align}  
  \dg^\nc(4)  \leq \dg^\nc_{4,\UB}. && 
  \label{eq:Deg_last}  
\end{align}

  \newpage
\subsection{Assigning Multiplicity} 
\label{sec:beta}

 We prepare an integer variable $\beta(e)$  
 for each edge $e$ in the scheme graph $\mathrm{SG}$ 
 to denote the bond-multiplicity of $e$ in a selected graph $G$ and
 include necessary constraints for the variables to satisfy in $G$. 
 \medskip

\noindent
{\bf variables: } \\
 ~~ $\bX(i)\in [0,3]$,   $i\in [2,\tX]$, $\mathrm{X}\in \{\mathrm{T,F}\}$:   
 the bond-multiplicity of edge  $\eX_{i}$;   \\ 
~~ $\bC(i)\in [0,3]$,     $i\in [\widetilde{\kC}+1,\mC]= \Iw\cup \Iz\cup\Iew$:    
     the bond-multiplicity of  \\
~~~~~   edge  $a_{i}\in \Ew\cup \Ez\cup\Eew$;   \\  
~~ $\bX(i,j)\in [0,3]$,  $i\in [1,\tX]$, $j\in [1,\nX]$
 ($\nC\!=\! \nC(i)$), 
  $\mathrm{X}\in \{\mathrm{C,T,F}\}$: \\
~~~~~ the bond-multiplicity of edge   $\eX_{i,j}=(\vX_{i,\prtX(j)},\vX_{i,j})$;   \\
~~
   $\beta^{+}(k), \beta^{-}(k)\in [0,3]$, $k\in [1, \kC]=\It\cup \Iw$: 
   the bond-multiplicity of the first  \\
~~~~~  (resp., last) edge of path $P_k$; \\  
~~
   $\beta^\inn(c)\in [0,3]$, $c\in [1,\cF]$: 
   the bond-multiplicity of the first edge of ${\rho}$-branch-subtree $T_c$\\
~~~~~  
   rooted at vertex $c$; \\  
~~  $\delbX(i,m)\in [0,1]$, $i\in [2,\tX]$,   $m\in[0,3]$, 
       $\mathrm{X}\in \{\mathrm{T,F}\}$:  
  $\delbX(i,m)=1$  $\Leftrightarrow$  $\bX(i)=m$; \\  
~~  $\delbC(i,m)\in [0,1]$,  
   $i\in [\widetilde{\kC},\mC]=\Iw\cup \Iz\cup\Iew$,  $m\in[0,3]$:  \\
~~~~~~   $\delbC(i,m)=1$  $\Leftrightarrow$  $\bC(i)=m$; \\  
~~ $\delbX(i,j,m)\in [0,1]$,  $i\in [1,\tX]$, $j\in [1,\nX]$
($\nC\!=\! \nC(i)$),  $m\in[0,3]$, 
$\mathrm{X}\in \{\mathrm{C,T,F}\}$:  \\
~~~~~~  $\delbX(i,j,m)=1$  $\Leftrightarrow$  $\bX(i,j)=m$; \\  
 ~~  
   $\delb^{+}(k,m), \delb^{-}(k,m)\in [0,3]$, $k\in [1, \kC]=\It\cup \Iw$,  $m\in[0,3]$:
   \\
~~~~~
     $\delb^+(k,m)=1$   (resp., $\delb^-(k,m)=1$)    $\Leftrightarrow$  
           $\beta^+(k)=m$ (resp., $\beta^-(k)=m$); \\  
~~
   $\delb^\inn(c,m)\in [0,3]$, $c\in [1,\cF]$,  $m\in[0,3]$: 
     $\delb^\inn(c,m)=1$ $\Leftrightarrow$  $\beta^\inn(c)=m$; \\  
~~\\
~~ $\bd(m)\in[0, \mUB]$, $m\in[1,3]$:
      the number of   edges with bond-multiplicity  $m$ in $G$; \\
~~ $\bd^{\co}(m)\in[0, \mUB^\co]$, $m\in[1,3]$:
     the number of  core-edges with bond-multiplicity  $m$ in $G$; \\
~~ $\bd^{\inn}(m)\in[0, \mUB^\nc]$, $m\in[1,3]$:
     the number of  ${\rho}$-internal edges with bond-multiplicity  $m$\\
~~~~~~ in $G$; \\
~~ $\bd^{\ex}(m)\in[0, n^*]$, $m\in[1,3]$:
  the number of  ${\rho}$-external edges  with bond-multiplicity  $m$ \\
~~~~~~   in $G$; \\
~~ $\bdX(m)\in [0,\mUB^\co],  \mathrm{X}\in \{\mathrm{C,T,CT,TC,}\}$,
      $\bdX(m)\in [0,\mUB^\nc], \mathrm{X}\in \{\mathrm{F,CF,TF}\}$, $m\in[1,3]$:  
    \\
~~~~~~  the number of  edges $\eX(i,0)$ with bond-multiplicity  $m$ in $G$; \\ 
~~ $\bdXp(m)\in [0,n^*], m\in[1,3], \mathrm{X}\in \{\mathrm{C,T,F}\}$:  
     the number of  edges $\eX(i,j)$ with $j\in\DsnX(p)$ \\
~~~~~~  and  bond-multiplicity  $m$ in $G$; \\ 

\noindent
{\bf constraints: } 
\begin{align}    
\eC(i)\leq \bC(i)\leq 3\eC(i), 
  i\in [\widetilde{\kC}+1,\mC]=\Iw\cup \Iz\cup\Iew, \label{eq:beta_first} 
\end{align}   

\begin{align}   
  \eX(i)\leq \bX(i)\leq 3 \eX(i), 
  &&    i\in [2,\tX],   \mathrm{X}\in \{\mathrm{T, F}\},    \label{eq:beta1}  
\end{align}   

\begin{align}   
  \vX(i,j)\leq \bX(i,j)\leq 3 \vX(i,j),   &&    i\in [1,\tX], j\in [1,\nX]
  ~(\nC\!=\! \nC(i)),
   \mathrm{X}\in \{\mathrm{C,T,F}\} ,     \label{eq:beta6}  
\end{align}    

\begin{align}   
\dclrT(k)\leq \beta^+(k)\leq 3 \dclrT(k), ~~~ 
\dclrT(k)\leq \beta^-(k)\leq 3 \dclrT(k), &&  k\in [1, \kC], \label{eq:beta8} \\ 
\dclrF(c)\leq \beta^\inn(c)\leq 3 \dclrF(c), &&   c\in [1,\cF], \label{eq:beta8}  
\end{align}

\begin{align} 
\sum_{m\in[0,3]} \delbX(i,m)=1,  ~~
\sum_{m\in[0,3]}m\cdot \delbX(i,m)=\bX(i), &&   i\in [2,\tX],  
  \mathrm{X}\in \{\mathrm{T,F}\},  \label{eq:beta10}    
\end{align}   

\begin{align} 
\sum_{m\in[0,3]} \delbC(i,m)=1,  ~~
\sum_{m\in[0,3]}m\cdot \delbC(i,m)=\bC(i), &&   i\in [\widetilde{\kC}+1,\mC],   \label{eq:beta11}    
\end{align}   

\begin{align}   
 \sum_{m\in[0,3]} \delbX(i,j,m)=1,   ~ 
 \sum_{m\in[0,3]} m\cdot \delbX(i,j,m)=\bX(i,j),   \hspace{2cm} \notag \\  
  i\in [1,\tX], j\in [1,\nX] (\nC\!=\! \nC(i)),  
   \mathrm{X}\in \{\mathrm{C,T,F}\} , 
 \label{eq:beta12}    
\end{align}   
 
\begin{align}   
\sum_{m\in[0,3]} \delb^+(k,m)=1,    ~~ 
\sum_{m\in[0,3]}m\cdot\delb^+(k,m)=\beta^+(k),
&&     k\in [1, \kC],  \notag \\      
\sum_{m\in[0,3]} \delb^-(k,m)=1,  ~~ 
\sum_{m\in[0,3]} m\cdot\delb^-(k,m)=\beta^-(k),  &&
      k\in [1, \kC], \notag \\
\sum_{m\in[0,3]} \delb^\inn(c,m)=1,  ~~ 
\sum_{m\in[0,3]} m\cdot\delb^\inn(c,m)=\beta^\inn(c),  &&    c\in [1,\cF],
   \label{eq:beta15}    
\end{align}   

\begin{align}   
\sum_{i\in [1,\tX], j\in\DsnX(p) } \delbX(i,j,m) = \bdXp(m), 
 &&  p\in[1,{\rho}],   \mathrm{X}\in \{\mathrm{C,T,F}\}, m\in [1,3],
     \label{eq:beta16} 
\end{align}   

\begin{align} 
 \sum_{i\in [\widetilde{\kC}+1,\mC]} \delbC(i,m) =\bdC(m), ~~
  \sum_{i\in [2,\tT]} \delbT(i,m)    =\bdT(m), ~~  && \notag \\
   \sum_{k\in [1, \kC]}\delb^+(k,m)=\bdCT(m), ~~
   \sum_{k\in [1, \kC]}\delb^-(k,m)=\bdTC(m), ~~  && \notag \\
 \bdC(m)+\bdT(m)+\bdCT(m)+\bdTC(m) = \bd^\co(m),  ~~ &&  \notag \\ 
\sum_{i\in [2,\tF]}\!\!\! \delbF(i,m) =\bdF(m), ~~
 \sum_{c\in [1,\widetilde{\tC}]} \delb^\inn(c,m)  =\bdCF(m), 
 ~~ &&  \notag \\ 
  \sum_{c\in [\widetilde{\tC}+1,\cF]}  \delb^\inn(c,m) =\bdTF(m), 
  \bdF(m)+\bdTF(m)+\bdCF(m) = \bd^\inn(m), ~~   &&  \notag \\ 
 \sum_{  p\in[1,{\rho}],   \mathrm{X}\in \{\mathrm{C,T,F}\}}
    \bdXp(m) = \bd^\ex(m) ,  ~ \notag&&  \\
\bd^\co(m) + \bd^\inn(m)+ \bd^\ex(m) = \bd(m),  
 && m\in [1,3],       \label{eq:beta_last} 
\end{align}

   \newpage
\subsection{Assigning Chemical Elements and  Valence Condition}
\label{sec:alpha}

We include constraints so that each vertex $v$ in a selected graph $H$
satisfies the valence condition; i.e., $\sum_{uv\in E(H)}\beta(uv)\leq  \val(\alpha(u))$. 
With these constraints, a chemical  ${\rho}$-lean  graph
 $G=(H,\alpha,\beta)$ on a selected subgraph $H$
will be constructed. 

 Let $\IVCs$ denote the set of indices $i$ of 
 vertices $\vC_{i,0}\in \VC^*$
 and $\alpha^*(i)$ denote $\alpha^*(\vC_{i,0})$, $i\in \IVCs$.

 \medskip
\noindent
{\bf constants: } \\ 
~~ A subsets
 $\Lambda^\co, \Lambda^\nc \subseteq \Lambda$ of chemical elements,
 where we denote by $[{\tt e}]$ (resp., $[{\tt e}]^\co$ and $[{\tt e}]^\nc$) \\
~~~~~ 
 of a standard encoding of an element ${\tt e}$ in of set $\Lambda$ 
 (resp.,    $\Lambda^\co_\epsilon$ and  $\Lambda^\nc_\epsilon$); \\
~~ A valence function: $\val: \Lambda \to [1,4]$; \\
~~ A function $\mathrm{mass}^*:\Lambda\to \mathbb{Z}$ 
(we let $\mathrm{mass}(\ta)$ denote  the observed mass of a chemical element \\
~~~~ 
$\ta\in \Lambda$, and define 
   $\mathrm{mass}^*(\ta)\triangleq
    \lfloor 10\cdot \mathrm{mass}(\ta)\rfloor$); \\

\noindent
{\bf constants for chemical specification $\sab$: } \\  
~~ 
 Subsets $\Lambda^*(i)\subseteq \Lambda^\co$, $i\in[1,\tC]$;\\ 
~~ 
 $\na_\LB(\ta),\na_\UB(\ta)\in [0,n^* ]$,  $\ta\in  \Lambda$:
lower and upper bounds on the number of vertices  $v$ \\
~~~~~~  with $\alpha(v)=\ta$; \\
~~
  $\na_\LB^\typ(\ta),\na_\UB^\typ(\ta)\in [0,n^* ]$,
 $\ta\in  \Lambda^\typ$,  $\typ\in \{\co,\nc\}$:
lower and upper bounds on the number \\
~~~~~~ of core-vertices (or non-core-vertices)
 $v$ with $\alpha(v)=\ta$; \\

\noindent
{\bf variables: } \\
~~ 
   $\bCT(i),\bTC(i)\in [0,3], i\in [1,\tT]$:
the bond-multiplicity of edge $\eCT_{j,i}$ (resp., $\eTC_{j,i}$)
if one exists; \\
~~ 
 $\bCF(i), \bTF(i)\in [0,3], i\in [1,\tF]$:
the bond-multiplicity of $\eCF_{j,i}$ (resp., $\eTF_{j,i}$)
if one exists; \\
~~\\
~~ $\aX(i,0)\in [\Lambda^\co_\epsilon ],
       \delaX(i,0,[\ta]^\co)\in [0,1],  \ta\in \Lambda^\co_\epsilon, i\in [1,\tX],
         \mathrm{X}\in \{\mathrm{C,T}\}$ \\  
~~ $\aF(i,j)\in [\Lambda^\nc_\epsilon ], 
      \delaF(i,j,[\ta]^\nc)\in [0,1],  \ta\in \Lambda^\nc_\epsilon,
          i\in [1,\tF], j\in [0,\nF]$: \\
~~ $\aX(i,j)\in [\Lambda^\nc_\epsilon ],
      \delaX(i,j,[\ta]^\nc)\in [0,1],  \ta\in  \Lambda^\nc_\epsilon ,
          i\in [1,\tX], j\in [1,\nX]~(\nC=\nC(i)), 
           \mathrm{X}\in \{\mathrm{C,T, F}\}$:  \\
~~~~~~   
$\aX(i,j)= [\ta]^\typ\geq 1$, $\typ\in\{\co,\nc\}$ (resp., $\aX(i,j)=0$)
  $\Leftrightarrow$ $\delaX(i,j,[\ta]^\typ)=1$ (resp., $\delaX(i,j, 0)=0$) \\
~~~~~~ 
  $\Leftrightarrow$ $\alpha(\vX_{i,j})= \ta\in \Lambda$ 
(resp., vertex $\vX_{i,j}$ is not used in $G$); \\
~~ $\mathrm{Mass}\in \mathbb{Z}_+$: 
 $\sum_{v\in V} \mathrm{mass}^*(\alpha(v))$; \\
~~ $\mathrm{n}_{\tt H}\in [0,4n^* ]$: 
 the number  of hydrogen atoms  to be included to  $G$; \\

\noindent
{\bf variables for  chemical specification $\sab$: } \\ 
~~   $\na([\ta])\in[\na_\LB(\ta),\na_\UB(\ta)]$,
 $\ta \in \Lambda$:
    the number  of vertices $v$ with $\alpha(v)=\ta$; \\ 
~~  $\na^{\co}([\ta]^\co),\naC([\ta]^\co),\naT([\ta]^\co)
\in[\na_\LB^\co(\ta),\na_\UB^\co(\ta)]$,
 $\ta \in \Lambda$:
    the number of core-vertices \\
~~~~~~~  $v\in V(G)$ 
    (resp., $v\in V(G)\cap \VC$ and $v\in V(G)\cap \VT$)
    with $\alpha(v)=\ta$; \\ 
 ~~  $\na^{\nc}([\ta]^\nc),\na^\inn([\ta]^\nc),\naXp([\ta]^\nc) 
 \in [\na_\LB^\nc(\ta),\na_\UB^\nc(\ta)]$,
 $\ta \in \Lambda$,   $\mathrm{X}\in \{\mathrm{C,T,F}\}$: \\
~~~~~~~ 
    the number of non-core-vertices $v\in V(G)$ 
    (resp., ${\rho}$-internal vertices $v\in V(G)\cap \VF$ \\
~~~~~~~  and  ${\rho}$-external-vertices $v\in V(G)\cap V(X_i)$ 
    with depth $p\in[1,{\rho}]$) such that  $\alpha(v)=\ta$; \\ 
    
\noindent
{\bf constraints: } 
\begin{align}    
 \beta^+(k)-3(\eT(i)-\chiT(i,k)+1) \leq 
\bCT(i)\leq \beta^+(k)+3(\eT(i)-\chiT(i,k)+1),  i\in [1,\tT], && \notag\\   
  \beta^-(k)-3(\eT(i+1)-\chiT(i,k)+1) \leq 
\bTC(i)\leq \beta^-(k)+3(\eT(i+1)-\chiT(i,k)+1),  i\in [1,\tT], && \notag\\ 
   k\in [1, \kC],   &&  \label{eq:alpha_first}  
\end{align}

\begin{align}   
 \beta^\inn(c)-3(\eF(i)-\chiF(i,c)+1) \leq 
\bCF(i)\leq \beta^\inn(c)+3(\eF(i)-\chiF(i,c)+1),   i\in [1,\tF], 
&&  c\in[1,\widetilde{\tC}] ,    \notag\\  
  \beta^\inn(c)-3(\eF(i)-\chiF(i,c)+1) \leq 
\bTF(i)\leq \beta^\inn(c)+3(\eF(i)-\chiF(i,c)+1),    i\in [1,\tF], 
&& c\in[\widetilde{\tC}+1,\cF] ,    \notag\\  
  \label{eq:alpha2}  
\end{align}

\begin{align}  
   \sum_{\ta\in \Lambda^\co_\epsilon} \delaX(i,0,[\ta]^\co)=1, ~~
   \sum_{\ta\in \Lambda^\co_\epsilon}
   [\ta]^\co\cdot\delaX(i,0,[\ta]^\co)=\aX(i,0), 
    \hspace{2cm}\notag \\ 
  ~~   i\in [1,\tX],  
  \mathrm{X}\in \{\mathrm{C,T}\},  
  \label{eq:alpha_first} 
\end{align}

\begin{align}  
   \sum_{\ta\in \Lambda^\nc_\epsilon} \delaF(i,0,[\ta]^\nc)=1, ~~
   \sum_{\ta\in \Lambda^\nc_\epsilon}
   [\ta]^\nc\cdot\delaF(i,0,[\ta]^\nc)=\aF(i,0), 
    \hspace{2cm}
  ~~   i\in [1,\tF], 
  \label{eq:alpha1} 
\end{align}

\begin{align}  
   \sum_{\ta\in \Lambda^\nc_\epsilon} \delaX(i,j,[\ta]^\nc)=1, ~~
   \sum_{\ta\in \Lambda^\nc_\epsilon}
   [\ta]^\nc\cdot\delaX(i,j,[\ta]^\nc)=\aX(i,j), 
    \hspace{2cm}\notag \\ 
  ~~   i\in [1,\tX], j\in [1,\nX]~(\nC\!=\! \nC(i)),  
  \mathrm{X}\in \{\mathrm{C,T,F}\},  
  \label{eq:alpha1b} 
\end{align}    


\begin{align}  
\sum_{j\in \EC(i)}\bC(j)  
+ \sum_{  k\in \It^+(i)\cup \Iw^+(i)} \beta^+(k)
+ \sum_{  k\in \It^-(i)\cup \Iw^-(i)} \beta^-(k)   &&  \notag\\
     + \beta^\inn(i)  +\sum_{h\in \CldC(0)}\bC(i,h)
   \leq \sum_{\ta\in \Lambda^\co}\val(\ta)\delaC(i,0,[\ta]^\co),  
 &&  i\in [1,\widetilde{\tC}],  \label{eq:alpha3} 
\end{align}

\begin{align}   
\sum_{j\in \EC(i)}\bC(j)  
+ \sum_{  k\in \It^+(i)\cup \Iw^+(i)} \beta^+(k)
+ \sum_{  k\in \It^-(i)\cup \Iw^-(i)} \beta^-(k)   &&  \notag\\
+\sum_{h\in \CldC(0)}\bC(i,h)
   \leq \sum_{\ta\in \Lambda^\co}\val(\ta)\delaC(i,0,[\ta]^\co),  
  &&  i\in [\widetilde{\tC}+1,\tC],   \label{eq:alpha3b} 
\end{align} 

\begin{align}  
 \bT(i)+\bT(i\!+\!1)   + \!\!\! \sum_{h\in \CldT(0)}\!\!\!  \bT(i,h) 
  + \bCT(i) + \bTC(i) + \beta^\inn(\widetilde{\tC}+i)
 \leq \sum_{\ta\in \Lambda^\co}\val(\ta)\delaT(i,0,[\ta]^\co), 
\hspace{1cm}    \notag \\
  i\in [1,\tT]~  (\bT(1)=\bT(\tT+1)=0),   \label{eq:alpha4} 
\end{align}
 
\begin{align} 
 \bF(i)+\bF(i\!+\!1) +\bCF(i) +\bTF(i)
     +\sum_{h\in \CldF(0)}\bF(i,h)   
     \leq \sum_{\ta\in \Lambda^\nc}\val(\ta)\delaF(j,0,[\ta]^\nc),  
\hspace{1cm}    \notag \\
  i\in [1,\tF] ~  (\bF(1)=\bF(\tF+1)=0),   \label{eq:alpha5} 
\end{align}   
  
\begin{align}   
 \bX(i,j) +\sum_{h\in \CldX(j)}\bX(i,h)
   \leq  \sum_{ \ta\in \Lambda^\nc} \val(\ta)\delaX(i,j,[\ta]^\nc),  
   \hspace{2cm} \notag \\
   i\in [1,\tX], j\in [1,\nX]~(\nC\!=\! \nC(i)),   
     \mathrm{X}\in \{\mathrm{C,T,F}\},
    \label{eq:alpha2}  
\end{align}


\begin{align}  
 \sum_{i\in [1,\tC] } \delaC(i,0,[\ta]^\co) = \naC([\ta]^\co)   , 
 \sum_{i\in [1,\tT] } \delaT(i,0,[\ta]^\co) = \naT([\ta]^\co)   ,  
 &&  \ta\in \Lambda^\co \notag \\ 
 \sum_{i\in [1,\tF] } \delaF(i,0,[\ta]^\nc) = \na^\inn([\ta]^\nc)   , 
 &&      \ta\in \Lambda^\nc,      \label{eq:alpha6} 
\end{align}    

\begin{align}  
     \sum_{i\in [1,\tX], j\in \DsnX(p)]} \delaX(i,j,[\ta]^\nc) = \naXp([\ta]^\nc),   
     && 
 p\in[1,{\rho}],  \mathrm{X}\in \{\mathrm{C, T,F}\},   \ta\in \Lambda^\nc,      \label{eq:alpha6} 
\end{align}    
       
\begin{align}  
       \naC([\ta]^\co)+ \naT([\ta]^\co)=  \na^\co([\ta]^\co),  
    &&   \ta\in \Lambda^\co,       \notag \\
  \sum_{ p\in[1,{\rho}],  \mathrm{X}\in \{\mathrm{C, T,F}\} }
   \naXp([\ta]^\nc)       =\na^\ex([\ta]^\nc), ~~
  \na^\inn([\ta]^\nc)  +  \na^\ex([\ta]^\nc) = \na^\nc([\ta]^\nc),  
  &&  \ta\in \Lambda^\nc,    \notag \\
  \na^\co([\ta]^\co) + \na^\nc([\ta]^\nc)=\na([\ta]),  
 &&   \ta\in \Lambda^\co\cap \Lambda^\nc,     \notag \\
  \na^\co([\ta]^\co)  =\na([\ta]),  
 &&      \ta\in \Lambda^\co \setminus \Lambda^\nc,     \notag \\ 
   \na^\nc([\ta]^\nc)  =\na([\ta]),  
 &&      \ta\in \Lambda^\nc \setminus \Lambda^\co,        
     \label{eq:alpha6} 
\end{align}

\begin{align}   
\sum_{ \ta\in\Lambda }\mathrm{mass}^*(\ta )\cdot \na([\ta])
 =\mathrm{Mass}, &&    \label{eq:alpha7} 
\end{align}  

\begin{align}   
\sum_{\ta\in\Lambda}\val(\ta)\cdot  \na([\ta])
   - 2\sum_{m\in[1,3], \typ\in\{\co,\inn,\ex\} }m\cdot \bd^\typ(m) 
    =\mathrm{n}_{\tt H}.     
  \label{eq:alpha7} 
\end{align}

 \begin{align}   
 \sum_{\ta\in \Lambda^*(i)} \delaC(i,0,[\ta]^\co) = 1,  
  &&  i\in [1,\tC],  
  \label{eq:alpha_last} 
\end{align}   

   \newpage
\subsection{Constraints for Bounds on the Number of Bonds}  
\label{sec:BDbond}

We include constraints for specification of lower and upper bounds
$\bd_\LB$ and $\bd_\UB$.

\bigskip
\noindent
{\bf constants for chemical specification $\sab$: } \\ 
~~ 
$\bd_{m, \LB}(i), \bd_{m, \UB}(i)\in [0,\cs_\UB]$, 
$i\in [1,\mC]$,  $m\in [2,3]$:  lower and upper bounds  \\
~~~~~ on the number of edges in $E(P_i)$ with bond-multiplicity $m$; \\

\noindent
{\bf variables for chemical specification $\sab$: } \\ 
~~ $\bdT(k,i,m)\in [0,1]$, $k\in [1, \kC]$, $i\in [2,\tT]$, $m\in [2,3]$: \\
~~~~~~ 
  $\bdT(k,i,m)=1$  $\Leftrightarrow$ path $P_k$ contains edge $\eT_i$ and
  $\beta(\eT_i)=m$; \\

\noindent
{\bf constraints: } 
\begin{align}    
\bd_{m,\LB}(i)\leq \delbC(i,m)\leq \bd_{m,\UB}(i), 
  i\in \Iew\cup \Iz, m\in [2,3], && 
  \label{eq:BDbond_first}  
\end{align}

\begin{align}   
\bdT(k,i,m)\geq \delbT(i,m)+\chiT(i,k)-1, 
~~~ k \in [1, \kC], i\in [2,\tT],  m\in [2,3], && 
 \label{eq:BDbond2}  
\end{align}

\begin{align}    
\sum_{j\in[2,\tT]}\delbT(j,m) \geq 
\sum_{k\in[1, \kC], i\in [2,\tT]}\!\!\!\! \bdT(k,i,m) , 
~~ m\in [2,3],  \label{eq:BDbond3}  
\end{align}    

\begin{align}    
 \bd_{m, \LB}(k) \leq 
   \sum_{i\in [2,\tT]}\bdT(k,i,m) +\delb^+(k,m)+\delb^-(k,m)    
   \leq   \bd_{m, \UB}(k), ~~~~  \notag \\
    k\in [1, \kC],   m\in [2,3]. ~~ 
     \label{eq:BDbond_last}  
\end{align}


  \newpage

\subsection{Descriptor for the Number of  Adjacency-configurations}  
\label{sec:AC}

We call a tuple $(\ta,\tb,m)\in \Lambda\times \Lambda\times[1,3]$
an {\em adjacency-configuration}.
The adjacency-configuration of an edge-configuration
$(\mu=\ta d, \mu'=\tb d', m)$ is defined to be
 $(\ta,\tb,m)$.
We include constraints to compute the frequency of each adjacency-configuration
in an inferred chemical graph $G$.

 \medskip
\noindent
{\bf constants: } \\ 
~~ Sets $\Gamma^\co$, $\Gamma^\inn$ and $\Gamma^\ex$ 
of edge-configurations,
where $\mu\leq \xi$ \\
~~~~ for any edge-configuration $\gamma=(\mu,\xi,m)
\in  \Gamma^\co$;\\
~~
Let  $\overline{\gamma}$ of an edge-configuration $\gamma=(\mu,\xi,m)$
denote the  edge-configuration $(\mu,\xi,m)$; \\
~~ Let $\Gamma_{<}^\co=\{(\mu,\xi,m)\in  \Gamma^\co\mid \mu < \xi \}$, 
$\Gamma_{=}^\co=\{(\mu,\xi,m)\in  \Gamma^\co\mid \mu= \xi \}$
and  \\
~~~~ $\Gamma_{>}^\co=\{\overline{\gamma}\mid 
    \gamma\in  \Gamma_{<}^\co  \}$;\\
~~ 
Let  \[\mbox{
$\Gac^\co$, $\Gac^\inn$, $\Gac^\ex$ 
$\Gacs^\co$, $\Gace^\co$ and $\Gacl^\co$ } \]
~~
 denote  the sets of the adjacency-configurations of
edge-configurations in the sets \\
~~ 
$\Gamma^\co$, $\Gamma^\inn$ , $\Gamma^\ex$ 
$\Gamma_{<}^\co$, $\Gamma_{=}^\co$ and $\Gamma_{>}^\co$
 respectively;\\
~~
Let  $\overline{\nu}$ of an adjacency-configuration $\nu=(\ta, \tb,m)$
denote the  adjacency-configuration $(\tb,\ta,m)$; \\
~~
 Prepare a coding of each of the three sets 
$\Gac^\co \cup \Gacl^\co$,
$\Gac^\inn$ and $\Gac^\ex$ and let 
$[\nu]^\co$ \\
~~ 
(resp., $[\nu]^\inn$ and $[\nu]^\ex$)  denote  
the coded integer of  an element $\nu$ in $\Gac^\co \cup \Gacl^\co$ \\
~~~~
(resp., $\Gac^\inn$ and $\Gac^\ex$); \\  
~~\\  
~~ 
Choose subsets \\
~~~~ $\tGacC,\tGacT,\tGacCT,\tGacTC \subseteq \Gac^\co\cup\Gacl^\co$; ~ 
$\tGacF, \tGacCF , \tGacTF \subseteq\Gac^\inn $;  ~
$\tGacex \subseteq \Gac^\ex $: 
 To compute the frequency \\ 
~~~~  of adjacency-configurations exactly,  set\\
~~~~
  $\tGacC:= \tGacT :=\tGacCT:=  \tGacTC :=\Gac^\co\cup\Gacl^\co$;
$\tGacF:=  \tGacCF := \tGacTF := \Gac^\inn $;  
$\tGacex :=\Gac^\ex $; \\
~~ \\
%
~~ $\ac_\LB^\co(\nu),  \ac_\UB^\co(\nu) \in [0,\mUB ], \nu\in \Gac^\co$, \\
~~ 
$\ac_\LB^\inn(\nu), \ac_\UB^\inn(\nu) \in [0, n^* ], \nu\in \Gac^\inn$, \\
~~
$\ac_\LB^\ex(\nu), \ac_\UB^\ex(\nu) \in [0, n^* ], \nu\in \Gac^\ex$:  \\
~~~~~ 
lower and upper bounds on the number of core-edges  $e=uv$ \\
 ~~~~~ (resp.,  ${\rho}$-internal edges and
 ${\rho}$-external edges  $e=(u,v)$) with $\alpha(u)=\ta$, 
 $\alpha(v)=\tb$ and $\beta(e)=m$; \\

\noindent
{\bf variables: } \\    
~~
$\ac^\co([\nu]^\co) \in [\ac_\LB^\co(\nu)\ac_\UB^\co(\nu)], \nu\in \Gac^\co$, \\
~~ 
$\ac^\inn([\nu]^\inn)\in [\ac_\LB^\inn(\nu),\ac_\UB^\inn(\nu)],
  \nu\in \Gac^\inn$,\\
~~ 
$\ac^\ex([\nu]^\ex)\in [\ac_\LB^\ex(\nu),\ac_\UB^\ex(\nu)],  \nu\in \Gac^\ex$:\\ 
~~~~~ 
the number of core-edges (resp., ${\rho}$-internal edges, 
${\rho}$-external edges and non-core-edges) \\
~~~~~  with  adjacency-configuration $\nu$; \\
~~
$\acC([\nu]^\co)\in [0,\mC],  \nu\in \tGacC$, 
$\acT([\nu]^\co)\in [0,\tT],   \nu\in \tGacT$, 
$\acF([\nu]^\inn)\in [0,\tF], \nu\in \tGacF$: \\
~~~~~ 
the number of core-edges $\eC\in \EC$ (resp.,  core-edges $\eT\in \ET$
and ${\rho}$-internal edges $\eF\in \EF$)  \\
~~~~~ with  adjacency-configuration $\nu$; \\
~~
$\acXp([\nu]^\ex)\in [1,\nX],   \nu\in \tGacex$,
$p\in[1,{\rho}],  \mathrm{X}\in \{\mathrm{C,T,F}\}$: 
the number of  ${\rho}$-external edges  \\
~~~~~ with depth $p$ and
adjacency-configuration $\nu$ in the ${\rho}$-fringe-tree in $X_i$; \\  
~~ 
$\acCT([\nu]^\co)\in [0, \min\{\kC,\tT\} ], \nu\in \tGacCT$,
$\acTC([\nu]^\co)\in [0,\min\{\kC,\tT\} ], \nu\in \tGacCT$, \\
~~ 
$\acCF([\nu]^\inn)\in [0,\widetilde{\tC}],  \nu\in \tGacCF$,
$\acTF([\nu]^\inn)\in [0,\tT], \nu\in \tGacTF$: 
the number of core-edges \\
~~~~~  $\eCT\in \ECT$  
(resp.,  core-edges $\eTC\in \ETC$
and ${\rho}$-internal edges $\eCF\in \ECF$ and $\eTF\in \ETF$)\\
~~~~~   with  adjacency-configuration $\nu$; \\ 
%
~~
$\dlacC(i,[\nu]^\co)\in [0,1], 
  i\in [\widetilde{\kC}+1,\mC]=\Iw\cup \Iz\cup\Iew, \nu\in \tGacC$, \\
~~
$\dlacT(i,[\nu]^\co)\in [0,1],  i\in [2,\tT],   \nu\in \tGacT$, \\
~~
$\dlacF(i,[\nu]^\inn)\in [0,1] , i\in [2,\tF],\nu\in \tGacF$:\\
~~~~~ $\dlacX(i,[\nu]^\typ)=1$  $\Leftrightarrow$
edge  $\eX_i$ has  adjacency-configuration $\nu$; \\ 
~~
$\dlacX(i,j,[\nu]^\ex)\in [0,1],   i\in [1,\tX], j\in[1,\nX], \nu\in \tGacex$,
 $\mathrm{X}\in \{\mathrm{C,T,F}\}$:\\
~~~~~ 
$\dlacX(i,j,[\nu]^\ex)=1$  $\Leftrightarrow$
${\rho}$-external edge  $\eX_{i,j}$ has  adjacency-configuration $\nu$; \\
~~ 
$\dlacCT(k,[\nu]^\co),\dlacTC(k,[\nu]^\co)\in [0,1],
k\in [1, \kC]=\It\cup \Iw,  \nu\in \tGacCT$:\\
~~~~~ 
$\dlacCT(k,[\nu]^\co)=1$   (resp., $\dlacTC(k,[\nu]^\co)=1$)  $\Leftrightarrow$
edge  $\eCT_{\tailC(k),j}$ (resp.,  $\eTC_{\hdC(k),j}$) \\
~~~~~   for some $j\in [1,\tT]$ has  adjacency-configuration $\nu$; \\
~~
$\dlacCF(c,[\nu]^\inn)\in [0,1],  c\in [1,\widetilde{\tC}],\nu\in \tGacCF$:\\
~~~~~ 
$\dlacCF(c,[\nu]^\inn)=1$    $\Leftrightarrow$
edge   $\eCF_{c,i}$  for some $i\in [1,\tF]$ has  adjacency-configuration $\nu$; \\
~~
  $\dlacTF(i,[\nu]^\inn)\in [0,1],  i\in [1,\tT],  \nu\in \tGacTF$:\\
~~~~~    $\dlacTF(i,[\nu]^\inn)=1$  $\Leftrightarrow$
edge   $\eTF_{i,j}$
 for some $j\in [1,\tF]$ has  adjacency-configuration $\nu$; \\
%
~~\\
~~
$\aCT(k),\aTC(k)\in [0, |\Lambda^\co|],   k\in [1, \kC]$: 
$\alpha(v)$  of the edge $(\vC_{\mathrm{tail}(k)},v)\in \ECT$ \\
~~~~~~ 
 (resp., $(v,\vC_{\mathrm{head}(k)})\in \ETC$) if any; \\
~~
$\aCF(c)\in [0, |\Lambda^\nc|], c\in [1,\widetilde{\tC}]$: 
 $\alpha(v)$  of
the edge $(\vC_{c,0},v)\in \ECF$   if any; \\
~~
$\aTF(i)\in [0, |\Lambda^\nc|], i\in [1,\tT]$: 
 $\alpha(v)$  of
the edge $(\vT_{i,0},v)\in \ETF$   if any; \\
~~ \\
~~
$\DlacCp(i),  \DlacCm(i), \in [0,|\Lambda^\co|], 
  i\in [\widetilde{\kC}+1,\mC]$, \\
~~
$\DlacTp(i),\DlacTm(i)\in [0,|\Lambda^\co|],  i\in [2,\tT]$, \\
~~
$\DlacFp(i),\DlacFm(i)\in [0,|\Lambda^\nc|] , i\in [2,\tF]$:\\
~~~~~ 
$\DlacXp(i)=\DlacXm(i)=0$ (resp., 
 $\DlacXp(i)=\alpha(u)$ and $\DlacXm(i)=\alpha(v)$) $\Leftrightarrow$
edge  $\eX_i=(u,v)\in \EX$ \\
~~~~~  is used in $G$ (resp., $\eX_i\not\in E(G)$); \\
~~
$\DlacXp(i,j),\DlacXm(i,j)\in [0,|\Lambda^\nc|], 
  i\in [1,\tX], j\in[1,\nX],  \mathrm{X}\in \{\mathrm{C,T,F}\}$:\\
~~~~~
$\DlacXp(i,j)=\DlacXm(i,j)=0$ (resp., 
 $\DlacXp(i,j)=\alpha(u)$ and $\DlacXm(i,j)=\alpha(v)$) $\Leftrightarrow$ \\
~~~~~ 
${\rho}$-external edge  $\eX_{i,j}=(u,v)$ is used in $G$
 (resp., $\eX_{i,j}\not\in E(G)$); \\
~~ 
$\DlacCTp(k),\DlacCTm(k)\in [0,|\Lambda^\co|],
k\in [1, \kC]=\It\cup \Iw$:\\
~~~~~ 
$\DlacCTp(k)=\DlacCTm(k) =0$ 
(resp.,  $\DlacCTp(k)=\alpha(u)$ and $\DlacCTm(k)=\alpha(v)$) 
 $\Leftrightarrow$ \\
~~~~~ 
edge  $\eCT_{\tailC(k),j}=(u,v)\in \ECT$   
  for some $j\in [1,\tT]$ is used in $G$ (resp., otherwise); \\
~~
$\DlacTCp(k),\DlacTCm(k)\in [0,|\Lambda^\co|],
k\in [1, \kC]=\It\cup \Iw$: 
Analogous with $\DlacCTp(k)$ and $\DlacCTm(k)$;\\
~~
$\DlacCFp(c)\in [0,|\Lambda^\co|],
\DlacCFm(c) \in [0,|\Lambda^\nc|],  c\in [1,\widetilde{\tC}]$:\\
~~~~~ 
$\DlacCFp(c)=\DlacCFm(c) =0$ (resp., 
 $\DlacCFp(c)=\alpha(u)$ and $\DlacCFm(c)=\alpha(v)$) 
 $\Leftrightarrow$ \\
~~~~~ 
edge  $\eCF_{c,i}=(u,v)\in \ECF$  
   for some $i\in [1,\tF]$ is used in $G$ (resp., otherwise); \\
~~
  $\DlacTFp(i)\in [0,|\Lambda^\co|],
  \DlacTFm(i)\in [0,|\Lambda^\nc|],  i\in [1,\tT]$:
  Analogous with $\DlacCFp(c)$ and $\DlacCFm(c)$;\\ 

\noindent
{\bf constraints: } 

\begin{align} 
 \acC([\nu]^\co) =0,  &&  \nu \in \Gac^\co\setminus \tGacC , \notag \\
 \acT([\nu]^\co) =0,  &&  \nu \in \Gac^\co\setminus \tGacT , \notag \\
 \acF([\nu]^\inn) =0,  &&  \nu \in \Gac^\inn\setminus \tGacF , \notag \\
 \acXp([\nu]^\ex) =0,  &&  \nu \in \Gac^\ex\setminus \tGacex , \notag \\
  &&  p\in[1,{\rho}],  \mathrm{X}\in \{\mathrm{C,T,F}\},\notag \\
 \acCT([\nu]^\co) =0,  &&  \nu \in \Gac^\co\setminus \tGacCT , \notag \\
 \acTC([\nu]^\co) =0,  &&  \nu \in \Gac^\co\setminus \tGacTC , \notag \\
 \acCF([\nu]^\inn) =0,  &&  \nu \in \Gac^\inn\setminus \tGacCF , \notag \\
 \acTF([\nu]^\inn) =0,  &&  \nu \in \Gac^\inn\setminus \tGacTF , \notag \\
 \label{eq:AC_first} 
\end{align}    

\begin{align} 
 \sum_{(\ta, \tb,m)=\nu\in  \Gac^\co}\acC([\nu]^\co) 
      =\sum_{i\in [\widetilde{\kC}+1,\mC]}\delbC(i,m),  &&   m\in [1,3]   , \notag \\
 \sum_{(\ta, \tb,m)=\nu\in  \Gac^\co}\acT([\nu]^\co) 
      =\sum_{i\in [2,\tT]}\delbT(i,m) ,  &&   m\in [1,3] , \notag \\
 \sum_{(\ta, \tb,m)=\nu\in \Gac^\inn}\acF([\nu]^\inn)
      =\sum_{i\in [2,\tF]}\delbF(i,m) ,  &&   m\in [1,3]  , \notag \\
 \sum_{(\ta, \tb,m)=\nu\in \Gac^\ex}\acXp([\nu]^\ex)
     =\sum_{i\in [1,\tX], j\in\DsnX(p)} \delbX(i,j,m),  &&   \notag \\
  &&  p\in[1,{\rho}],  \mathrm{X}\in \{\mathrm{C,T,F}\}, m\in [1,3] ,\notag \\
 \sum_{(\ta, \tb,m)=\nu\in \Gac^\co}\acCT([\nu]^\co)
     =\sum_{k\in [1, \kC]} \delb^+(k,m),  &&   m\in [1,3]  , \notag \\
 \sum_{(\ta, \tb,m)=\nu\in \Gac^\co}\acTC([\nu]^\co)
    =\sum_{k\in [1, \kC]} \delb^-(k,m),  &&   m\in [1,3]  , \notag \\
 \sum_{(\ta, \tb,m)=\nu\in \Gac^\inn}\acCF([\nu]^\inn)
    =\sum_{c\in [1,\widetilde{\tC}]} \delb^\inn(c,m),  &&   m\in [1,3]  , \notag \\
 \sum_{(\ta, \tb,m)=\nu\in \Gac^\inn}\acTF([\nu]^\inn) 
    =\sum_{c\in [\widetilde{\tC}+1, \cF]} \delb^\inn(c,m),  &&   m\in [1,3]  , \notag \\
 \label{eq:AC_first2} 
\end{align}    

\begin{align} 
\sum_{\nu=(\ta,\tb,m) \in \tGacC }\!\!\!\! m\cdot \dlacC(i, [\nu]^\co) 
=\bC(i), && \notag  \\  
\DlacCp(i) +\sum_{\nu=(\ta,\tb,m) \in \tGacC }\!\!\!\! [\ta]^\co\dlacC(i, [\nu]^\co) 
=\aC(\tailC(i),0),  && \notag  \\
\DlacCm(i) +\sum_{\nu=(\ta,\tb,m) \in \tGacC }\!\!\!\! [\tb]^\co\dlacC(i, [\nu]^\co) 
=\aC(\hdC(i),0),  &&\notag  \\
\DlacCp(i)+\DlacCm(i) \leq 2|\Lambda^\co|(1 - \eC(i)),
&& i\in [\widetilde{\kC}+1,\mC],  \notag  \\
\sum_{i\in [\widetilde{\kC}+1,\mC]}\!\!\!\! \dlacC(i, [\nu]^\co) =\acC([\nu]^\co),  
&&  \nu \in \tGacC ,   \label{eq:AC1}   
\end{align}

\begin{align} 
\sum_{\nu=(\ta,\tb,m) \in \tGacT }\!\!\!\! m\cdot \dlacT(i, [\nu]^\co) 
=\bT(i), && \notag  \\  
\DlacTp(i) +\sum_{\nu=(\ta,\tb,m) \in \tGacT }\!\!\!\! [\ta]^\co\dlacT(i, [\nu]^\co) 
=\aT(i-1,0),  && \notag  \\
\DlacTm(i) +\sum_{\nu=(\ta,\tb,m) \in \tGacT }\!\!\!\! [\tb]^\co\dlacT(i, [\nu]^\co) 
=\aT(i,0),  &&\notag  \\
\DlacTp(i)+\DlacTm(i) \leq 2|\Lambda^\co|(1 - \eT(i)),
&& i\in [2,\tT],   \notag  \\
 \sum_{ i\in [2,\tT]} \!\! \dlacT(i, [\nu]^\co) =\acT([\nu]^\co),    
&& \nu \in \tGacT ,    \label{eq:AC2} 
\end{align}    

\begin{align} 
\sum_{\nu=(\ta,\tb,m) \in \tGacF }\!\!\!\! m\cdot \dlacF(i, [\nu]^\inn) 
=\bF(i), && \notag  \\  
\DlacFp(i) +\sum_{\nu=(\ta,\tb,m) \in \tGacF }\!\!\!\! [\ta]^\nc\dlacF(i, [\nu]^\inn) 
=\aF(i-1,0),  && \notag  \\
\DlacFm(i) +\sum_{\nu=(\ta,\tb,m) \in \tGacF }\!\!\!\! [\tb]^\nc\dlacF(i, [\nu]^\inn) 
=\aF(i,0),  &&\notag  \\
\DlacFp(i)+\DlacFm(i) \leq 2|\Lambda^\nc|(1 - \eF(i)),
&& i\in [2,\tF],   \notag  \\
  \sum_{ i\in [2,\tF]} \!\! \dlacF(i, [\nu]^\inn) =\acF([\nu]^\inn),  
 &&  \nu \in \tGacF ,    \label{eq:AC3} 
\end{align}

\begin{align} 
\sum_{\nu=(\ta,\tb,m) \in \tGacex }\!\!\!\! m\cdot \dlacX(i,j, [\nu]^\ex) 
=\bX(i,j), && \notag  \\  
\DlacXp(i,j) +\sum_{\nu=(\ta,\tb,m) \in \tGacex }\!\!\!\! [\ta]^\nc\dlacX(i,j, [\nu]^\ex) 
=\aX(i,\prtX(j)),  && \notag  \\
\DlacXm(i,j) +\sum_{\nu=(\ta,\tb,m) \in \tGacex }\!\!\!\! [\tb]^\nc\dlacX(i,j, [\nu]^\ex) 
=\aX(i,j),  &&\notag  \\
\DlacXp(i,j)+\DlacXm(i,j) \leq 2|\Lambda^\nc|(1 - \vX(i,j)),
&&  i\in [1,\tX],  j\in[1,\nX], \notag  \\
\sum_{ i\in [1,\tX],j\in\DsnX(p)}\!\!\!\!  \!\! \dlacX(i, j,[\nu]^\ex) 
=\acXp([\nu]^\ex),  
  &&   \nu \in \tGacex ,  p\in[1,{\rho}],  \notag  \\
  &&    \mathrm{X}\in \{\mathrm{C,T,F}\},    \label{eq:AC4} 
\end{align}

\begin{align} 
 \aT(i,0)+|\Lambda^\co|(1-\chiT(i,k)+\eT(i))\geq \aCT(k),  && \notag  \\  
\aCT(k)\geq \aT(i,0)- |\Lambda^\co|(1-\chiT(i,k)+\eT(i)), 
&&  i\in [1,\tT],     \notag  \\  
\sum_{\nu=(\ta,\tb,m) \in \tGacCT }\!\!\!\! m\cdot \dlacCT(k, [\nu]^\co) 
=\beta^{+}(k), && \notag  \\  
\DlacCTp(k) +\sum_{\nu=(\ta,\tb,m) \in \tGacCT }\!\!\!\! [\ta]^\co\dlacCT(k, [\nu]^\co) 
=\aC(\tailC(k),0),  && \notag  \\
\DlacCTm(k) +\sum_{\nu=(\ta,\tb,m) \in \tGacCT }\!\!\!\! [\tb]^\co\dlacCT(k, [\nu]^\co) 
=  \aCT(k),  &&\notag  \\
\DlacCTp(k)+\DlacCTm(k) \leq 2|\Lambda^\co|(1 - \dclrT(k)),
&& k\in [1, \kC],  \notag  \\
\sum_{k\in [1, \kC]}\!\! \dlacCT(k, [\nu]^\co) =\acCT([\nu]^\co),  
 && \nu \in  \tGacCT ,  \label{eq:AC5} 
\end{align}

\begin{align} 
 \aT(i,0)+|\Lambda^\co|(1-\chiT(i,k)+\eT(i+1))\geq \aTC(k),   && \notag  \\  
\aTC(k)\geq \aT(i,0)- |\Lambda^\co|(1-\chiT(i,k)+\eT(i+1)), 
&&  i\in [1,\tT],    \notag  \\  
\sum_{\nu=(\ta,\tb,m) \in \tGacTC }\!\!\!\! m\cdot \dlacTC(k, [\nu]^\co) 
=\beta^{-}(k), && \notag  \\  
\DlacTCp(k) +\sum_{\nu=(\ta,\tb,m) \in \tGacTC }\!\!\!\! [\ta]^\co\dlacTC(k, [\nu]^\co) 
=  \aTC(k),  &&\notag  \\
\DlacTCm(k) +\sum_{\nu=(\ta,\tb,m) \in \tGacTC }\!\!\!\! [\tb]^\co\dlacTC(k, [\nu]^\co) 
=\aC(\hdC(k),0),  &&   \notag  \\
\DlacTCp(k)+\DlacTCm(k) \leq 2|\Lambda^\co|(1 - \dclrT(k)), && k\in [1, \kC],  \notag  \\  
\sum_{k\in [1, \kC]}\!\! \dlacTC(k, [\nu]^\co) =\acTC([\nu]^\co),  
 && \nu \in  \tGacTC ,  \label{eq:AC5} 
\end{align}

\begin{align} 
\aF(i,0)+|\Lambda^\nc|(1-\chiF(i,c)+\eF(i))\geq \aCF(c),  && \notag  \\  
\aCF(c)\geq \aF(i,0)- |\Lambda^\nc|(1-\chiF(i,c)+\eF(i)), 
   &&   i\in [1,\tF], \notag  \\   
\sum_{\nu=(\ta,\tb,m) \in \tGacCF }\!\!\!\! m\cdot \dlacCF(c, [\nu]^\inn) 
=\beta^\inn(c), && \notag  \\  
\DlacCFp(c) +\sum_{\nu=(\ta,\tb,m) \in \tGacCF }\!\!\!\! [\ta]^\co\dlacCF(c, [\nu]^\inn) 
= \aC(\hdC(c),0),  &&\notag  \\
\DlacCFm(c) +\sum_{\nu=(\ta,\tb,m) \in \tGacCF }\!\!\!\! [\tb]^\nc\dlacCF(c, [\nu]^\inn) 
=\aCF(c) ,   &&\notag  \\
\DlacCFp(c)+\DlacCFm(c) \leq 2\max\{|\Lambda^\co|,|\Lambda^\nc|\}(1 - \dclrF(c)), 
&& c\in [1,\widetilde{\tC}],  \notag  \\
\sum_{c\in [1,\widetilde{\tC}]}\!\! \dlacCF(c, [\nu]^\inn) =\acCF([\nu]^\inn),  
 && \nu \in  \tGacCF ,  \label{eq:AC6} 
\end{align}

\begin{align} 
\aF(j,0)+|\Lambda^\nc|(1-\chiF(j,i+\widetilde{\tC})+\eF(j))\geq \aTF(i), 
   && \notag  \\  
\aTF(i)\geq \aF(j,0)- |\Lambda^\nc|(1-\chiF(j,i+\widetilde{\tC})+\eF(j)), 
  &&  j\in [1,\tF],  \notag  \\   
\sum_{\nu=(\ta,\tb,m) \in \tGacTF }\!\!\!\! m\cdot \dlacTF(i, [\nu]^\inn) 
=\beta^\inn(i+\widetilde{\tC}), && \notag  \\  
\DlacTFp(i) +\sum_{\nu=(\ta,\tb,m) \in \tGacTF }\!\!\!\! [\ta]^\co\dlacTF(i, [\nu]^\inn) 
= \aT(i,0),  &&\notag  \\
\DlacTFm(i) +\sum_{\nu=(\ta,\tb,m) \in \tGacTF }\!\!\!\! [\tb]^\nc\dlacTF(i, [\nu]^\inn) 
=\aTF(i) ,   
&& \notag  \\
\DlacTFp(i)+\DlacTFm(i) \leq 2\max\{|\Lambda^\co|,|\Lambda^\nc|\}
(1 - \dclrF(i+\widetilde{\tC})), 
&& i\in [1,\tT],  \notag  \\
\sum_{i\in [1,\tT]}\!\! \dlacTF(i, [\nu]^\inn) =\acTF([\nu]^\inn),  
 && \nu \in  \tGacTF ,  \label{eq:AC5} 
\end{align}

\begin{align} 
\sum_{\mathrm{X}\in\{\mathrm{C,T,CT,TC}\}}(\acX([\nu]^\co)+\acX([\overline{\nu}]^\co)) 
 =\ac^\co([\nu]^\co) , && \nu\in \Gacs^\co,  \notag \\  
\sum_{\mathrm{X}\in\{\mathrm{C,T,CT,TC}\}} \acX([\nu]^\co)
 =\ac^\co([\nu]^\co) , && \nu\in \Gace^\co,  \notag \\ 
\sum_{\mathrm{X}\in\{\mathrm{F,CF,TF}\}}\acX([\nu]^\co) =\ac^\inn([\nu]^\co), 
 &&  \nu\in \Gac^\inn,  \notag \\
\sum_{p\in[1,{\rho}],    \mathrm{X}\in\{\mathrm{T,C,F}\}}
 \acXp([\nu]^\co) = \ac^\ex([\nu]^\co),  
&&  \nu\in \Gac^\ex.
   \label{eq:AC_last} 
\end{align}

  \newpage

\subsection{Descriptor for the Number of Chemical Symbols}  
\label{sec:CS}

We include constraints for computing
 the frequency of each chemical symbol in $\Ldg$.
 Let $\cs(v)$ denote the chemical symbol of a vertex $v$ in 
 a chemical graph $G$ to be inferred; i.e.,
 $\cs(v)=\mu=\ta d\in \Ldg$ such that $\alpha(v)=\ta$ and
 $\deg_G(v)=d$. 

 \medskip
\noindent
{\bf constants: } \\ 
~~ Sets $\Ldg^\co$ and $\Ldg^\nc$ of chemical symbols;\\ 
~~ 
 Prepare a coding of each of the two sets 
$\Ldg^\co$  and $\Ldg^\nc$ and let $[\mu]^\co$  
(resp., $[\mu]^\nc$)  denote  \\
~~~~~ 
the coded integer of  an element $\mu \in \Ldg^\co$
(resp., $\Ldg^\nc$); \\  
~~\\  
~~ 
Choose subsets \\
~~~~ $\tLdgC, \tLdgT  \subseteq \Ldg^\co$; 
 $\tLdgF,\tLdgCnc,\tLdgTnc,\tLdgFnc \subseteq \Ldg^\nc$:  \\ 
~~~~  
 To compute the frequency of chemical symbols exactly,  set\\
~~~~ $\tLdgC:= \tLdgT := \Ldg^\co$; 
 $\tLdgF := \tLdgCnc:=\tLdgTnc:=\tLdgFnc:=  \Ldg^\nc$; \\
~~ \\

\noindent
{\bf variables: } \\ 
 ~~  $\ns^\co([\mu]^\co )\in[0,\cs_\UB]$,  $\mu\in \Ldg^\co$: 
      the number of core-vertices $v$  with $\cs(v)=\mu$; \\
 ~~  $\ns^\nc([\mu]^\nc)\in[0,n^*-\cs_\LB]$,  $\mu\in \Ldg^\nc$: 
 the number of non-core-vertices $v$   with $\cs(v)=\mu$; \\
~~
   $\dlnsX(i,0,[\mu]^\co)\in [0,1]$, $ i\in [1,\tX], j\in [0,\nX],\mu\in \Ldg^\co$, 
   $\mathrm{X}\in \{\mathrm{C,T}\}$, \\
~~
   $\dlnsF(i,0,[\mu]^\nc)\in [0,1], i\in [1,\tF], \mu\in \Ldg^\nc$, \\
~~ 
   $\dlnsX(i,j,[\mu]^\nc)\in [0,1]$, $ i\in [1,\tX], j\in [1,\nX]$   
   $~(\nC\!=\! \nC(i))$,  $\mu=\ta d \in \Ldg^\nc$, 
  $\mathrm{X}\in \{\mathrm{C,T,F}\}$: \\
~~~~~~    $\dlnsX(i,j,[\mu]^\nc)=1$   $\Leftrightarrow$
    $\alpha(\vX_{i,j})=\ta$ and  $\deg_G(\vX_{i,j})=d$; \\

\noindent
{\bf constraints: } 
\begin{align}  
   \sum_{\mu\in \tLdgX\cup\{\epsilon\} } \dlnsX(i,0,[\mu]^\co)=1, ~~ 
   \sum_{\mu=\ta d\in \tLdgX }[\ta]^\co\cdot\dlnsX(i,0,[\mu]^\co)=\aX(i,0), 
   \notag \\
   \sum_{\mu=\ta d\in \tLdgX }d\cdot\dlnsX(i,0,[\mu]^\co)=\degX(i,0), 
    \hspace{2cm}\notag \\ 
  ~~   i\in [1,\tX],   
  \mathrm{X}\in \{\mathrm{C,T}\},  
  \label{eq:CS_first} 
\end{align}

\begin{align}  
   \sum_{\mu\in \tLdgF\cup\{\epsilon\} } \dlnsF(i,0,[\mu]^\nc)=1, ~~ 
   \sum_{\mu=\ta d\in \tLdgF }[\ta]^\nc\cdot\dlnsF(i,0,[\mu]^\nc)=\aF(i,0), 
   \notag \\
   \sum_{\mu=\ta d\in \tLdgF }d\cdot\dlnsF(i,0,[\mu]^\nc)=\degF(i,0), 
    \hspace{2cm}\notag \\ 
  ~~   i\in [1,\tF],    
  \label{eq:CS2} 
\end{align}

\begin{align}  
   \sum_{\mu\in \tLdgXnc\cup\{\epsilon\} } \dlnsX(i,j,[\mu]^\nc)=1, ~~ 
   \sum_{\mu=\ta d\in \tLdgXnc }[\ta]^\nc\cdot\dlnsX(i,j,[\mu]^\nc)=\aX(i,j), 
   \notag \\
   \sum_{\mu=\ta d\in \tLdgXnc }d\cdot\dlnsX(i,j,[\mu]^\nc)=\degX(i,j), 
    \hspace{2cm}\notag \\ 
  ~~   i\in [1,\tX], j\in [1,\nX]~(\nC\!=\! \nC(i)),  
  \mathrm{X}\in \{\mathrm{C,T,F}\},  
  \label{eq:CS3} 
\end{align}    

\begin{align}  
   \sum_{i\in [1,\tC]} \dlnsC(i,0,[\mu]^\co)
   +    \sum_{i\in [1,\tT]} \dlnsT(i,0,[\mu]^\co) =\ns^\co([\mu]^\co),
    && \mu\in \Ldg^\co, \notag\\    
   \sum_{i\in [1,\tF]} \dlnsF(i,0,[\mu]^\nc)
  +    \sum \limits_{\substack{  i\in [1,\tX], j\in [1,\nX], \\
               \mathrm{X}\in \{\mathrm{C,T,F}\}  }}                  
     \dlnsX(i,j,[\mu]^\nc) =\ns^\nc([\mu]^\nc),
   &&  \mu\in \Ldg^\nc.
  \label{eq:CS_last} 
\end{align}

  \newpage

\subsection{Descriptor for the Number of Edge-configurations}  
\label{sec:EC}

We include constraints to compute the frequency of each edge-configuration
in an inferred chemical graph $G$.

 \medskip
\noindent
{\bf constants: } \\ 
~~ Sets $\Gamma^\co$, $\Gamma^\inn$ and $\Gamma^\ex$ 
of edge-configurations,
where $\mu\leq \xi$ \\
~~~~ for any edge-configuration $\gamma=(\mu,\xi,m)
\in  \Gamma^\co$;\\
~~ Let $\Gamma_{<}^\co=\{(\mu,\xi,m)\in  \Gamma^\co\mid \mu < \xi \}$, 
$\Gamma_{=}^\co=\{(\mu,\xi,m)\in  \Gamma^\co\mid \mu= \xi \}$
and  \\
~~~~ $\Gamma_{>}^\co=\{(\xi,\mu,m)\mid 
    (\mu,\xi,m)\in  \Gamma_{<}^\co  \}$;\\
~~ 
 Prepare a coding of each of the three sets 
$\Gamma^\co \cup \Gamma_{>}^\co$,
$\Gamma^\inn$ and $\Gamma^\ex$ and let 
$[\gamma]^\co$ \\
~~ 
(resp., $[\gamma]^\inn$ and $[\gamma]^\ex$)  denote  
the coded integer of  an element $\gamma$ in $\Gamma^\co \cup \Gamma_{>}^\co$
(resp., $\Gamma^\inn$ and $\Gamma^\ex$); \\  
~~\\  
~~ 
Choose subsets \\
~~~~ $\tGecC,\tGecT,\tGecCT,\tGecTC \subseteq \Gamma^\co\cup\Gamma_{>}^\co$; ~ 
$\tGecF, \tGecCF , \tGecTF \subseteq\Gamma^\inn $;  ~
$\tGecex \subseteq \Gamma^\ex $: 
 To compute the frequency \\ 
~~~~  of edge-configurations exactly,  set\\
~~~~
  $\tGecC:= \tGecT :=\tGecCT:=  \tGecTC :=\Gamma^\co\cup\Gamma_{>}^\co$;
$\tGecF:=  \tGecCF := \tGecTF := \Gamma^\inn $;  
$\tGecex :=\Gamma^\ex $; \\
~~ \\
~~ $\ec_\LB^\co(\gamma),  \ec_\UB^\co(\gamma) \in [0,\mUB ], \gamma\in \Gamma^\co$, \\
~~ $\ec_\LB^\inn(\gamma), \ec_\UB^\inn(\gamma) \in [0, n^* ], \gamma\in \Gamma^\inn$, \\
~~
$\ec_\LB^\ex(\gamma), \ec_\UB^\ex(\gamma)\in [0, n^* ], \gamma\in \Gamma^\ex$:  \\
~~~~~ 
lower and upper bounds on the number of core-edges  $e=uv$ \\
 ~~~~~   (resp.,  ${\rho}$-internal edges and
 ${\rho}$-external edges  $e=(u,v)$)  with $\alpha(u)=\ta$, 
 $\alpha(v)=\tb$ and $\beta(e)=m$; \\

\noindent
{\bf variables: } \\   
~~
$\ec^\co([\gamma]^\co) \in [\ec_\LB^\co(\gamma),\ec_\UB^\co(\gamma)], 
\gamma\in \Gamma^\co$,\\
~~  
$\ec^\inn([\gamma]^\inn)\in [\ec_\LB^\inn(\gamma),\ec_\UB^\inn(\gamma)],
  \gamma\in \Gamma^\inn$,\\
~~ 
$\ec^\ex([\gamma]^\ex)\in [\ec_\LB^\ex(\gamma),\ec_\UB^\ex(\gamma)],  
\gamma\in \Gamma^\ex$:\\ 
~~~~~ 
the number of core-edges (resp., ${\rho}$-internal edges, 
${\rho}$-external edges and non-core-edges) \\
~~~~~  with  edge-configuration $\gamma$; \\
~~
$\ecC([\gamma]^\co)\in [0,\mC],  \gamma\in \tGecC$, 
$\ecT([\gamma]^\co)\in [0,\tT],   \gamma\in \tGecT$, 
$\ecF([\gamma]^\inn)\in [0,\tF], \gamma\in \tGecF$: \\
~~~~~ 
the number of core-edges $\eC\in \EC$ (resp.,  core-edges $\eT\in \ET$
and ${\rho}$-internal edges $\eF\in \EF$)  \\
~~~~~ with  edge-configuration $\gamma$; \\ 
~~
$\ecXp([\gamma]^\ex)\in [1,\nX],   \gamma\in \tGecex$,
$p\in[1,{\rho}],  \mathrm{X}\in \{\mathrm{C,T,F}\}$:\\
~~~~~ 
the number of  ${\rho}$-external edges with depth $p$ in a rooted tree $X_i$; \\ 
~~ 
$\ecCT([\gamma]^\co)\in [0, \min\{\kC,\tT\} ], \gamma\in \tGecCT$,
$\ecTC([\gamma]^\co)\in [0,\min\{\kC,\tT\} ], \gamma\in \tGecCT$, \\
~~ 
$\ecCF([\gamma]^\inn)\in [0,\widetilde{\tC}],  \gamma\in \tGecCF$,
$\ecTF([\gamma]^\inn)\in [0,\tT], \gamma\in \tGecTF$: 
the number of core-edges $\eCT\in \ECT$  \\
~~~~~   (resp.,  core-edges $\eTC\in \ETC$
and ${\rho}$-internal edges $\eCF\in \ECF$ and $\eTF\in \ETF$)\\
~~~~~   with  edge-configuration $\gamma$; \\ 
~~
$\dlecC(i,[\gamma]^\co)\in [0,1], 
  i\in [\widetilde{\kC}+1,\mC]=\Iw\cup \Iz\cup\Iew, \gamma\in \tGecC$, \\
~~
$\dlecT(i,[\gamma]^\co)\in [0,1],  i\in [2,\tT],   \gamma\in \tGecT$, \\
~~
$\dlecF(i,[\gamma]^\inn)\in [0,1] , i\in [2,\tF],\gamma\in \tGecF$:\\
~~~~~ 
$\dlecX(i,[\gamma]^\typ)=1$  $\Leftrightarrow$
edge  $\eX_i$ has  edge-configuration $\gamma$; \\
~~
$\dlecX(i,j,[\gamma]^\ex)\in [0,1],   i\in [1,\tX], j\in[1\nX], \gamma\in \tGecex$,
 $\mathrm{X}\in \{\mathrm{C,T,F}\}$:\\
~~~~~ 
$\dlecX(i,j,[\gamma]^\ex)=1$  $\Leftrightarrow$
${\rho}$-external edge  $\eX_{i,j}$ has  edge-configuration $\gamma$; \\
~~
$\dlecCT(k,[\gamma]^\co),\dlecTC(k,[\gamma]^\co)\in [0,1],
k\in [1, \kC]=\It\cup \Iw,  \gamma\in \tGecCT$:\\
~~~~~ 
$\dlecCT(k,[\gamma]^\co)=1$   (resp., $\dlecTC(k,[\gamma]^\co)=1$)  $\Leftrightarrow$
edge  $\eCT_{\tailC(k),j}$ (resp.,  $\eTC_{\hdC(k),j}$) \\
~~~~~   for some $j\in [1,\tT]$ has  edge-configuration $\gamma$; \\
~~
$\dlecCF(c,[\gamma]^\inn)\in [0,1],  c\in [1,\widetilde{\tC}],\gamma\in \tGecCF$:\\
~~~~~ 
$\dlecCF(c,[\gamma]^\inn)=1$    $\Leftrightarrow$
edge   $\eCF_{c,i}$  for some $i\in [1,\tF]$ has  edge-configuration $\gamma$; \\
~~ 
  $\dlecTF(i,[\gamma]^\inn)\in [0,1],  i\in [1,\tT],  \gamma\in \tGecTF$:\\
~~~~~  $\dlecTF(i,[\gamma]^\inn)=1$) $\Leftrightarrow$
edge     $\eTF_{i,j}$ for some $j\in [1,\tF]$ has  edge-configuration $\gamma$; \\
~~ \\
~~
$\degCTT(k),\degTCT(k)\in [0, 4],   k\in [1, \kC]$: 
$\deg_G(v)$  of
the edge $(\vC_{\mathrm{tail}(k)},v)\in \ECT$ \\
~~~~~ 
 (resp., $(v,\vC_{\mathrm{head}(k)})\in \ETC$)  if any; \\
~~
$\degCFF(c)\in [0, 4], c\in [1,\widetilde{\tC}]$: 
 $\deg_G(v)$  of
the edge $(\vC_{c,0},v)\in \ECF$   if any; \\
~~
$\degTFF(i)\in [0, 4], i\in [1,\tT]$: 
 $\deg_G(v)$  of
the edge $(\vT_{i,0},v)\in \ETF$   if any; \\ 
~~ \\
~~
$\DlecCp(i),  \DlecCm(i), \in [0,4], 
  i\in [\widetilde{\kC}+1,\mC]$, \\
~~
$\DlecTp(i),\DlecTm(i)\in [0,4],  i\in [2,\tT]$, \\
~~
$\DlecFp(i),\DlecFm(i)\in [0,4] , i\in [2,\tF]$:\\
~~~~~ 
$\DlecXp(i)=\DlecXm(i)=0$ (resp., 
 $\DlecXp(i)=\deg_G(u)$
  and $\DlecXm(i)=\deg_G(v)$) $\Leftrightarrow$ \\
~~~~~ 
edge  $\eX_i=(u,v)\in \EX$  is used in $G$ (resp., $\eX_i\not\in E(G)$); \\
~~
$\DlecXp(i,j),\DlecXm(i,j)\in [0,4], 
  i\in [1,\tX], j\in[1,\nX],  \mathrm{X}\in \{\mathrm{C,T,F}\}$:\\
~~~~~
$\DlecXp(i,j)=\DlecXm(i,j)=0$ (resp., 
 $\DlecXp(i,j)=\deg_G(u)$ 
 and $\DlecXm(i,j)=\deg_G(v)$) $\Leftrightarrow$ \\
~~~~~ 
${\rho}$-external edge  $\eX_{i,j}=(u,v)$ is used in $G$
 (resp., $\eX_{i,j}\not\in E(G)$); \\
~~ 
$\DlecCTp(k),\DlecCTm(k)\in [0,4],
k\in [1, \kC]=\It\cup \Iw$:\\
~~~~~ 
$\DlecCTp(k)=\DlecCTm(k) =0$ 
(resp.,  $\DlecCTp(k)=\deg_G(u)$
 and $\DlecCTm(k)=\deg_G(v)$) 
 $\Leftrightarrow$ \\
~~~~~ 
edge  $\eCT_{\tailC(k),j}=(u,v)\in \ECT$   
  for some $j\in [1,\tT]$ is used in $G$ (resp., otherwise); \\
~~
$\DlecTCp(k),\DlecTCm(k)\in [0,4],
k\in [1, \kC]=\It\cup \Iw$: 
Analogous with $\DlecCTp(k)$ and $\DlecCTm(k)$;\\
~~
$\DlacCFp(c), \DlecCFm(c) \in [0,4],  c\in [1,\widetilde{\tC}]$:\\
~~~~~ 
$\DlecCFp(c)=\DlecCFm(c) =0$ (resp., 
 $\DlecCFp(c)=\deg_G(u)$ 
 and $\DlecCFm(c)=\deg_G(v)$) 
 $\Leftrightarrow$ \\
~~~~~ 
edge  $\eCF_{c,j}=(u,v)\in \ECF$  
   for some $j\in [1,\tF]$ is used in $G$ (resp., otherwise); \\
~~
  $\DlecTFp(i),  \DlecTFm(i)\in [0,4],  i\in [1,\tT]$:
  Analogous with $\DlecCFp(c)$ and $\DlecCFm(c)$;\\

\noindent
{\bf constraints: } 

\begin{align} 
 \ecC([\gamma]^\co) =0,  &&  \gamma \in \Gamma^\co\setminus \tGecC , \notag \\
 \ecT([\gamma]^\co) =0,  &&  \gamma \in \Gamma^\co\setminus \tGecT , \notag \\
 \ecF([\gamma]^\inn) =0,  &&  \gamma \in \Gamma^\inn\setminus \tGecF , \notag \\
 \ecXp([\gamma]^\ex) =0,  &&  \gamma \in \Gamma^\ex\setminus \tGecex , \notag \\
  &&  p\in[1,{\rho}],  \mathrm{X}\in \{\mathrm{C,T,F}\},\notag \\
 \ecCT([\gamma]^\co) =0,  &&  \gamma \in \Gamma^\co\setminus \tGecCT , \notag \\
 \ecTC([\gamma]^\co) =0,  &&  \gamma \in \Gamma^\co\setminus \tGecTC , \notag \\
 \ecCF([\gamma]^\inn) =0,  &&  \gamma \in \Gamma^\inn\setminus \tGecCF , \notag \\
 \ecTF([\gamma]^\inn) =0,  &&  \gamma \in \Gamma^\inn\setminus \tGecTF , \notag \\
 \label{eq:EC_first} 
\end{align}

\begin{align} 
 \sum_{(\mu, \mu',m)=\gamma\in  \Gamma^\co}\ecC([\gamma]^\co) 
      =\sum_{i\in [\widetilde{\kC}+1,\mC]}\delbC(i,m),  &&   m\in [1,3]   , \notag \\
 \sum_{(\mu, \mu',m)=\gamma\in  \Gamma^\co}\ecT([\gamma]^\co) 
      =\sum_{i\in [2,\tT]}\delbT(i,m) ,  &&   m\in [1,3] , \notag \\
 \sum_{(\mu, \mu',m)=\gamma\in \Gamma^\inn}\ecF([\gamma]^\inn)
      =\sum_{i\in [2,\tF]}\delbF(i,m) ,  &&   m\in [1,3]  , \notag \\
 \sum_{(\mu, \mu',m)=\gamma\in \Gamma^\ex}\ecXp([\gamma]^\ex)
     =\sum_{i\in [1,\tX], j\in\DsnX(p)} \delbX(i,j,m),  &&  \notag \\
  &&  p\in[1,{\rho}],  \mathrm{X}\in \{\mathrm{C,T,F}\},  m\in [1,3] ,\notag \\
 \sum_{(\mu, \mu',m)=\gamma\in \Gamma^\co}\ecCT([\gamma]^\co)
     =\sum_{k\in [1, \kC]} \delb^+(k,m),  &&   m\in [1,3]  , \notag \\
 \sum_{(\mu, \mu',m)=\gamma\in \Gamma^\co}\ecTC([\gamma]^\co)
    =\sum_{k\in [1, \kC]} \delb^-(k,m),  &&   m\in [1,3]  , \notag \\
 \sum_{(\mu, \mu',m)=\gamma\in \Gamma^\inn}\ecCF([\gamma]^\inn)
    =\sum_{c\in [1,\widetilde{\tC}]} \delb^\inn(c,m),  &&   m\in [1,3]  , \notag \\
 \sum_{(\mu, \mu',m)=\gamma\in \Gamma^\inn}\ecTF([\gamma]^\inn) 
    =\sum_{c\in [\widetilde{\tC}+1, \cF]} \delb^\inn(c,m),  &&   m\in [1,3]  , \notag \\
 \label{eq:EC_first2} 
\end{align}

\begin{align}  
\sum_{\gamma=(\ta d,\tb d',m) \in \tGecC }\!\!\!\! [(\ta,\tb,m)]^\co\cdot \dlecC(i, [\gamma]^\co) 
= \sum_{\nu \in \tGacC } [\nu]^\co\cdot \dlacC(i, [\nu]^\co) , && \notag  \\  
\DlecCp(i) +\sum_{\gamma=(\ta d,\xi,m) \in \tGecC }\!\!\!\! 
  d\cdot \dlecC(i, [\gamma]^\co) 
=\degC(\tailC(i),0),  && \notag  \\
\DlecCm(i) +\sum_{\gamma=(\mu,\tb d,m) \in \tGecC }\!\!\!\!
  d\cdot\dlecC(i, [\gamma]^\co) 
= \degC(\hdC(i),0),  &&\notag  \\
\DlecCp(i)+\DlecCm(i) \leq 8(1 - \eC(i)),
&& i\in [\widetilde{\kC}+1,\mC],  \notag  \\
\sum_{i\in [\widetilde{\kC}+1,\mC]}\!\!\!\! \dlecC(i, [\gamma]^\co) =\ecC([\gamma]^\co),  
&&  \gamma \in \tGecC ,   \label{eq:EC1}   
\end{align}

\begin{align} 
\sum_{\gamma=(\ta d,\tb d',m) \in \tGecT }\!\!\!\! [(\ta,\tb,m)]^\co\cdot \dlecT(i, [\gamma]^\co) 
= \sum_{\nu \in \tGacT} [\nu]^\co\cdot \dlacT(i, [\nu]^\co) , && \notag  \\  
\DlecTp(i) +\sum_{\gamma=(\ta d,\xi,m) \in \tGecT }\!\!\!\!
  d\cdot  \dlecT(i, [\gamma]^\co) 
   =\degT(i-1 ,0),  && \notag  \\
\DlecTm(i) +\sum_{\gamma=(\mu,\tb d,m) \in \tGecT }\!\!\!\! 
 d\cdot \dlecT(i, [\gamma]^\co) 
 =\degT(i,0),     &&\notag  \\
\DlecTp(i)+\DlecTm(i) \leq 8(1 - \eT(i)),
&& i\in [2,\tT],   \notag  \\
 \sum_{ i\in [2,\tT]} \!\! \dlecT(i, [\gamma]^\co) =\ecT([\gamma]^\co),    
&& \gamma \in \tGecT ,    \label{eq:EC2} 
\end{align}     

\begin{align} 
\sum_{\gamma=(\ta d,\tb d',m) \in \tGecF }\!\!\!\! 
[(\ta,\tb,m)]^\inn\cdot \dlecF(i, [\gamma]^\inn) 
= \sum_{\nu \in \tGacF } [\nu]^\inn\cdot \dlacF(i, [\nu]^\inn) , && \notag  \\  
\DlecFp(i) +\sum_{\gamma=(\ta d,\xi,m) \in \tGecF }\!\!\!\! 
  d\cdot \dlecF(i, [\gamma]^\inn) 
=\degF(i-1 ,0),    && \notag  \\
\DlecFm(i) +\sum_{\gamma=(\mu,\tb d, m) \in \tGecF }\!\!\!\! 
 d\cdot  \dlecF(i, [\gamma]^\inn) 
=\degF(i,0 ),    &&\notag  \\
\DlecFp(i)+\DlecFm(i) \leq 8(1 - \eF(i)),
&& i\in [2,\tF],   \notag  \\
  \sum_{ i\in [2,\tF]} \!\! \dlecF(i, [\gamma]^\inn) =\ecF([\gamma]^\inn),  
 &&  \gamma \in \tGecF ,    \label{eq:EC3} 
\end{align}    
  
\begin{align} 
\sum_{\gamma=(\ta d,\tb d',m) \in \tGecex }\!\!\!\! 
[(\ta,\tb,m)]^\ex\cdot \dlecX(i,j, [\gamma]^\ex) 
= \sum_{\nu \in \tGacex } [\nu]^\ex\cdot \dlacX(i,j, [\nu]^\ex) , && \notag  \\  
\DlecXp(i,j) +\sum_{\gamma=(\ta d,\xi,m) \in \tGecex }\!\!\!\! 
 d\cdot \dlecX(i,j, [\gamma]^\ex) 
=\degX(i, \prtX(j)),    && \notag  \\
\DlecXm(i,j) +\sum_{\gamma=(\mu,\tb d,m) \in \tGecex }\!\!\!\! 
  d\cdot \dlecX(i,j, [\gamma]^\ex) 
=\degX(i, j),    && \notag  \\
\DlecXp(i,j)+\DlecXm(i,j) \leq 8(1 - \vX(i,j)),
&&  i\in [1,\tX],  j\in[1,\nX], \notag  \\
\sum_{ i\in [1,\tX],j\in\DsnX(p)}\!\!\!\!  \!\! \dlecX(i, j,[\gamma]^\ex) 
=\ecXp([\gamma]^\ex),  
  &&   \gamma \in \tGecex ,  p\in[1,{\rho}],  \notag  \\
  &&    \mathrm{X}\in \{\mathrm{C,T,F}\},    \label{eq:EC4} 
\end{align}

\begin{align} 
\degT(i,0)+4(1-\chiT(i,k)+\eT(i))\geq \degCTT(k),  
 && \notag  \\  
\degCTT(k)\geq \degT(i,0)- 4(1-\chiT(i,k)+\eT(i)), &&  i\in [1,\tT],    \notag  \\  
\sum_{\gamma=(\ta d,\tb d',m) \in \tGecCT }\!\!\!\! 
[(\ta,\tb,m)]^\co\cdot \dlecCT(k, [\gamma]^\co) 
= \sum_{\nu \in \tGacCT} [\nu]^\co\cdot \dlacCT(k, [\nu]^\co) , && \notag  \\  
\DlecCTp(k) +\sum_{\gamma=(\ta d,\xi,m) \in \tGecCT }\!\!\!\! 
  d\cdot \dlecCT(k, [\gamma]^\co) 
=\degC(\tailC(k),0),    && \notag  \\
\DlecCTm(k) +\sum_{\gamma=(\mu,\tb d, m) \in \tGecCT }\!\!\!\! 
 d\cdot  \dlecCT(k, [\gamma]^\co) 
= \degCTT(k),     &&\notag  \\
\DlecCTp(k)+\DlecCTm(k) \leq 8(1 - \dclrT(k)),
&& k\in [1, \kC],  \notag  \\
\sum_{k\in [1, \kC]}\!\! \dlecCT(k, [\gamma]^\co) =\ecCT([\gamma]^\co),  
 && \gamma \in  \tGecCT ,  \label{eq:EC5} 
\end{align}

\begin{align} 
 \degT(i,0)+4(1-\chiT(i,k)+\eT(i+1))\geq \degTCT(k),  
   && \notag  \\  
\degTCT(k)\geq \degT(i,0)- 4(1-\chiT(i,k)+\eT(i+1)), 
&&  i\in [1,\tT],    \notag  \\  
\sum_{\gamma=(\ta d,\tb d',m) \in \tGecTC }\!\!\!\! 
[(\ta,\tb,m)]^\co\cdot \dlecTC(k, [\gamma]^\co) 
= \sum_{\nu \in \tGacTC} [\nu]^\co\cdot \dlacTC(k, [\nu]^\co) , && \notag  \\  
\DlecTCp(k) +\sum_{\gamma=(\ta d,\xi,m) \in \tGecTC }\!\!\!\! 
  d\cdot \dlecTC(k, [\gamma]^\co) 
= \degTCT(k),     &&\notag  \\
\DlecTCm(k) +\sum_{\gamma=(\mu,\tb d, m) \in \tGecTC }\!\!\!\! 
 d\cdot  \dlecTC(k, [\gamma]^\co) 
=\degC(\hdC(k),0),    && \notag  \\
\DlecTCp(k)+\DlecTCm(k) \leq 8(1 - \dclrT(k)),
&& k\in [1, \kC],  \notag  \\
\sum_{k\in [1, \kC]}\!\! \dlecTC(k, [\gamma]^\co) =\ecTC([\gamma]^\co),  
 && \gamma \in  \tGecTC ,  \label{eq:EC5} 
\end{align}

\begin{align} 
\degF(i,0)+4(1-\chiF(i,c)+\eF(i))\geq \degCFF(c), 
   && \notag  \\  
\degCFF(c)\geq \degF(i,0)- 4(1-\chiF(i,c)+\eF(i)), 
 && i\in [1,\tF],   \notag  \\   
\sum_{\gamma=(\ta d,\tb d',m) \in \tGecCF }\!\!\!\! 
[(\ta,\tb,m)]^\inn\cdot \dlecCF(c, [\gamma]^\inn) 
= \sum_{\nu \in \tGacCF} [\nu]^\inn\cdot \dlacCF(c, [\nu]^\inn) , && \notag  \\  
\DlecCFp(c) +\sum_{\gamma=(\ta d,\xi,m) \in \tGecCF }\!\!\!\! 
  d\cdot \dlecCF(c, [\gamma]^\inn) 
=\degC(c,0),    && \notag  \\
\DlecCFm(c) +\sum_{\gamma=(\mu,\tb d, m) \in \tGecCF }\!\!\!\! 
 d\cdot  \dlecCF(c, [\gamma]^\inn) 
= \degCFF(c),     &&\notag  \\
\DlecCFp(c)+\DlecCFm(c) \leq 8(1 - \dclrF(c)),
&& c\in [1,\widetilde{\tC}],  \notag  \\
\sum_{c\in [1,\widetilde{\tC}]}\!\! \dlecCF(c, [\gamma]^\co) =\ecCF([\gamma]^\co),  
 && \gamma \in  \tGecCF ,  \label{eq:EC6} 
\end{align}

\begin{align} 
\degF(j,0)+4(1-\chiF(j,i+\widetilde{\tC})+\eF(j))\geq \degTFF(i), 
  && \notag  \\  
\degTFF(i)\geq \degF(j,0)- 4(1-\chiF(j,i+\widetilde{\tC})+\eF(j)), 
 && j\in [1,\tF],   \notag  \\ 
\sum_{\gamma=(\ta d,\tb d',m) \in \tGecTF }\!\!\!\! 
[(\ta,\tb,m)]^\inn\cdot \dlecTF(i, [\gamma]^\inn) 
= \sum_{\nu \in \tGacTF} [\nu]^\inn\cdot \dlacTF(i, [\nu]^\inn) , && \notag  \\  
\DlecTFp(i) +\sum_{\gamma=(\ta d,\xi,m) \in \tGecTF }\!\!\!\! 
  d\cdot \dlecTF(i, [\gamma]^\inn) 
=\degT(i,0),    && \notag  \\
\DlecTFm(i) +\sum_{\gamma=(\mu,\tb d, m) \in \tGecTF }\!\!\!\! 
 d\cdot  \dlecTF(i, [\gamma]^\inn) 
= \degTFF(i),     &&\notag  \\
\DlecTFp(i)+\DlecTFm(i) \leq 8(1 - \dclrF(i+\widetilde{\tC})),
&& i\in [1,\tT],  \notag  \\
\sum_{i\in [1,\tT]}\!\! \dlecTF(i, [\gamma]^\co) =\ecTF([\gamma]^\co),  
 && \gamma \in  \tGecTF ,  \label{eq:EC6} 
\end{align}

\begin{align} 
\sum_{\mathrm{X}\in\{\mathrm{C,T,CT,TC}\}}(\ecX([\gamma]^\co)
+\ecX([\overline{\gamma}]^\co)) 
 =\ec^\co([\gamma]^\co) , &&  \gamma\in \Gamma_{<}^\co,  \notag \\  
\sum_{\mathrm{X}\in\{\mathrm{C,T,CT,TC}\}} \ecX([\gamma]^\co) 
 =\ec^\co([\gamma]^\co) , && \gamma\in \Gamma_{=}^\co,  \notag \\ 
\sum_{\mathrm{X}\in\{\mathrm{F,CF,TF}\}}\ecX([\gamma]^\co) =\ec^\inn([\gamma]^\co), 
 && \gamma\in \Gamma^\inn,  \notag \\
\sum_{p\in[1,{\rho}],    \mathrm{X}\in\{\mathrm{T,C,F}\}}
 \ecXp([\gamma]^\co) = \ec^\ex([\gamma]^\co),  
&&  \gamma\in \Gamma^\ex .
   \label{eq:EC_last} 
\end{align}

\end{document}